\documentclass[trackchanges,twocolumn]{aastex7}
\usepackage{graphicx}
\usepackage[table, svgnames, dvipsnames]{xcolor}
\definecolor{darkgreen}{RGB}{0,100,0}
\definecolor{teal}{RGB}{0,128,128}
\definecolor{purple}{RGB}{128,0,128}
\definecolor{burntsienna}{RGB}{233,116,81}

\usepackage{colortbl}
\usepackage{enumitem}
\usepackage{subfigure}
\usepackage{dirtytalk}
\usepackage{multirow}
\usepackage{comment}
\usepackage{caption}

\usepackage{txfonts}

\newcommand{\appropto}{\mathrel{\vcenter{
  \offinterlineskip\halign{\hfil$##$\cr
    \propto\cr\noalign{\kern2pt}\sim\cr\noalign{\kern-2pt}}}}}

\begin{document}

\title{Multiphase gas offsets in the atmospheres of central galaxies and their consequences for SMBH activation \\ I. The hot and warm ionized gas phases}

\author[orcid=0000-0001-5338-4472]{Francesco Ubertosi}
\affiliation{Dipartimento di Fisica e Astronomia, Università di Bologna, via Gobetti 93/2, I-40129 Bologna, Italy}
\affiliation{Istituto Nazionale di Astrofisica - Istituto di Radioastronomia (IRA), via Gobetti 101, I-40129 Bologna, Italy}
\email[show]{francesco.ubertosi2@unibo.it}  

\author[orcid=0000-0001-9807-8479]{Fabrizio Brighenti}
\affiliation{Dipartimento di Fisica e Astronomia, Università di Bologna, via Gobetti 93/2, I-40129 Bologna, Italy}
\affiliation{University of California Observatories/Lick Observatory, Department of Astronomy and Astrophysics, Santa Cruz, CA 95064, USA}
\email{fabrizio.brighenti@unibo.it}  

\author[orcid=0000-0002-5671-6900]{Ewan O'Sullivan}
\affiliation{Center for Astrophysics $|$ Harvard \& Smithsonian, 60 Garden Street, Cambridge, MA 02138, USA}
\email{eosullivan@cfa.harvard.edu}  

\author[orcid=0000-0002-4962-0740]{Gerrit Schellenberger}
\affiliation{Center for Astrophysics $|$ Harvard \& Smithsonian, 60 Garden Street, Cambridge, MA 02138, USA}
\email{gerrit.schellenberger@cfa.harvard.edu} 

\author[orcid=0000-0002-0843-3009]{Myriam Gitti}
\affiliation{Dipartimento di Fisica e Astronomia, Università di Bologna, via Gobetti 93/2, I-40129 Bologna, Italy}
\affiliation{Istituto Nazionale di Astrofisica - Istituto di Radioastronomia (IRA), via Gobetti 101, I-40129 Bologna, Italy}
\email{myriam.gitti@unibo.it}  

\author[orcid=0000-0002-1634-9886]{Simona Giacintucci}
\affiliation{Naval Research Laboratory, 4555 Overlook Avenue SW, Code 7213, Washington, DC 20375, USA}
\email{simona.giacintucci.civ@us.navy.mil}

\author[orcid=0000-0002-8341-342X]{Pasquale Temi}
\affiliation{NASA Ames Research Center, MS 245-6, Moffett Field, CA 94035-1000, USA}
\email{pasquale.temi@nasa.gov}  

\author[orcid=0009-0003-9413-6901]{Laurence P. David}
\affiliation{Center for Astrophysics $|$ Harvard \& Smithsonian, 60 Garden Street, Cambridge, MA 02138, USA}
\email{ldavid@cfa.harvard.edu}

\author[orcid=0009-0007-0318-2814]{Jan Vrtilek}
\affiliation{Center for Astrophysics $|$ Harvard \& Smithsonian, 60 Garden Street, Cambridge, MA 02138, USA}
\email{jvrtilek@cfa.harvard.edu}

\author[orcid=0000-0002-8476-6307]{Tiziana Venturi}
\affil{Istituto Nazionale di Astrofisica - Istituto di Radioastronomia (IRA), via Gobetti 101, I-40129 Bologna, Italy}
\email{tventuri@ira.inaf.it}

\author[orcid=0000-0003-0995-5201]{Elisabetta Liuzzo}
\affil{Istituto Nazionale di Astrofisica - Istituto di Radioastronomia (IRA), via Gobetti 101, I-40129 Bologna, Italy}
\email{liuzzo@ira.inaf.it}

\author[orcid=0000-0002-0375-8330]{Marcella Massardi}
\affil{Istituto Nazionale di Astrofisica - Istituto di Radioastronomia (IRA), via Gobetti 101, I-40129 Bologna, Italy}
\email{massardi@ira.inaf.it}

\author[orcid=0000-0001-7509-2972]{Kamlesh Rajpurohit}
\affiliation{Center for Astrophysics $|$ Harvard \& Smithsonian, 60 Garden Street, Cambridge, MA 02138, USA}
\email{kamlesh.rajpurohit@cfa.harvard.edu}

\begin{abstract}

We investigate the spatial relationships between multi-phase gas components and supermassive black hole (SMBH) activity in a sample of 25 cool core galaxy groups and clusters. Using high angular resolution observations from \textit{Chandra}, VLT/MUSE, and VLBA, we robustly locate the position, respectively, of the X-ray peak of the intracluster medium (ICM), of the H$\alpha$ peak of the warm ionized gas, and of the SMBH radio core on parsec scales. We identify spatial offsets between the X-ray peak of the hot gas and the SMBH in 80\% of the systems, with an average displacement of $\langle\Delta^{\text{SMBH}}_{\text{X-ray}}\rangle = 4.8$~kpc (dispersion of $3.8$~kpc). In contrast, the peak of warm ionized gas traced by H$\alpha$ exhibits much smaller offsets ($\langle\Delta^{\text{SMBH}}_{\text{H}\alpha}\rangle = 0.6$~kpc; dispersion of $1.4$~kpc) and a lower incidence of displacement (15\%). Our findings suggest that hot gas sloshing primarily drives the observed spatial offsets, with AGN-driven uplift  contributing in some systems. 
Importantly, systems with H$\alpha$ -- SMBH offsets of $\geq$1~kpc uniformly lack detectable radio cores on VLBA scales, with upper limits on the 5~GHz power of $P_{5\,\text{GHz}} \leq 10^{21-22}$~W~Hz$^{-1}$, while those without such offsets exhibit radio powerful AGN with pc-scale radio emission up to $P_{5\,\text{GHz}} \sim 10^{24-25}$~W~Hz$^{-1}$. This correlation indicates that centrally concentrated warm gas is critical for sustaining radio-loud SMBH activity, possibly supporting scenarios of cold-mode accretion. Overall, our results highlight the importance of high-angular-resolution, multi-wavelength observations for understanding the interplay between multiphase gas cooling and AGN fueling in central galaxies.

\end{abstract}



\section{Introduction}\label{sec:intro}

A key component of the baryon cycle in galaxy groups and clusters is the gaseous halo of these systems. The so-called intracluster or intragroup medium (ICM/IGrM) can lose energy by thermally radiating in the X-ray band, potentially leading to the condensation of the hot gas into cooler phases. In turn, these phases can support star formation at the center of the system, the cool core (e.g., for reviews, \citealt{2003ARA&A..41..191M,2004cgpc.symp..143D,2022PhR...973....1D}). This process does not proceed undisturbed: the brightest cluster/group galaxy (BCG/BGG) typically hosts a supermassive black hole (SMBH) that, when active, drives powerful jets through the surrounding medium, thereby depositing energy and reducing the efficiency of the cooling process (e.g., for reviews, \citealt{2007ARA&A..45..117M,2012ARA&A..50..455F,2012AdAst2012E...6G,2012NJPh...14e5023M,2020NatAs...4...10G,2021Univ....7..142E}). This active galactic nucleus (AGN) feedback mechanism is vital to explain the long-term evolution of inflow and outflow processes in groups and clusters. 
\par In virialized, perfectly relaxed galaxy groups and clusters, the BCG/BGG is expected to reside at the center of the potential well of its host cluster, coinciding with the peaks of the hot ICM, of the warm ionized gas, and of the cold molecular phase (e.g., \citealt{vanderbosh2005,cui2016}). Indeed, numerous studies have demonstrated that the hot, warm, and molecular gas phases are interconnected over long timescales, as evidenced by, for example, correlations between the hot gas surface brightness, entropy, or cooling time and H$\alpha$ luminosity (e.g., \citealt{sanderson2009,2009ApJS..182...12C}), spatial alignment between the warm and molecular gas phases with the X-ray cooling region (e.g., \citealt{2010ApJ...721.1262M,olivares2019,2019MNRAS.490.3025R,olivares2025}), and relations linking molecular gas mass to X-ray gas mass and H$\alpha$ emission \citep{edge2001,pulido2018,olivares2022}. 
\par However, in some cases the X-ray and H$\alpha$ emission peaks tracing gas cooling are separated from the BCG core \citep{hamer2012,pasini2021,rosignoli2024}. For example, in a sample of 65 X-ray selected clusters, \cite{sanderson2009} found that all of the BCGs/BGGs with warm gas detections (i.e., in cool core systems) are separated from the X-ray peak by less than about 15~kpc (see also \citealt{2010A&A...513A..37H}). The origin of the observed offsets is generally attributed to gravitational disturbances in the cluster potential, often triggered by minor mergers that induce sloshing motions of the gas. Sloshing consists in the gas peak being offset from the bottom of the potential well, and subsequently falling back (e.g., \citealt{ascasibar2006,markevitch2007,zuhone2016}). Ram pressure slows the motion of the gas, thus creating a temporary offset between the galaxy and its gaseous halo, as well as offsets among the different gas phases (e.g., \citealt{million2010,hamer2012}). Another plausible mechanism for displacing the ICM/IGrM from the BCG/BGGs is mechanical uplift by AGN jets, which can distort the otherwise centrally-concentrated configuration of the cool core (e.g., \citealt{kirkpatrick2015,mcnamara2016}).
While AGN-driven outflows are capable of lifting portions of the X-ray emitting gas (e.g., \citealt{werner2010,gitti2011}), 
existing studies suggest that jets are not capable of removing the entire dense, cool gas reservoir from the galaxy \citep{rosignoli2024}.
\par Past studies have explored the consequences of these spatial offsets on ICM/IGrM cooling and on AGN feedback. 
The results indicate that clusters with a larger BCG - X-ray peak separation have a weaker cool core \citep{sanderson2009}; that cooling (traced by the detection of molecular gas) persists in hot and warm gas peaks even when these are offset from the BCG \citep{hamer2012}; and that over long timescales, AGN feedback remains effective despite the temporary displacement of the gaseous atmosphere \citep{pasini2021,rosignoli2024}. 
\par However, the consequences of these spatial offsets on AGN feeding (i.e., the process through which gas sinks at the center of the galaxy and fuels the SMBH activity) remain mostly unexplored. A key implication of a multiphase AGN feeding scenario is that if a particular gas phase directly fuels the AGN, then a sustained spatial offset between this gas phase and the BCG should correspond to a quiescent SMBH. We have thus started a project to explore multi-phase gas offsets in central galaxies, and their connection to the SMBH activation. In this first work (paper I) we focus on the hot and warm ionized gas properties, by combining high spatial resolution radio, X-ray, and optical observations of 25 galaxy groups and clusters. We defer to a future work (paper II, Ubertosi et al. in preparation) the exploration of the cold molecular phase. We define the sample and present our data analysis techniques in Section \ref{sec:sampledata}, and we present the results in Section \ref{sec:results}, focusing first on the X-ray gas offsets (Sec.~\ref{subsec:Xray}), and then on the H$\alpha$ gas offsets (Sec.~\ref{subsec:Halpha}). We discuss our results in the context of the configuration of cool cores (Sec.~\ref{subsec:fragm}), of the timescales of spatial offset formation (Sec.~\ref{subsec:timescales}), and of the connection to the SMBH activation (Sec.~\ref{subsec:connectsmbh}). We then summarize our conclusions in Section~\ref{sec:summary}.
\par Throughout this work, we assume a $\Lambda$CDM cosmology with $H_{0} = 70$~km/s/Mpc, $\Omega_{m} = 0.3$, and $\Omega_{\Lambda} = 0.7$. Uncertainties are reported at 1$\sigma$ unless otherwise stated.


\begin{figure*}[ht!]
    \centering
    \includegraphics[width=0.495\linewidth, trim={0.8cm 0 1cm 0}]{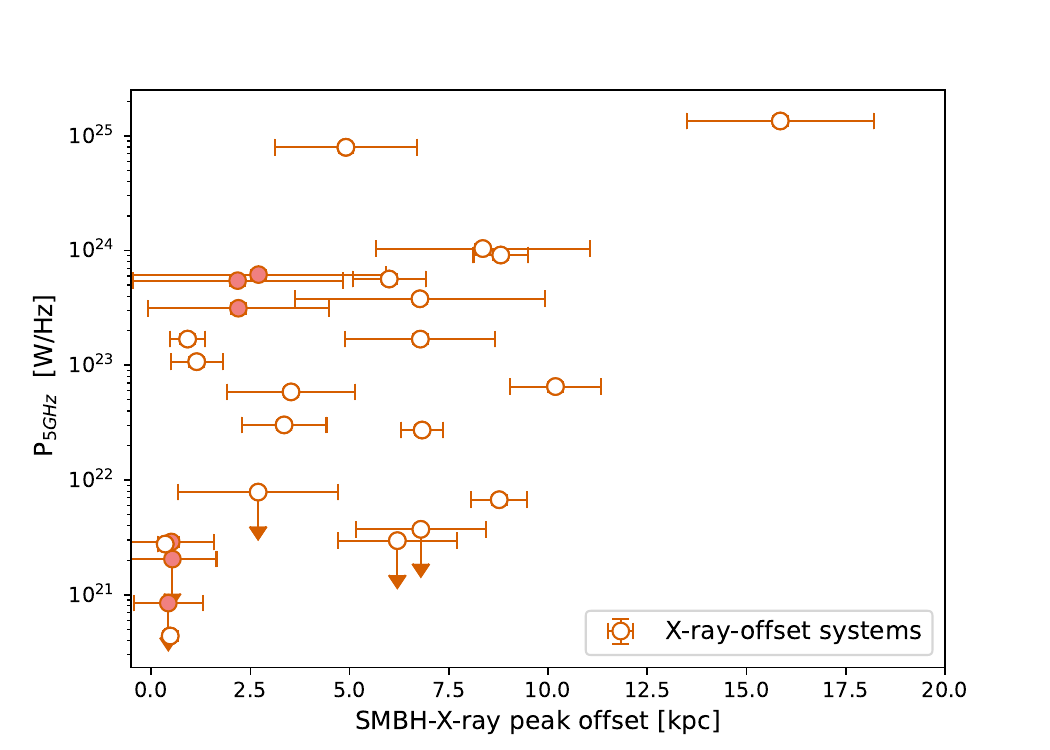}
    \includegraphics[width=0.495\linewidth, trim={0.8cm 0 1cm 0}]{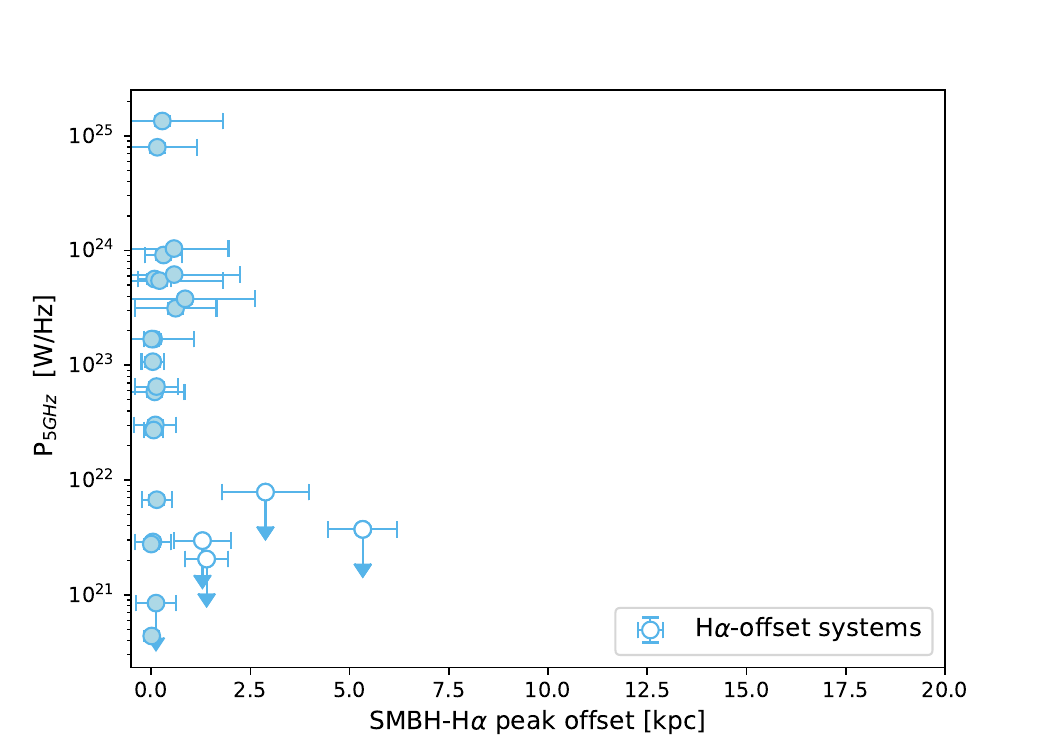}
    \caption{The link between multiphase gas offsets and SMBH activation. Left: Radio power at 5~GHz (from pc-scale VLBA radio observations) of the 25 BCGs in our sample vs the distance between the position of the SMBH and of the hot gas peak (from X-ray emission in \textit{Chandra} data). Right: Radio power at 5~GHz vs the distance between the position of the SMBH and of the warm gas peak (from the H$\alpha$ emission line in VLT/MUSE data). In both panels, empty points represent systems with significant offsets (that is, for which the distance $\Delta$ between the SMBH and the gas peak is larger than the uncertainty $\delta\Delta$), and arrows represent upper limits on the radio power.}
    \label{fig:offset1}
\end{figure*}
\begin{figure*}[ht!]
    \centering
    \includegraphics[width=0.45\linewidth, trim={0.2cm 0 1cm 1cm}]{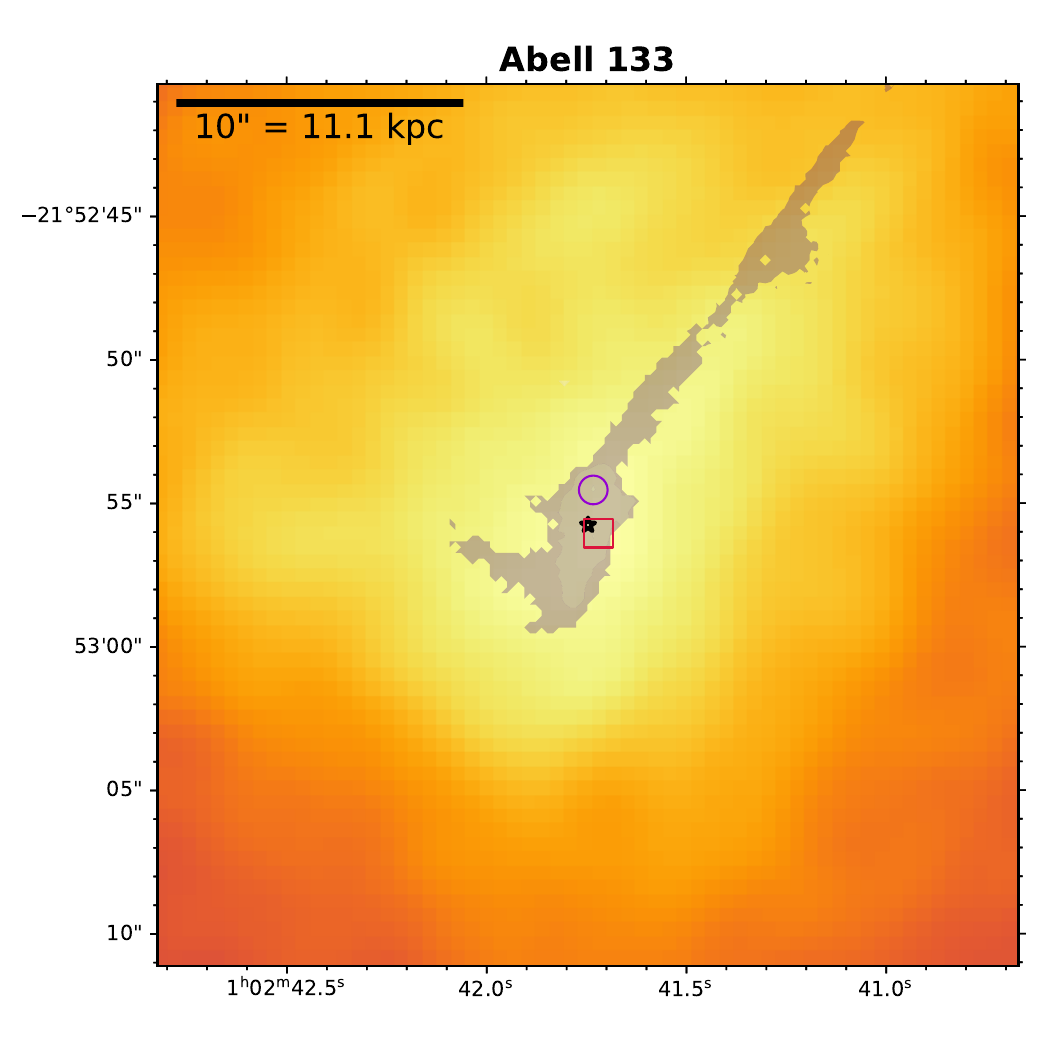}
    \includegraphics[width=0.45\linewidth, trim={0.2cm 0 1cm 1cm}]{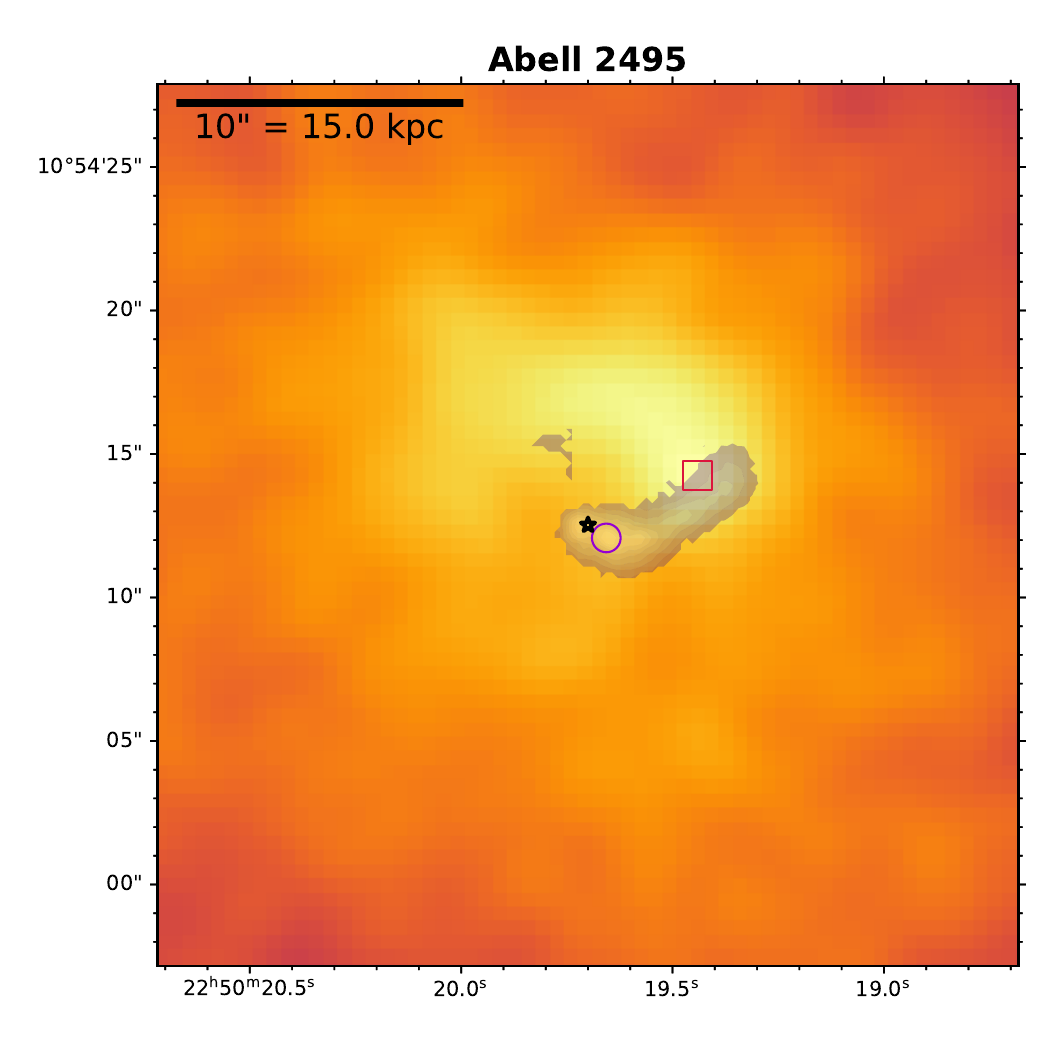}\\
    \includegraphics[width=0.45\linewidth, trim={0.2cm 0 1cm 1.6cm}]{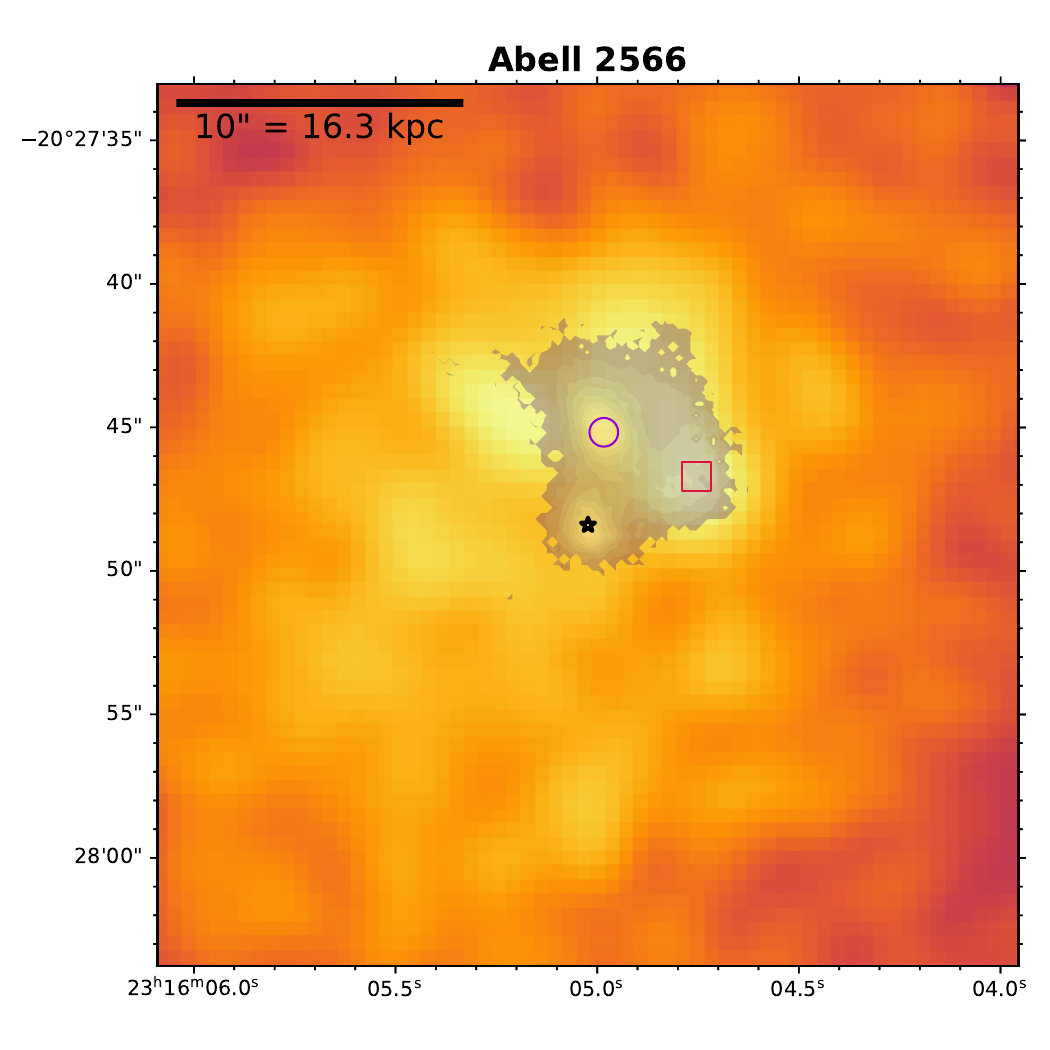}
    \includegraphics[width=0.45\linewidth, trim={0.2cm 0 1cm 1.6cm}]{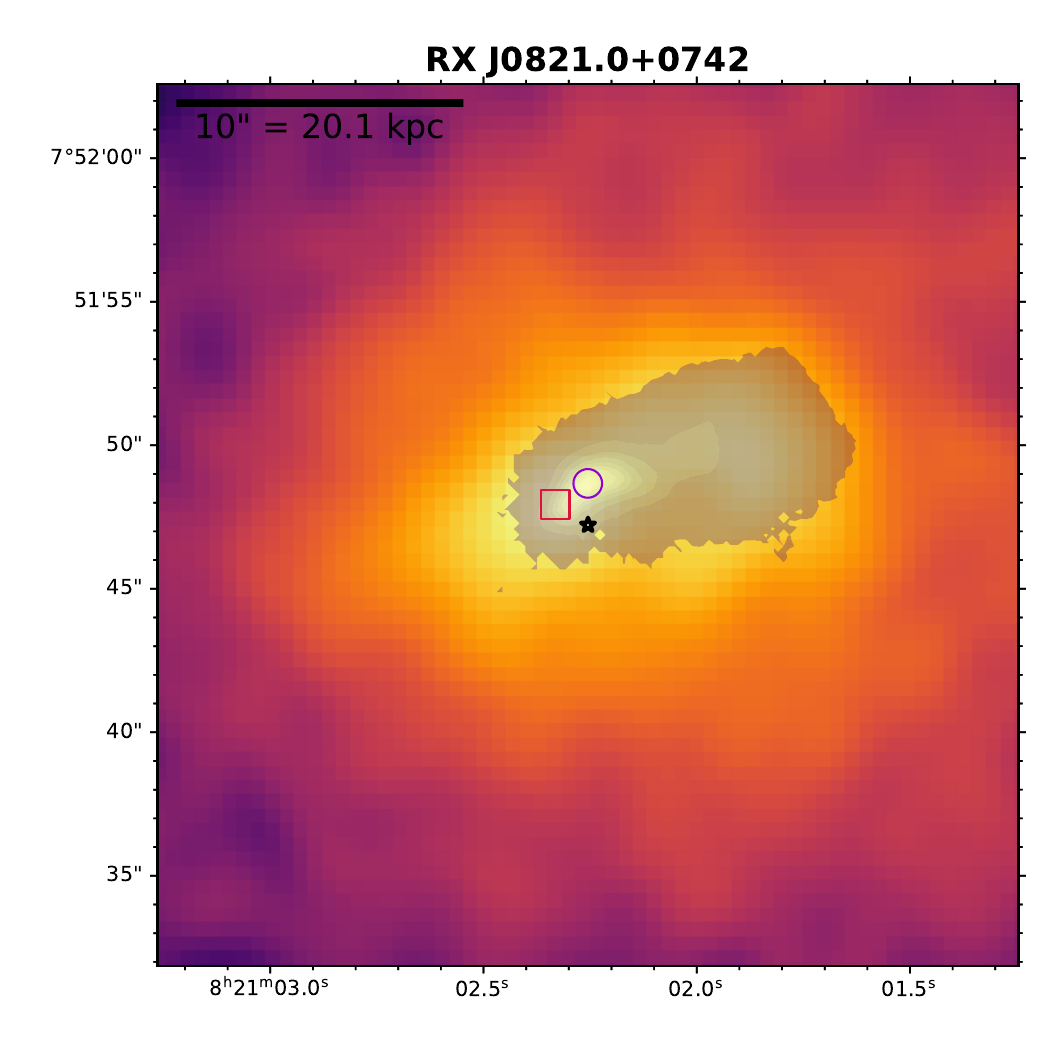}
    \caption{Multiwavelength view of the four systems with a significant H$\alpha$ peak - SMBH offset. The panels show smoothed X-ray {\it Chandra} maps of the hot gas, with grayscale contours of the H$\alpha$ line from MUSE data. The red square marks the position of the X-ray peak, the purple circle shows the location of the H$\alpha$ peak, and the black star shows the position of the SMBH (see Tab.~\ref{tab:completeinfo}).}
    \label{fig:multiexample}
\end{figure*}

\begin{figure*}[ht!]
    \centering
    \includegraphics[width=0.3\linewidth, trim={0.2cm 0 1cm 1.6cm}]{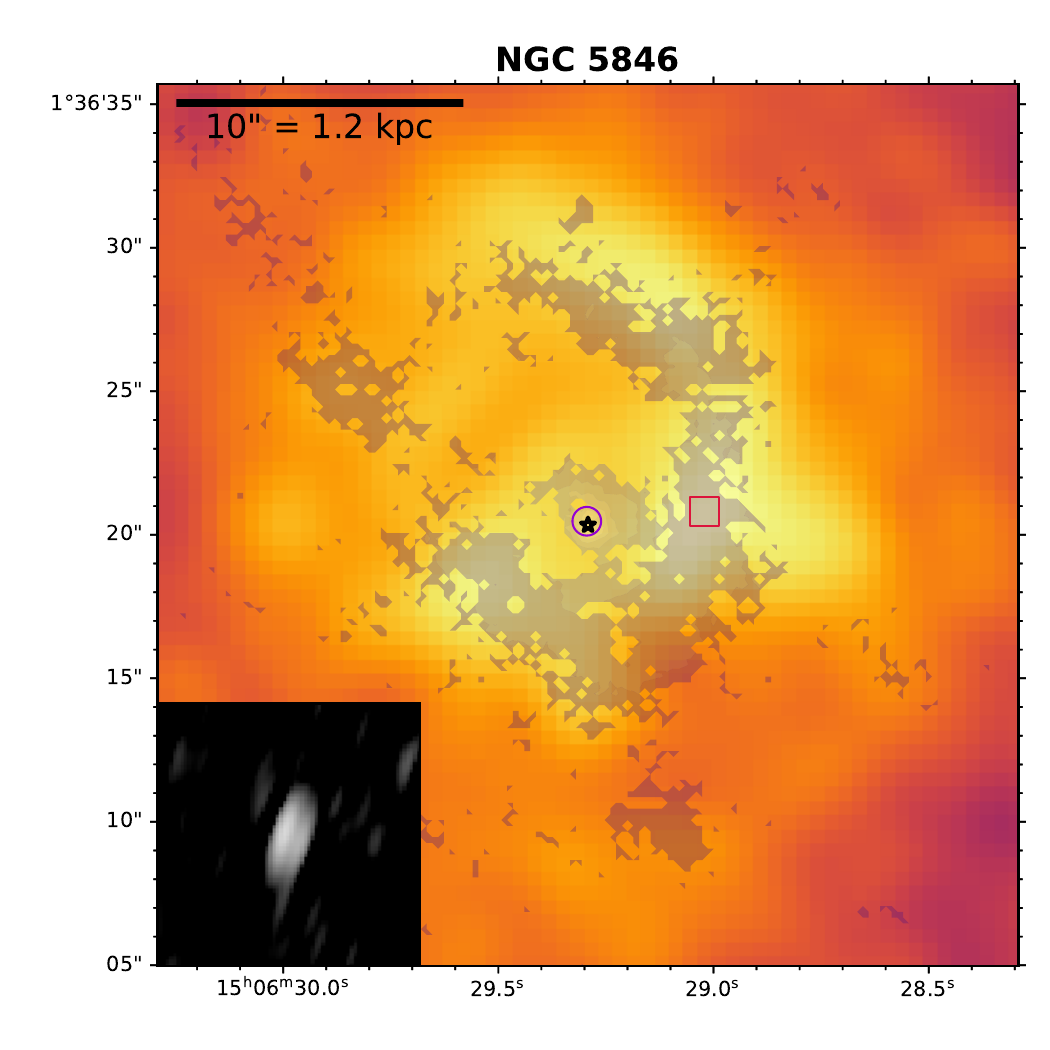}
    \includegraphics[width=0.3\linewidth, trim={0.2cm 0 1cm 1.6cm}]{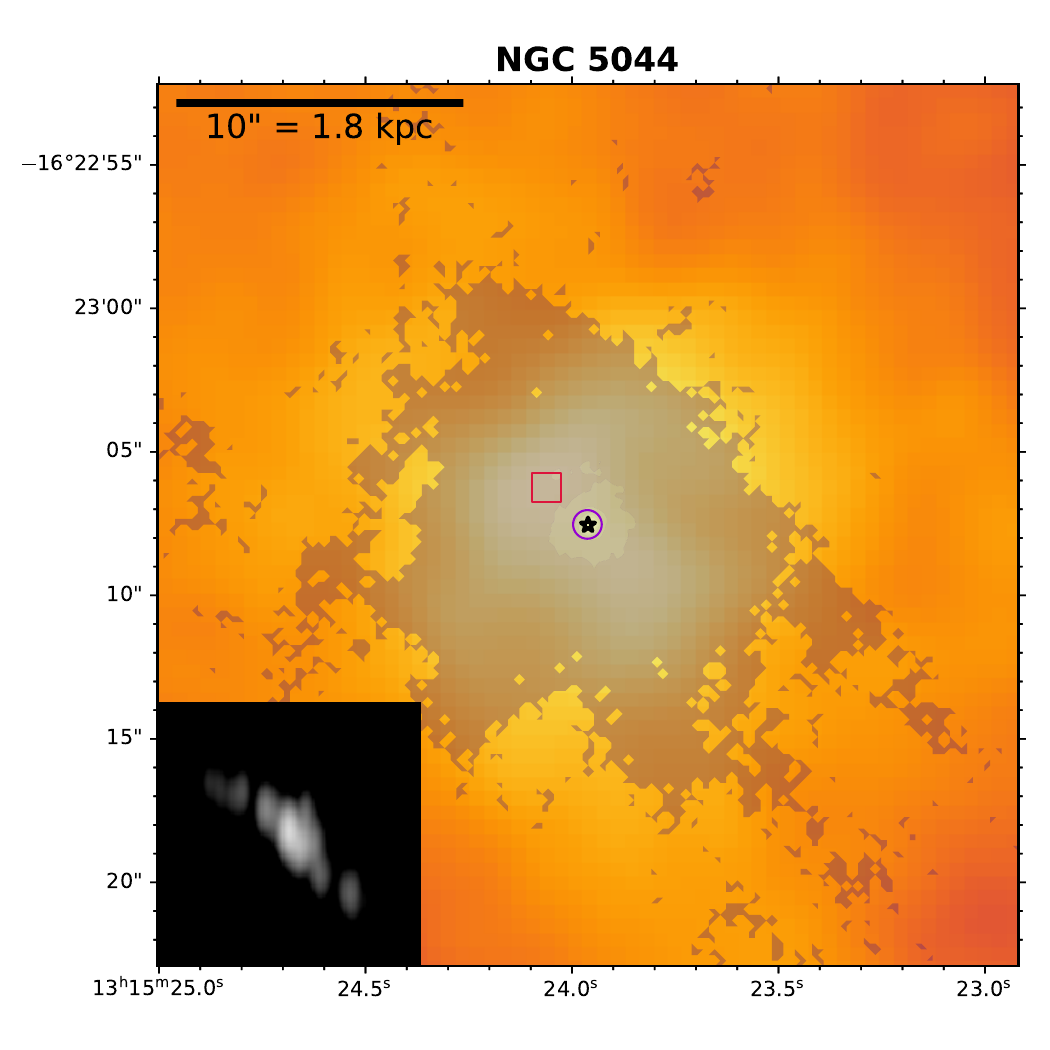}
    \includegraphics[width=0.3\linewidth, trim={0.2cm 0 1cm 1.6cm}]{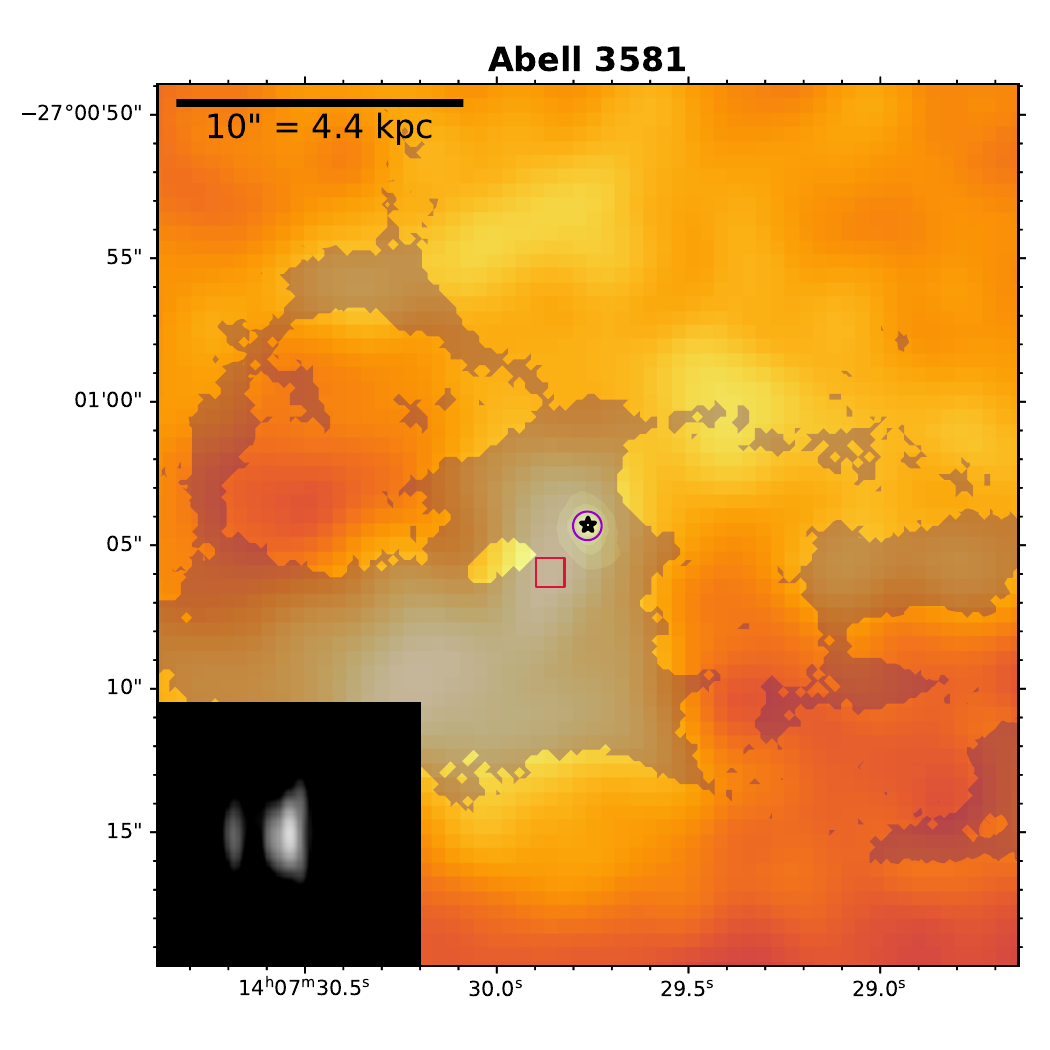}\\
    \includegraphics[width=0.3\linewidth, trim={0.2cm 0 1cm 1.6cm}]{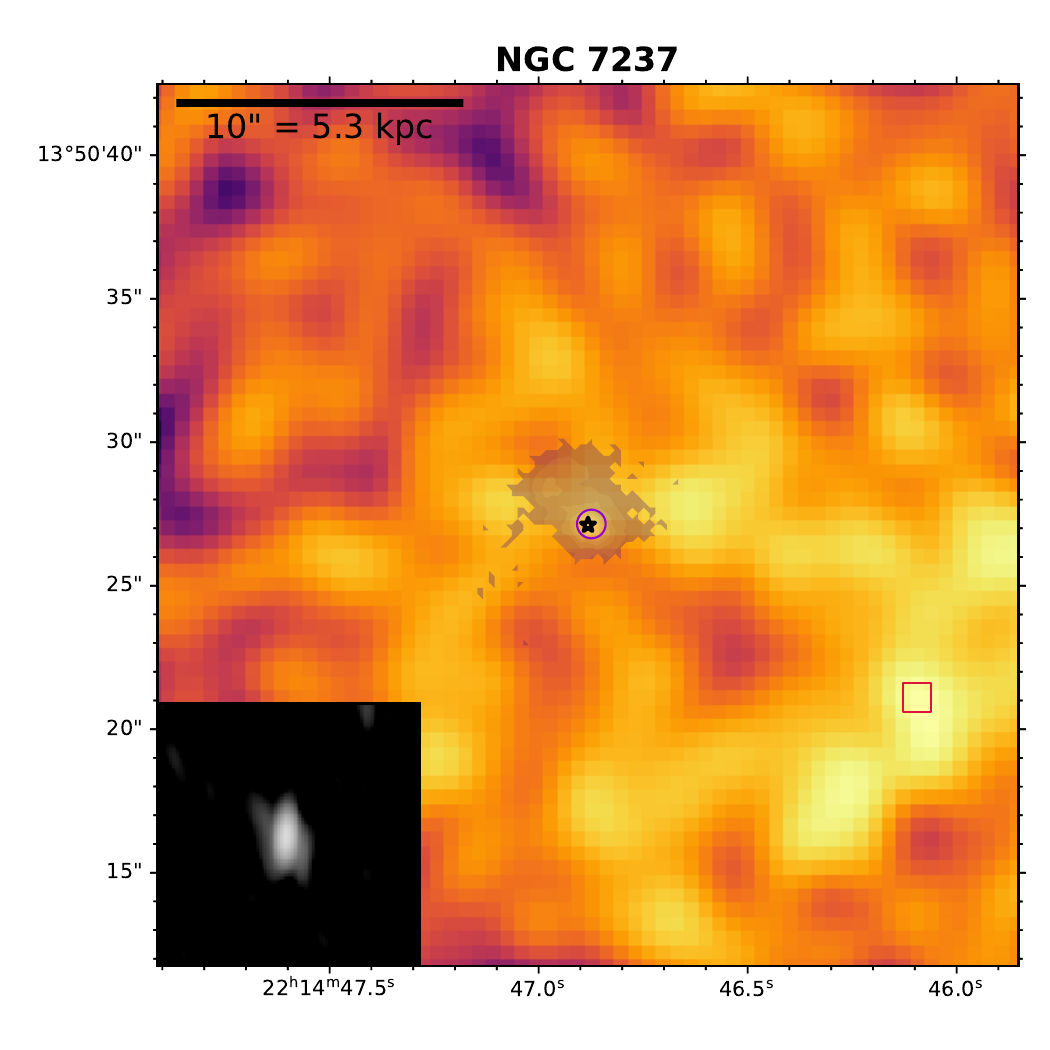}
    \includegraphics[width=0.3\linewidth, trim={0.2cm 0 1cm 1.6cm}]{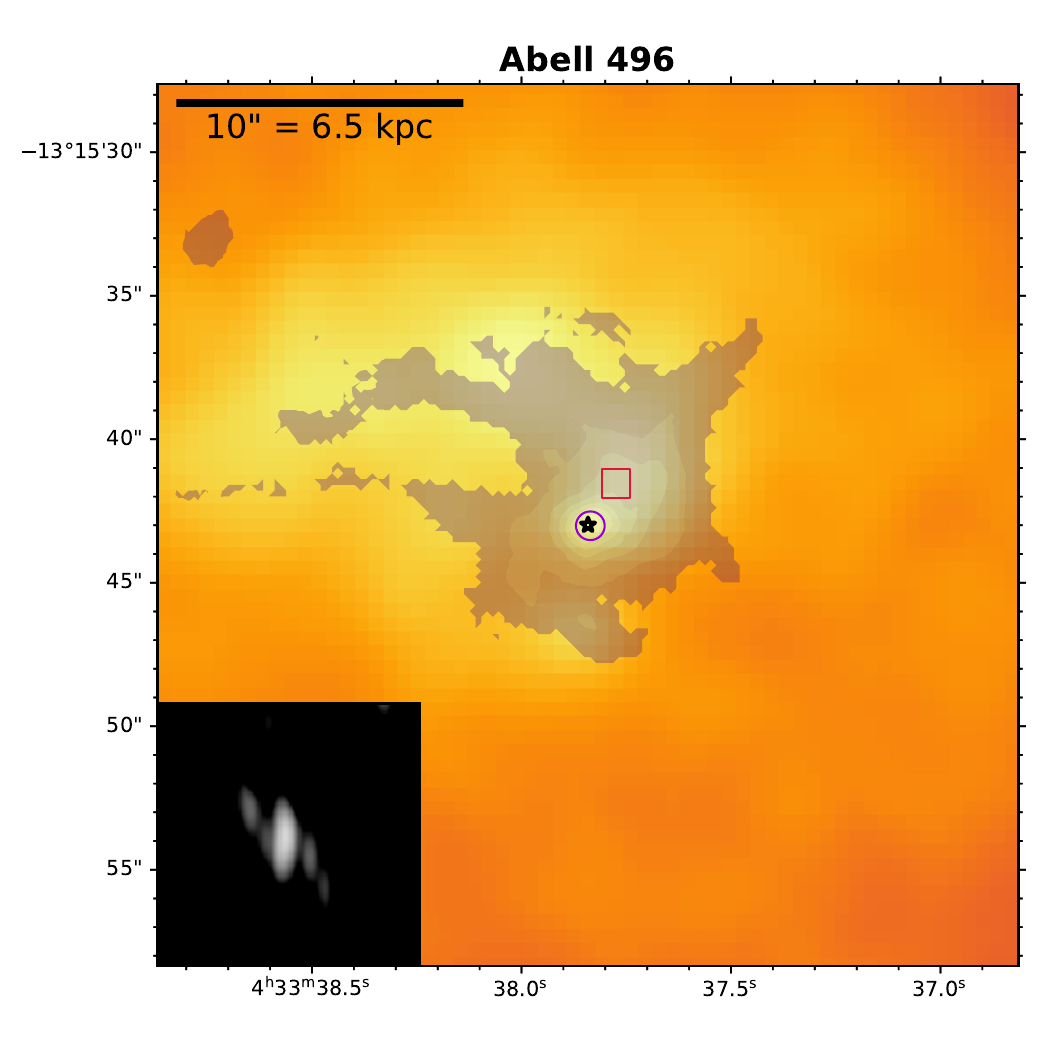}
    \includegraphics[width=0.3\linewidth, trim={0.2cm 0 1cm 1.6cm}]{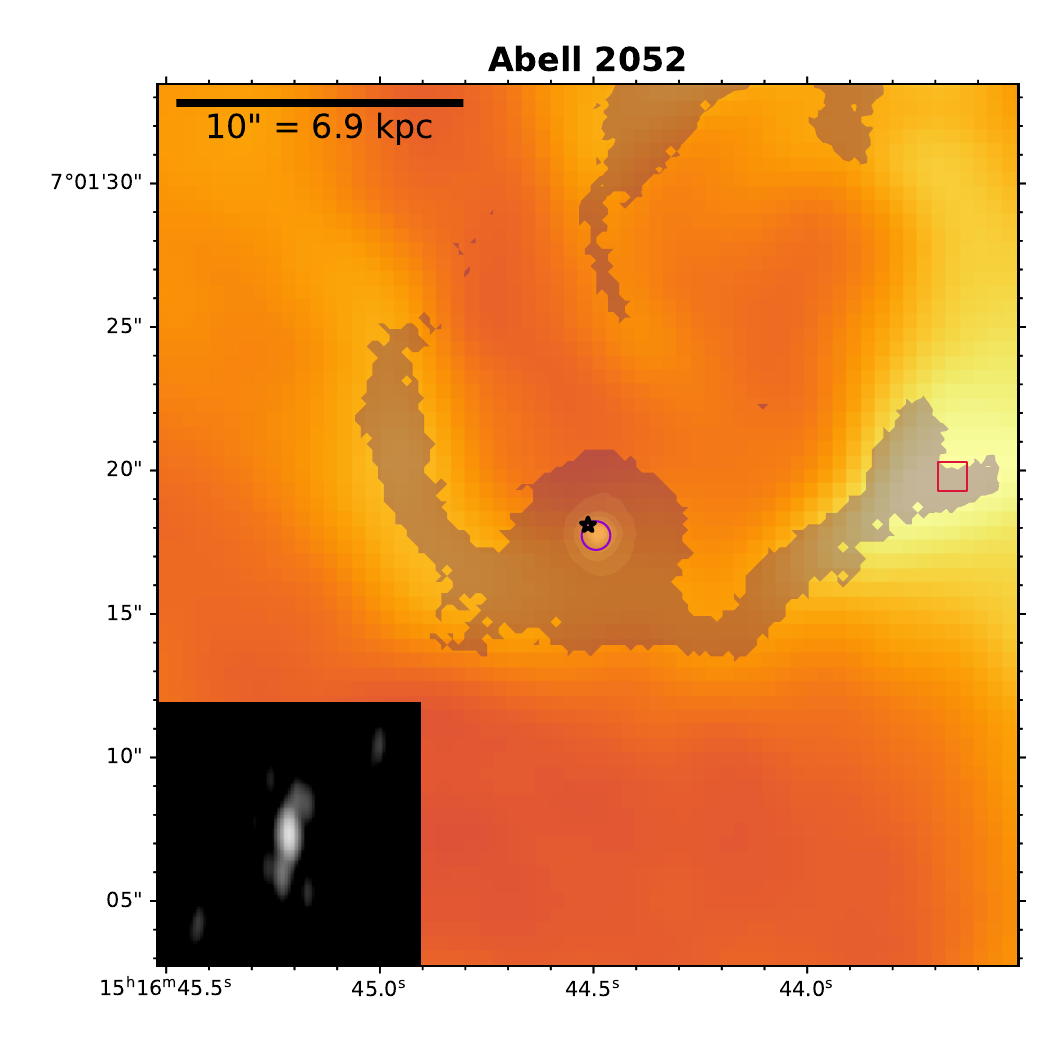}\\
    \includegraphics[width=0.3\linewidth, trim={0.2cm 0 1cm 1.6cm}]{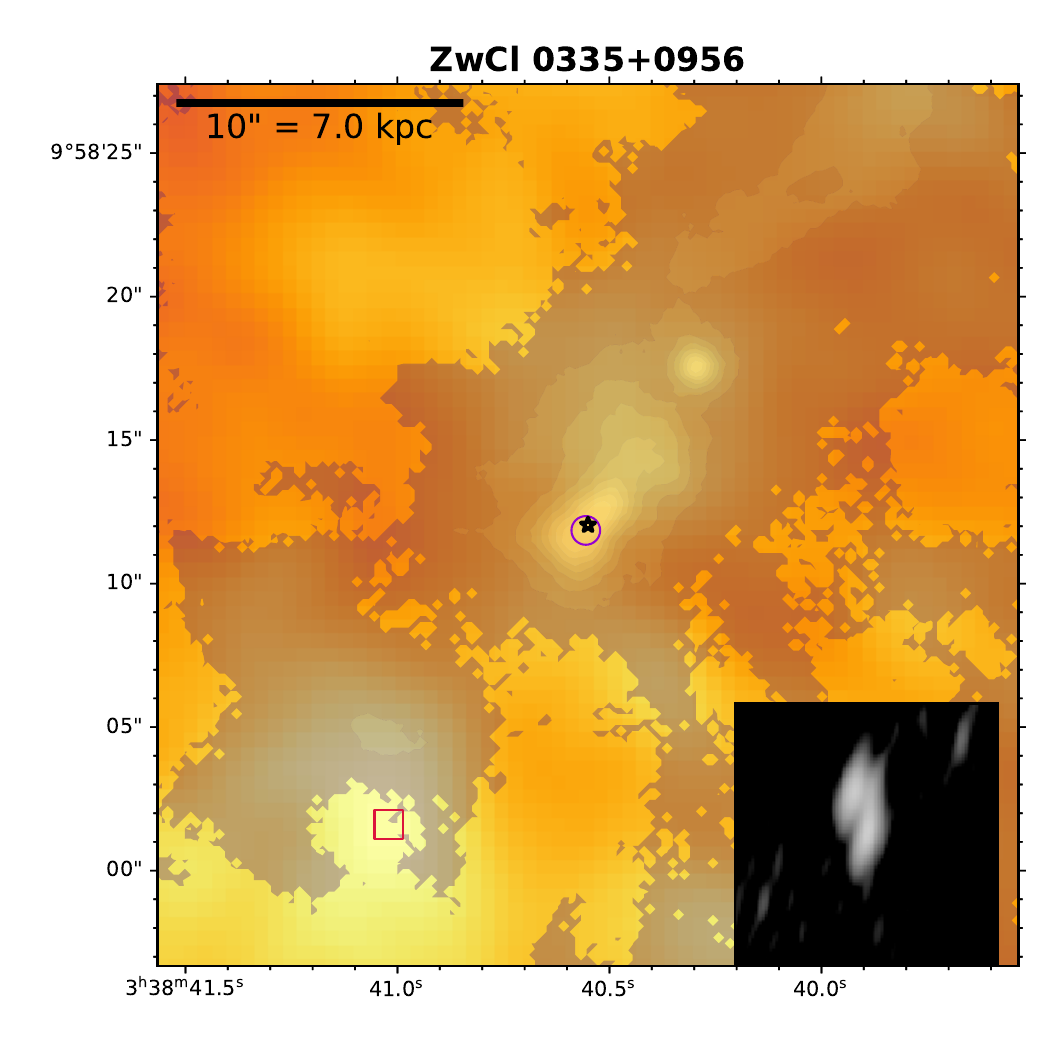}
    \includegraphics[width=0.3\linewidth, trim={0.2cm 0 1cm 1.6cm}]{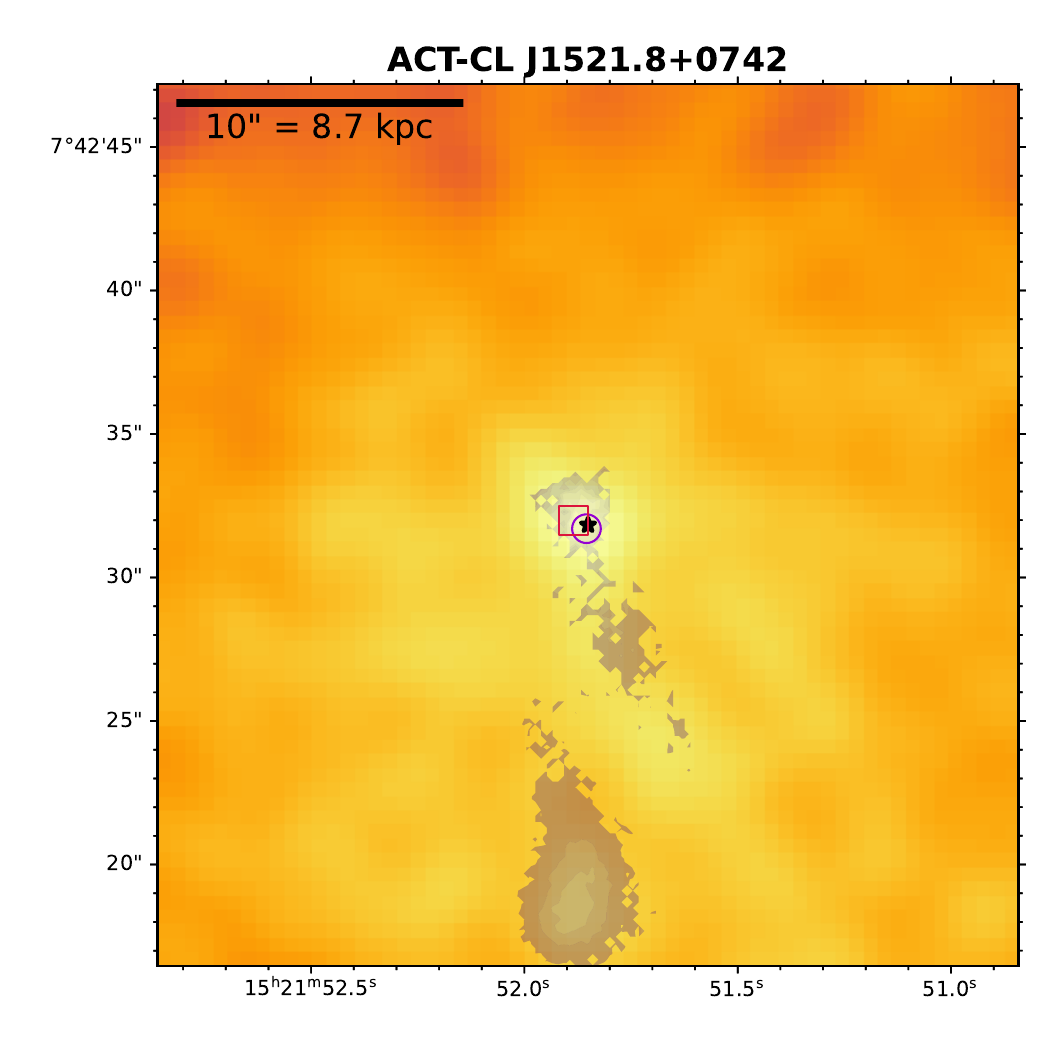}
    \includegraphics[width=0.3\linewidth, trim={0.2cm 0 1cm 1.6cm}]{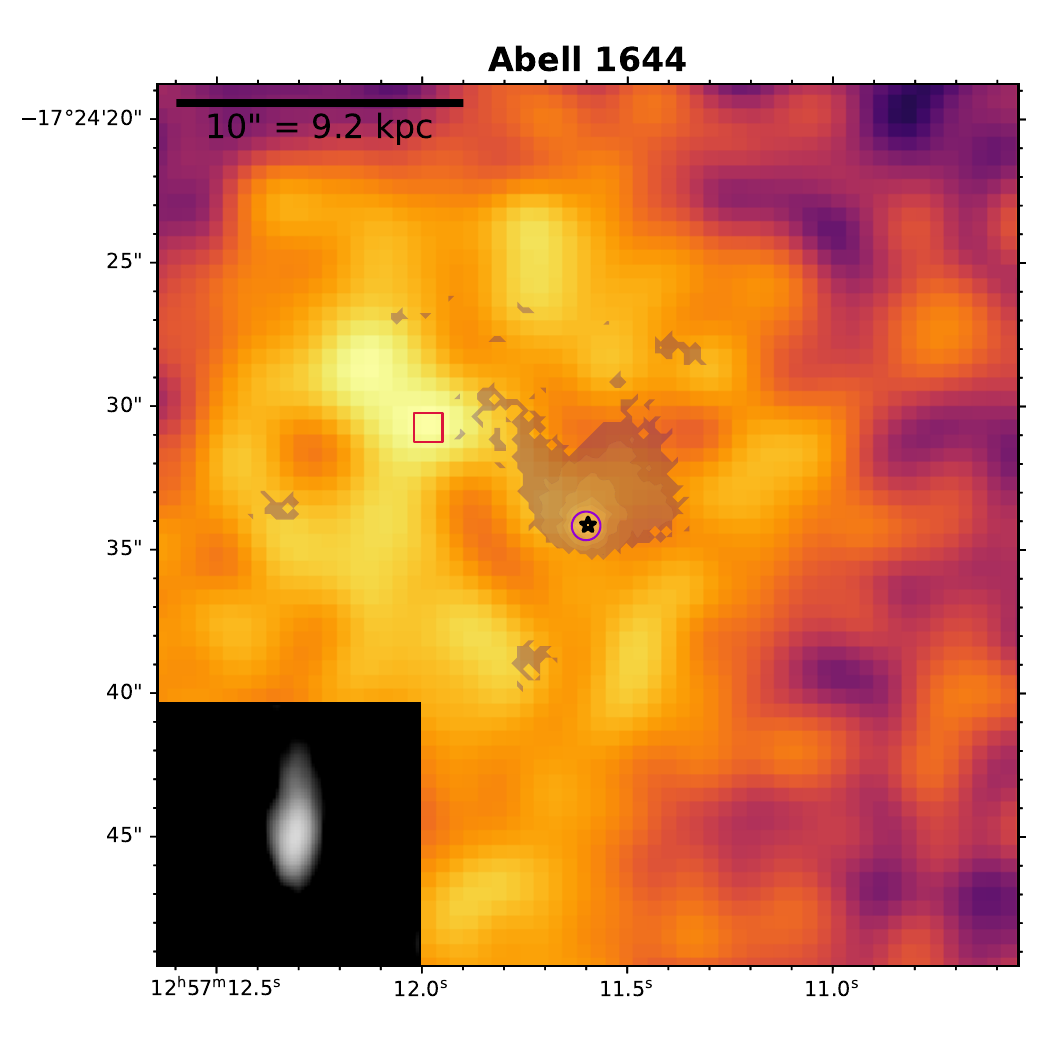}
    \caption{Multiwavelength view of the systems in our sample. The panels show smoothed X-ray {\it Chandra} maps of the hot gas, with grayscale contours of the H$\alpha$ line from MUSE data. The red square marks the position of the X-ray peak, the purple circle shows the location of the H$\alpha$ peak, and the black star shows the position of the SMBH (see Tab.~\ref{tab:completeinfo}). When detected, the VLBA image of the pc-scale radio core is shown in the insets.}
    \label{fig:multi1}
\end{figure*}
\begin{figure*}[ht!]
    \centering
    \ContinuedFloat
    \includegraphics[width=0.3\linewidth, trim={0.2cm 0 1cm 1.6cm}]{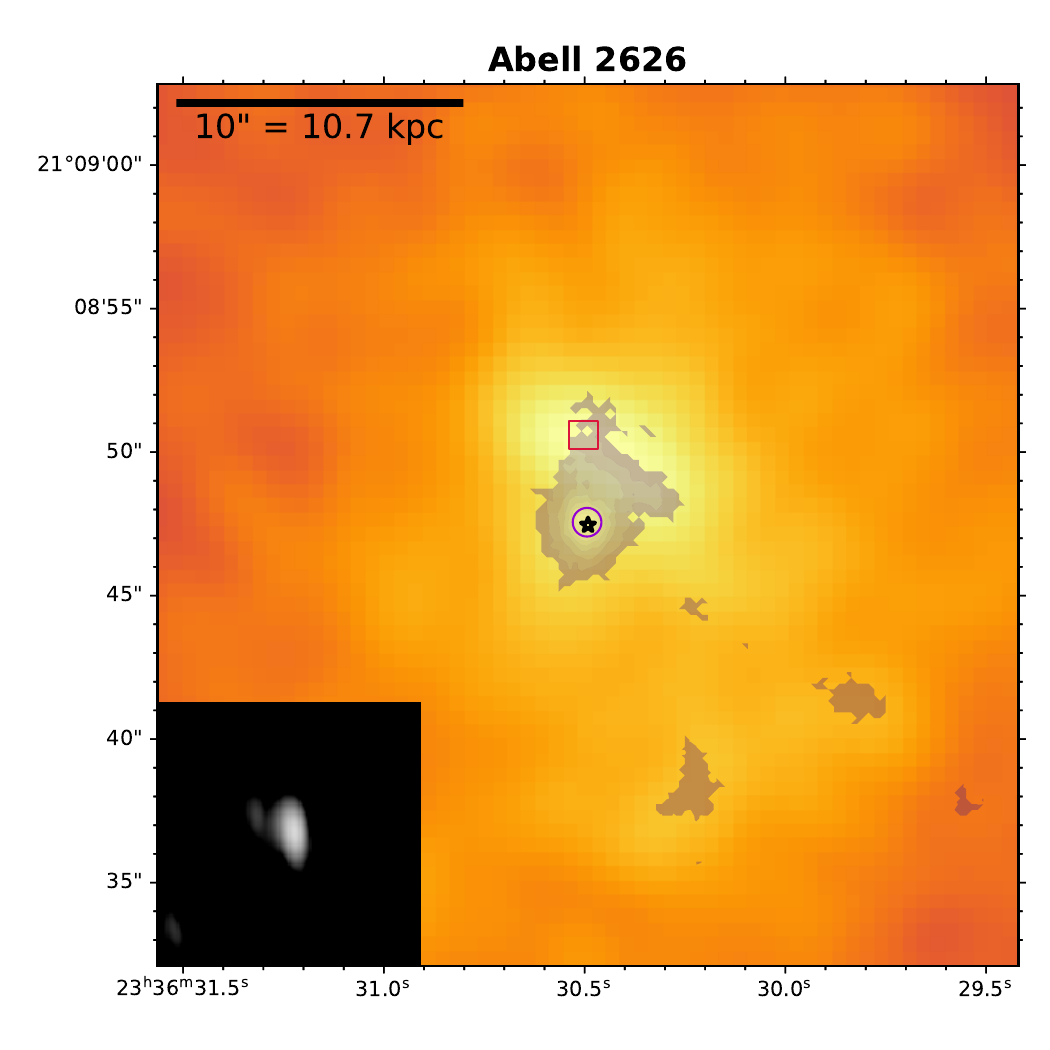}
    \includegraphics[width=0.3\linewidth, trim={0.2cm 0 1cm 1.6cm}]{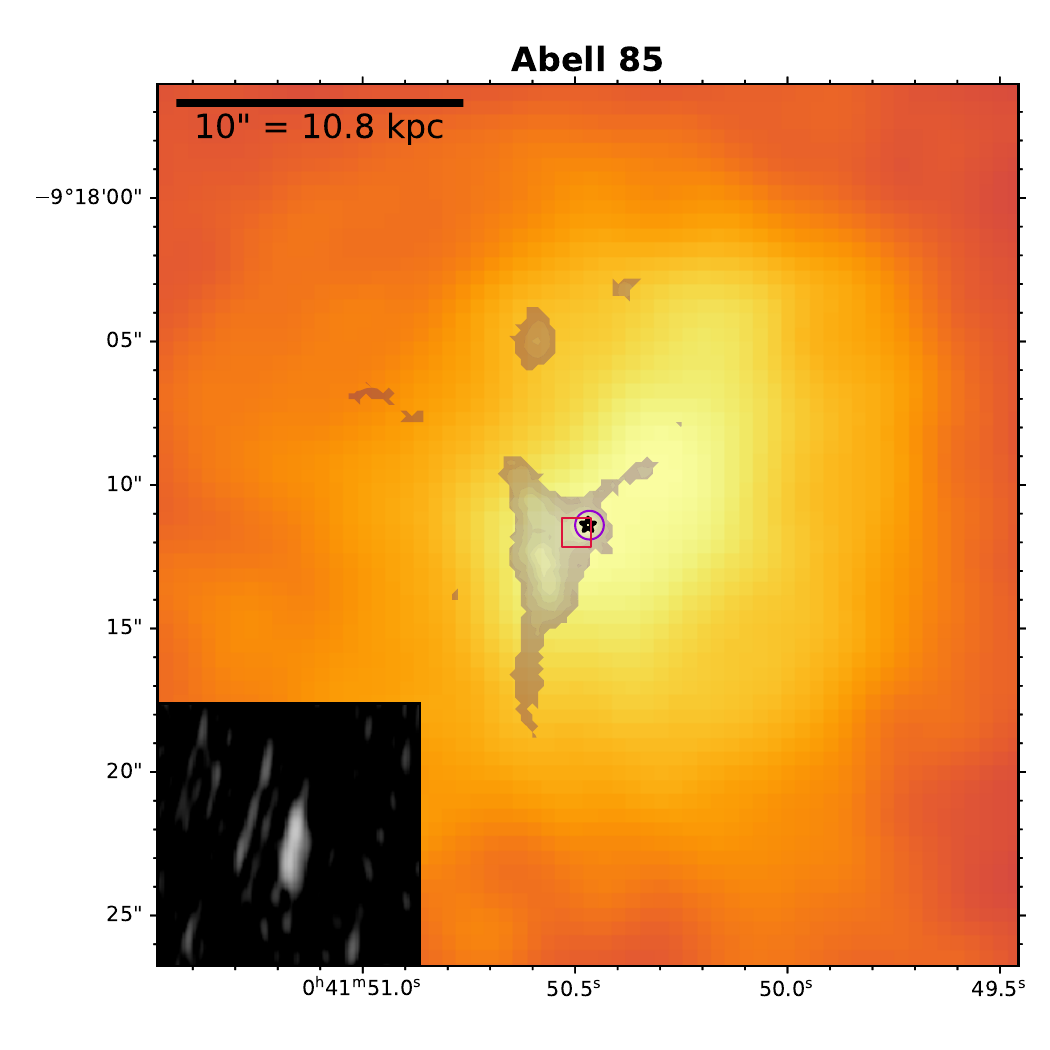}
    \includegraphics[width=0.3\linewidth, trim={0.2cm 0 1cm 1.6cm}]{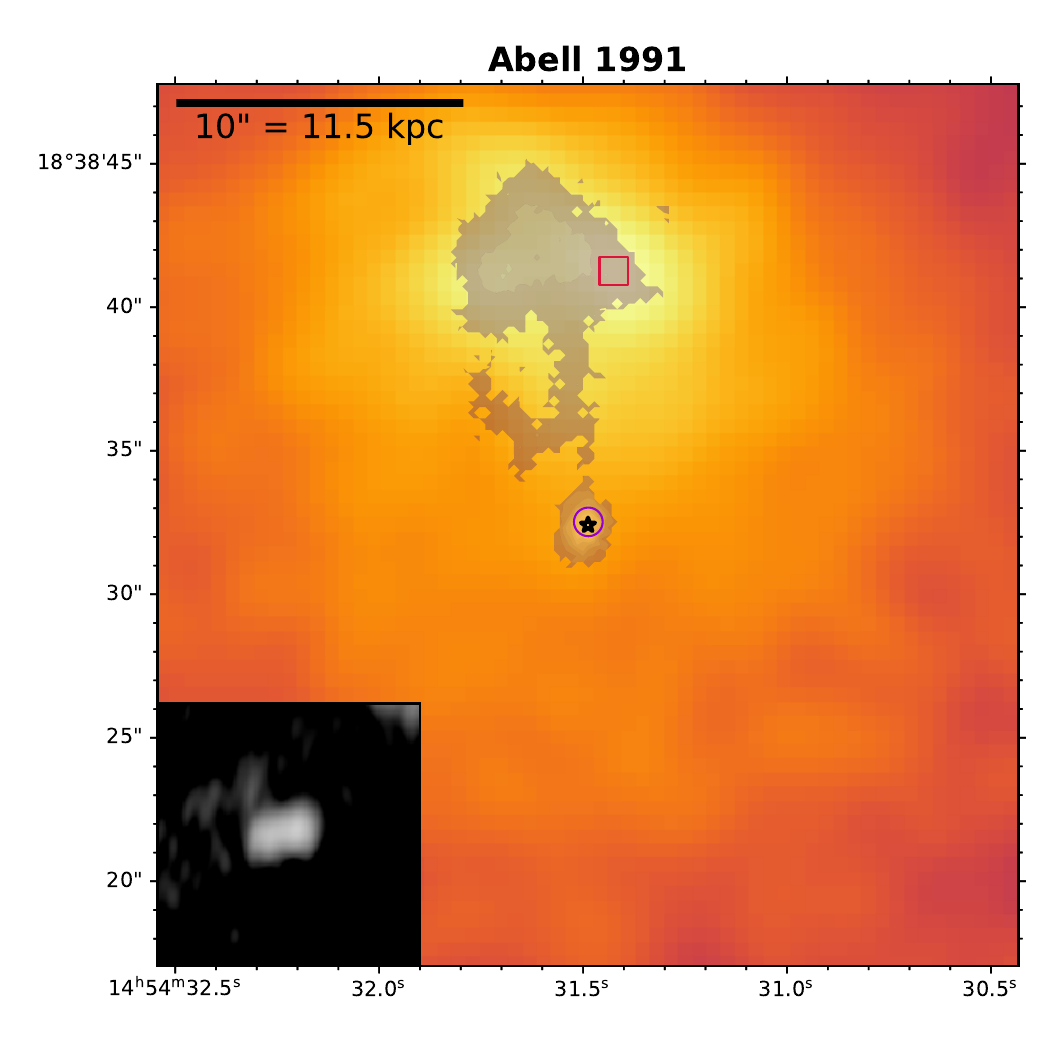}\\
    \includegraphics[width=0.3\linewidth, trim={0.2cm 0 1cm 1.6cm}]{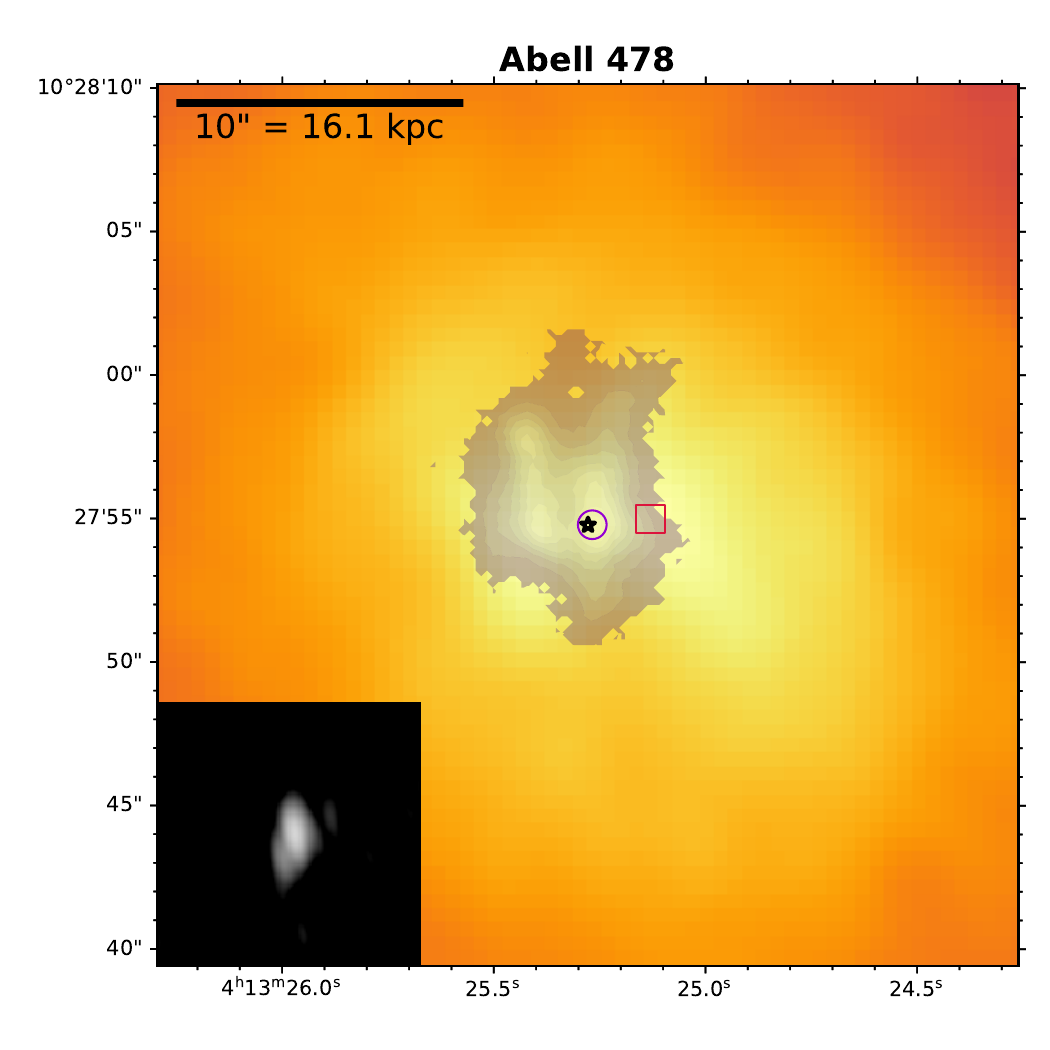}
    \includegraphics[width=0.3\linewidth, trim={0.2cm 0 1cm 1.6cm}]{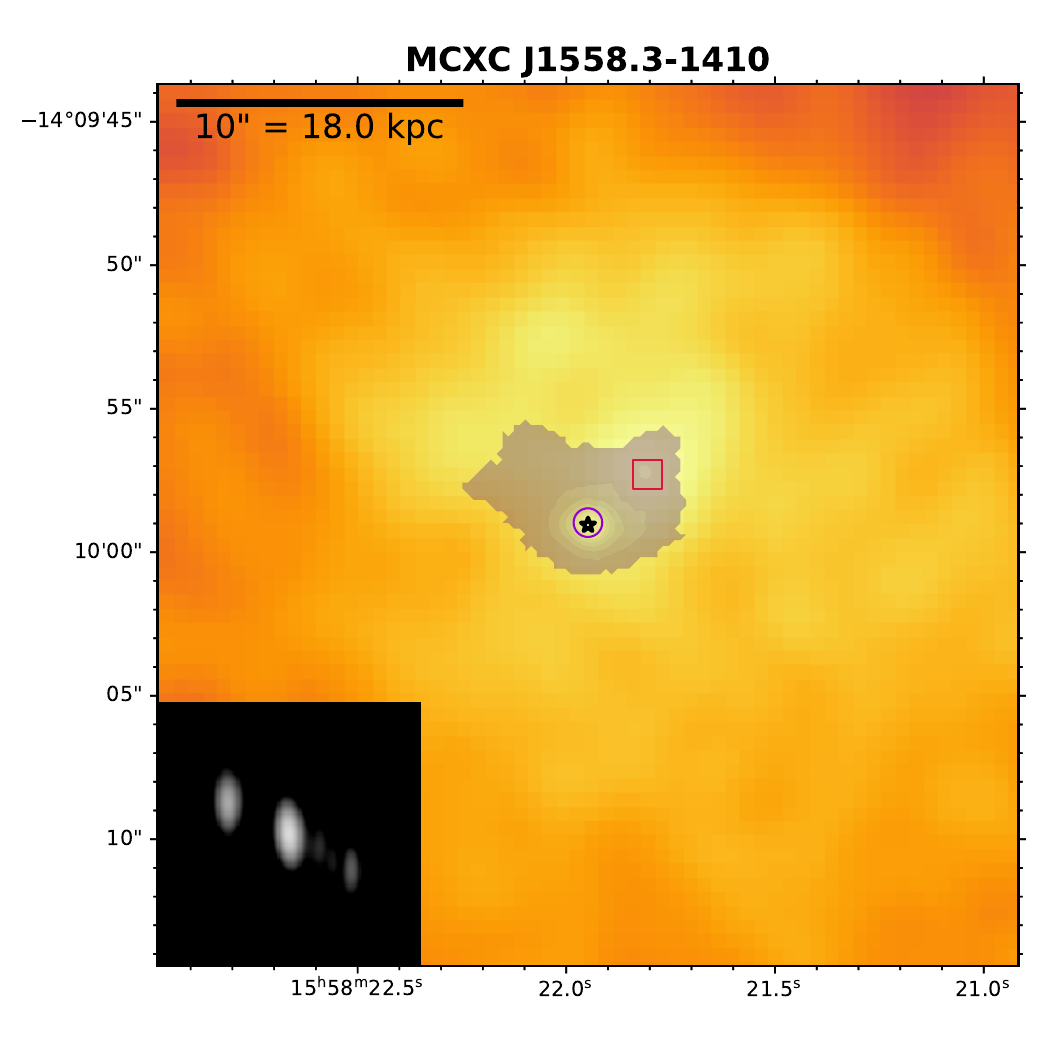}
    \includegraphics[width=0.3\linewidth, trim={0.2cm 0 1cm 1.6cm}]{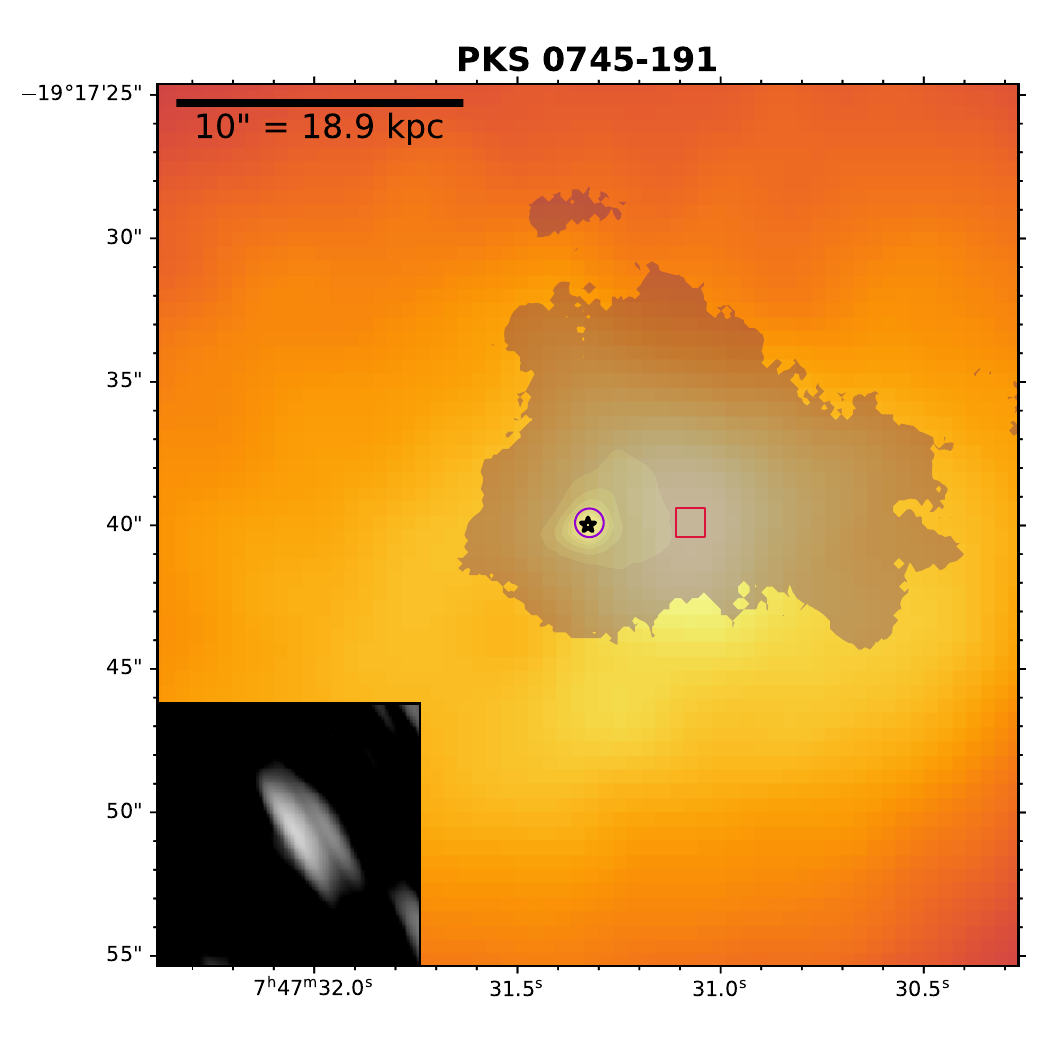}
    \includegraphics[width=0.3\linewidth, trim={0.2cm 0 1cm 1.6cm}]{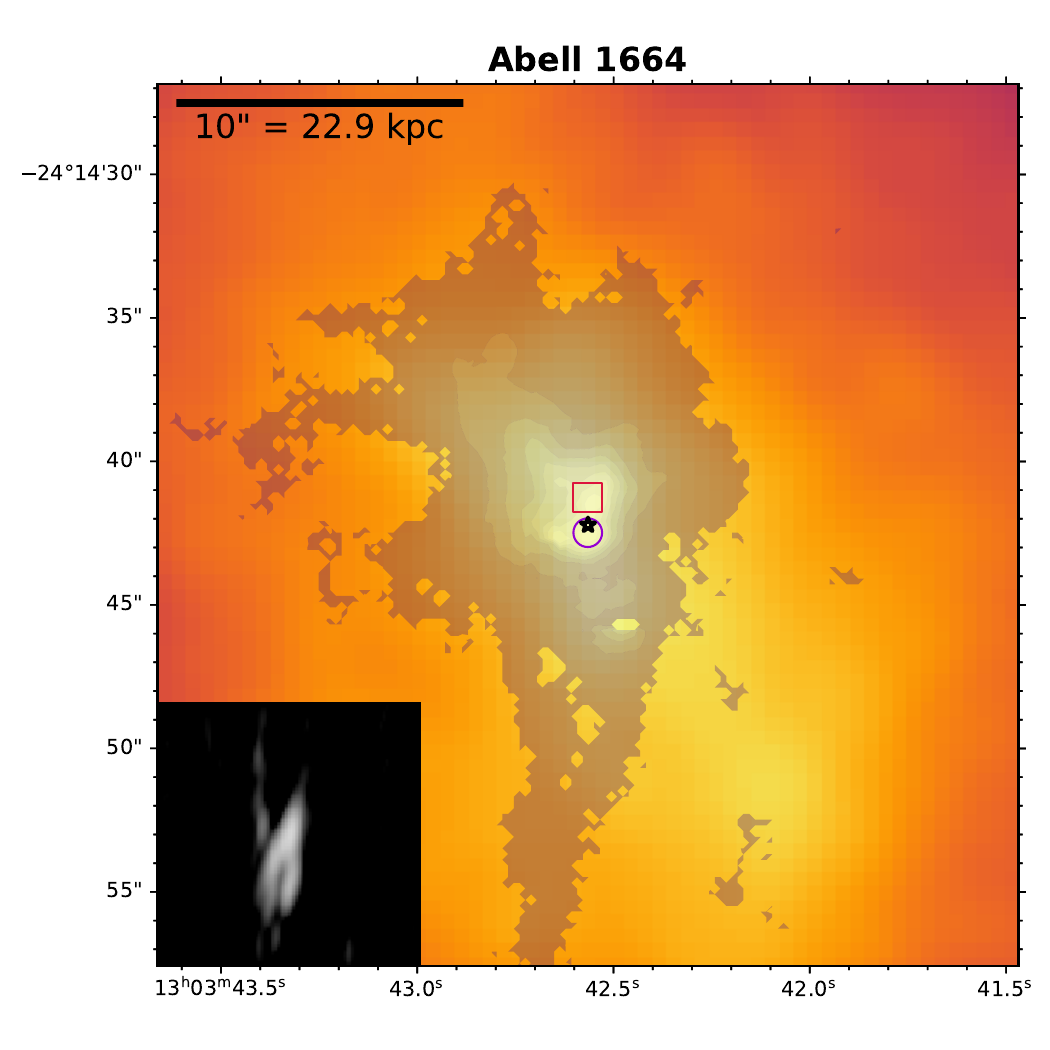}
    \includegraphics[width=0.3\linewidth, trim={0.2cm 0 1cm 1.6cm}]{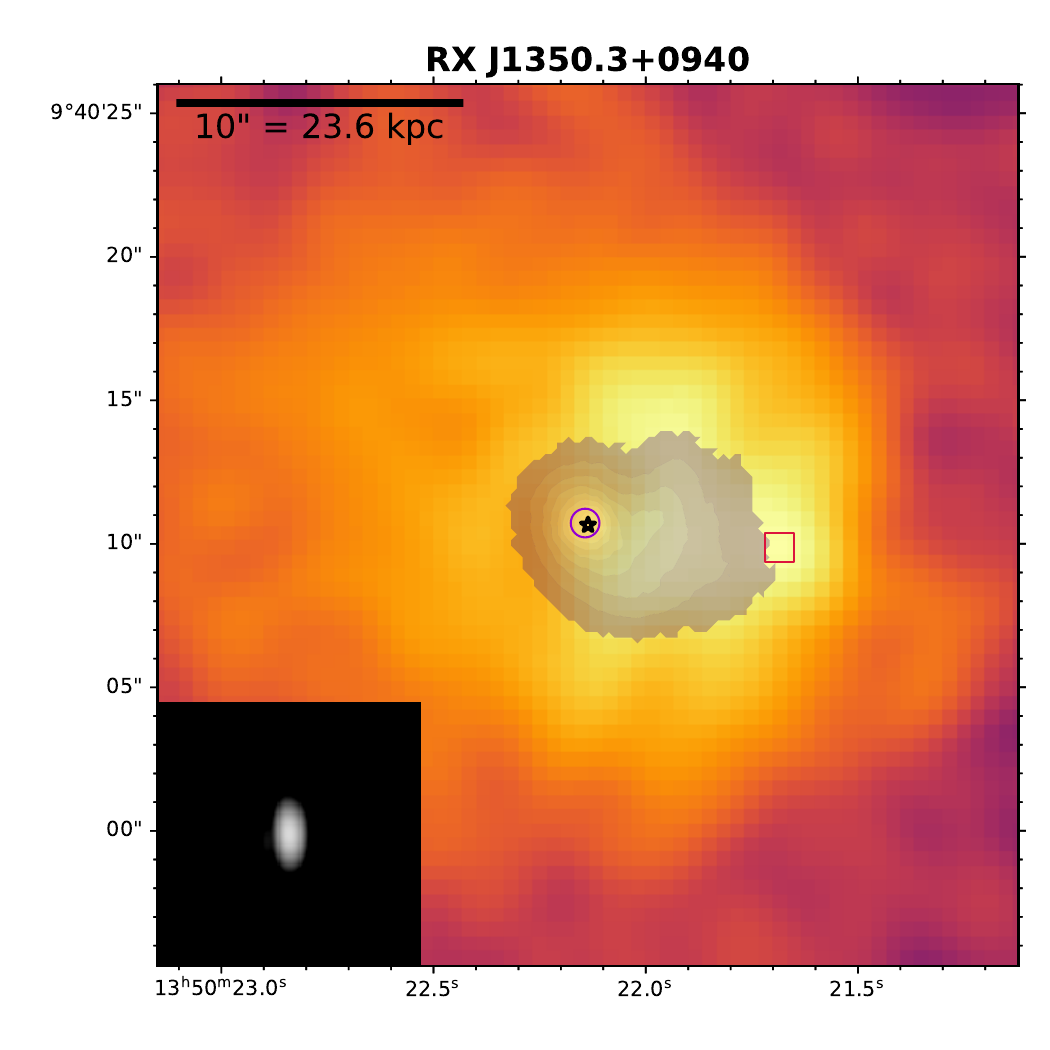}
    \includegraphics[width=0.3\linewidth, trim={0.2cm 0 1cm 1.6cm}]{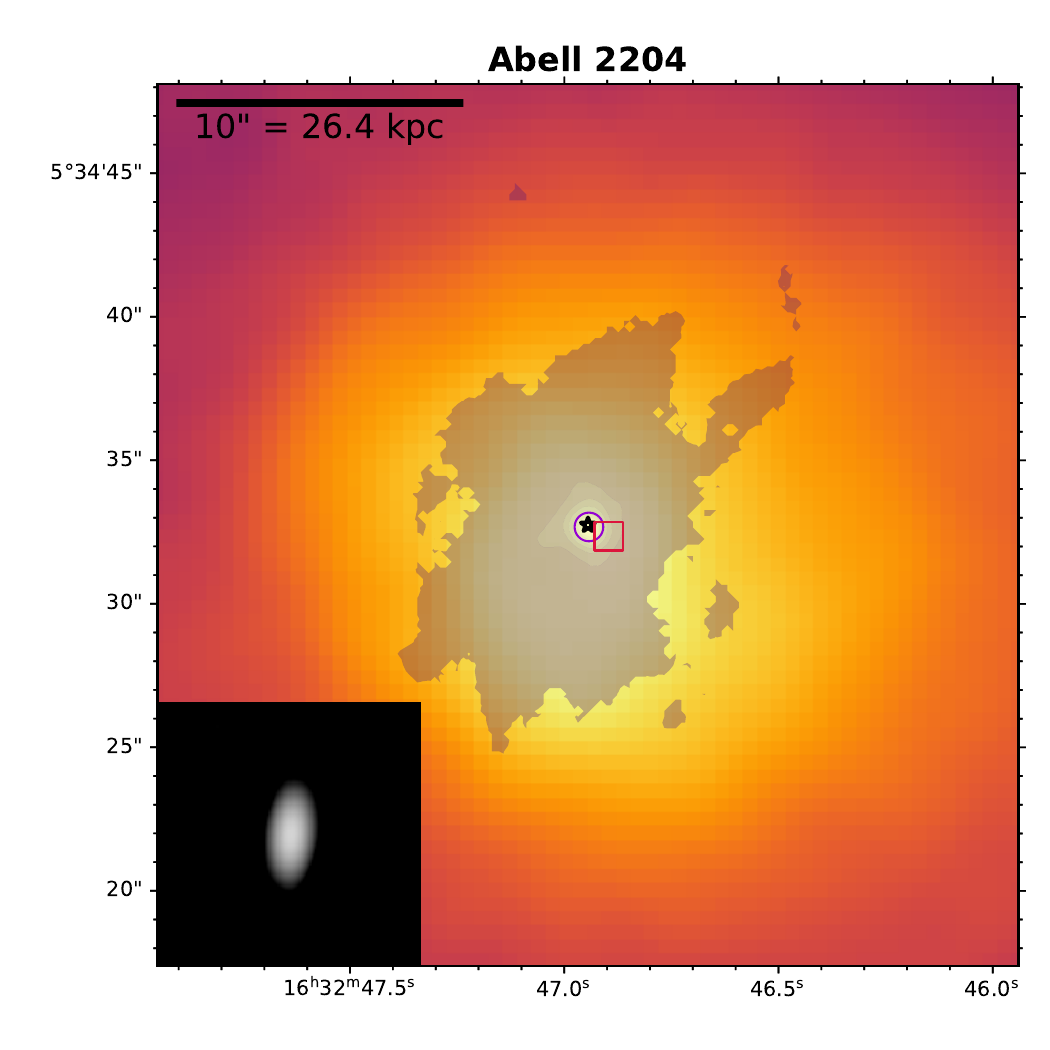}\\
    \includegraphics[width=0.3\linewidth, trim={0.2cm 0 1cm 1.6cm}]{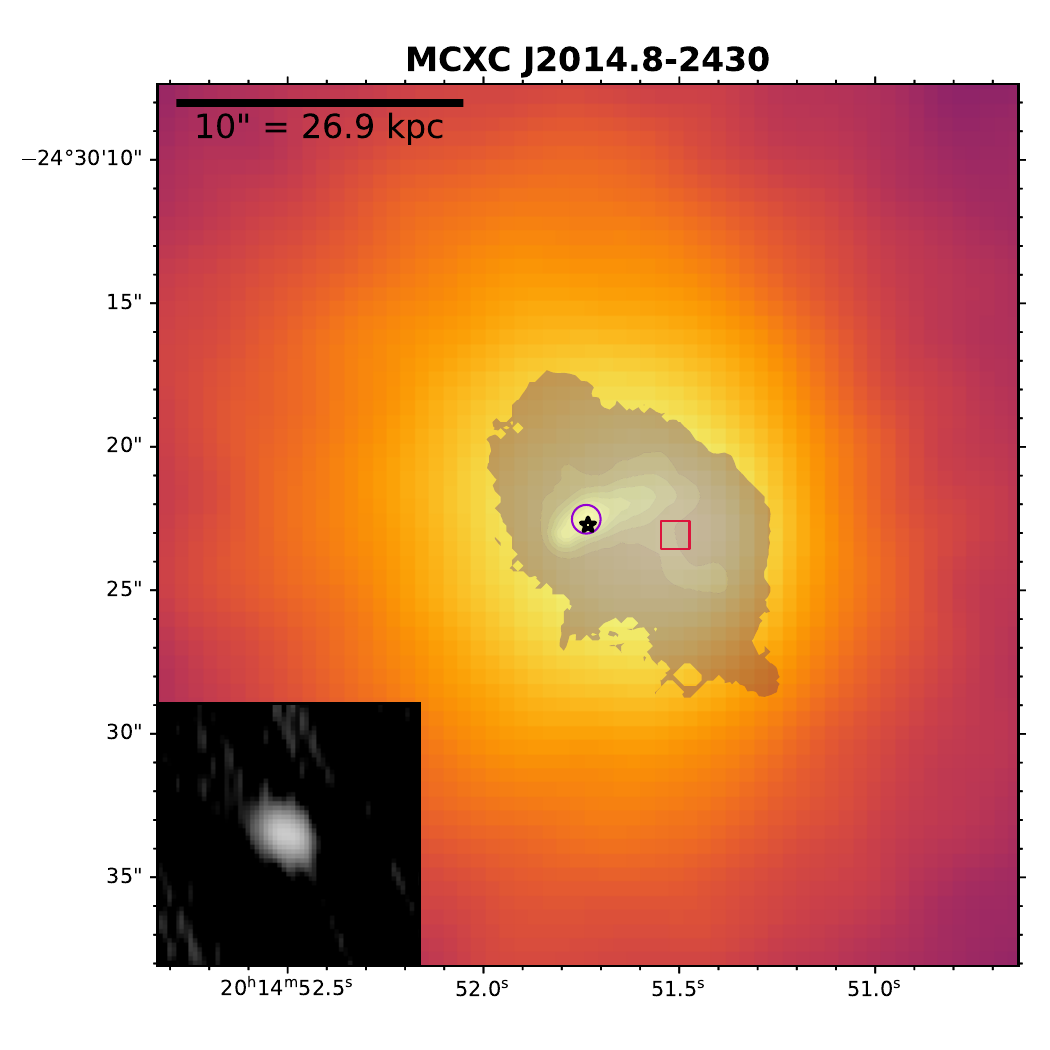}
    \includegraphics[width=0.3\linewidth, trim={0.2cm 0 1cm 1.6cm}]{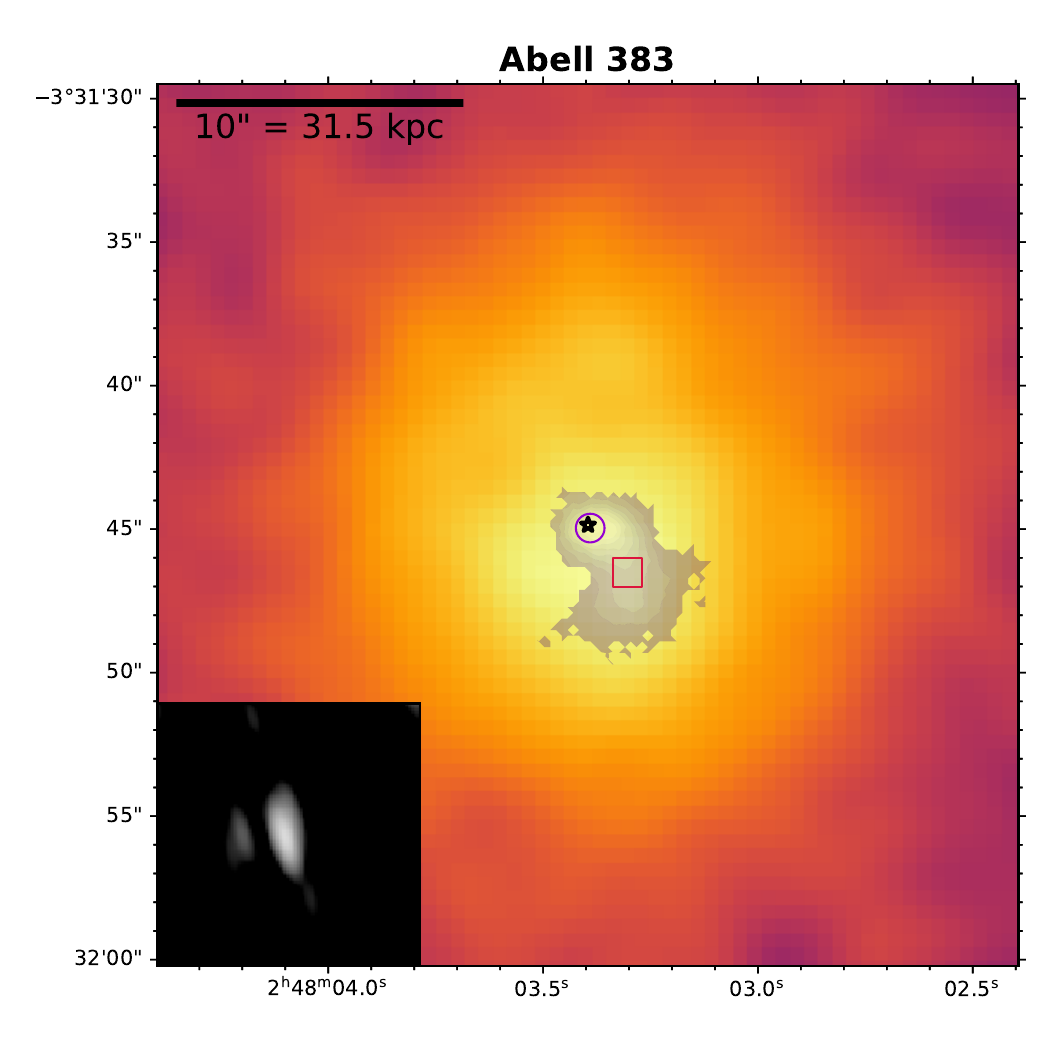}
    \includegraphics[width=0.3\linewidth, trim={0.2cm 0 1cm 1.6cm}]{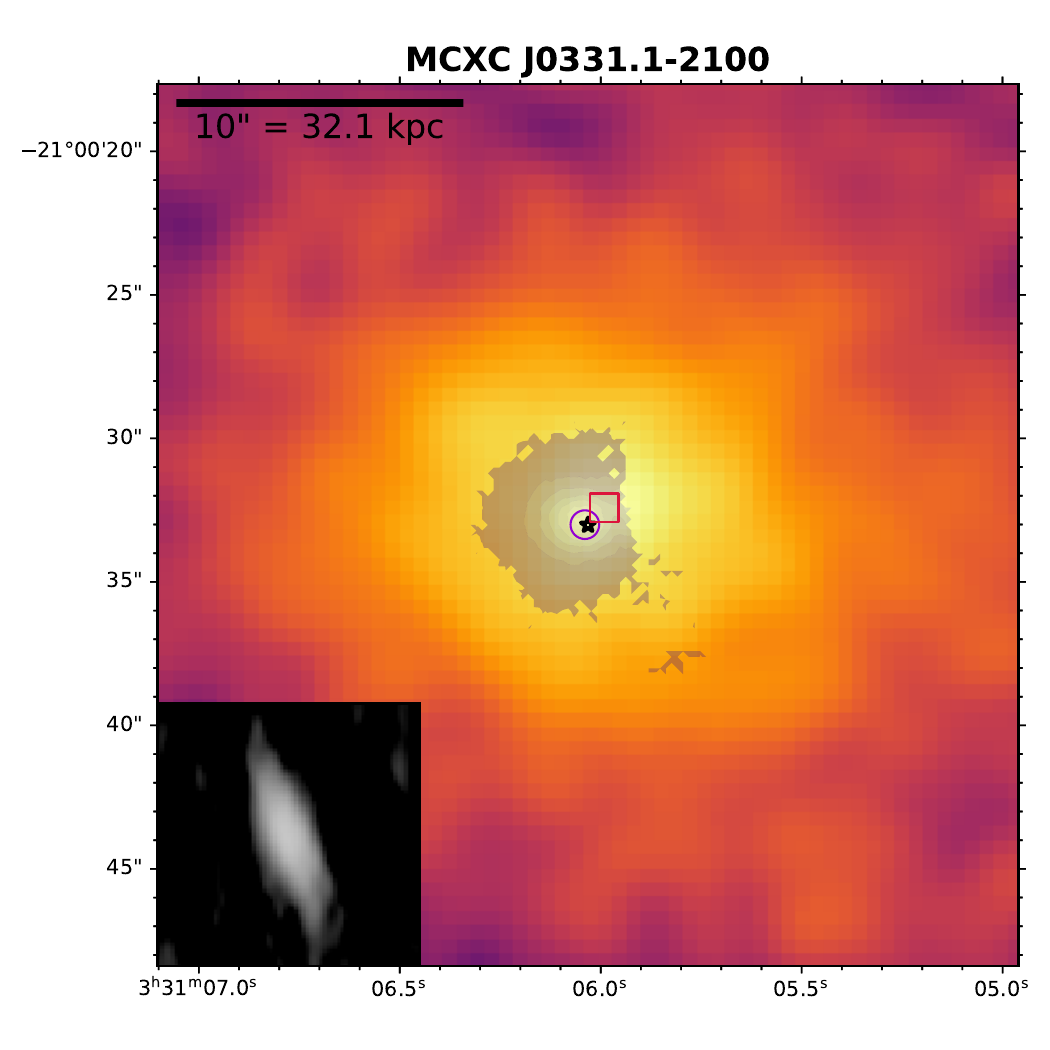}
    \caption{Continued.}
    \label{fig:multi2}
\end{figure*} 

\section{Sample selection and data analysis}\label{sec:sampledata}
Our primary aim is to study the presence (or absence) of an active SMBH in the presence (or absence) of spatial offsets between the core of the BCG, the X-ray peak of the hot gas, and the emission line (H$\alpha$) peak of the warm gas.  As spatial resolution is key, we based our selection on a combination of Very Long Baseline Array (VLBA), \textit{Chandra}, and Very Large Telescope (VLT) Multi Unit Spectroscopic Explorer (MUSE) data. We first select systems based on the existence of VLBA observations of central galaxies in galaxy clusters and groups. The milli-arcsecond VLBA resolution can unambiguously pinpoint the presence of an active SMBH on parsec scales. Radio observations performed with arcsec-like resolution can also pinpoint AGN activity, but the $\sim$kpc spatial scale sampled by these observations implies that the radio emission could come from cosmic rays accumulated on such scales over relatively long timescales, without tracing ongoing jet ejection (e.g., \citealt{liuzzo2009,middleberg2013,cheng2025}). 
\par We thus consider the VLBA data from projects BE056, BE063, BE065, and BE069, which comprise 5 GHz snapshot observations with a fairly uniform sensitivity and spatial resolution (a few mas) of 197 systems (the data for 59 of these were previously presented in \citealt{Hogan15a,Hogan15b}). Then, we cross-matched the VLBA targets with the {\it Chandra} and the VLT/MUSE archives, obtaining a sample of 57 systems with archival observations from these three facilities. By further excluding objects for which H$\alpha$ emission is not detected in the MUSE cubes,
we obtain a compilation of 51 systems.
Finally, we apply a redshift cut, limiting this study to systems at $z<0.2$\footnote{This cut is motivated by the $0.5\arcsec - 1\arcsec$ angular resolution of {\it Chandra} and MUSE. Existing studies have shown that offsets between warm or hot gas and the BCG are typically on the order of a few kpc in relaxed systems \citep{hamer2012,pasini2021,rosignoli2024}. At higher redshifts, the typical angular uncertainty becomes too large to reliably study kpc-scale offsets.}, which brings us to our final sample of 25 systems (see Tab.~\ref{tab:completeinfo} in Appendix~\ref{app:sample}).
\par We note that, as our selection criteria are primarily based on the availability of VLBA, {\it Chandra}, and VLT/MUSE data, the resulting sample is unlikely to be statistically complete. However, being targets of radio surveys and dedicated X-ray and optical spectroscopic observations, the 25 systems are likely representative of galaxy groups and clusters where an active radio source, a bright extended X-ray halo, and ionized warm gas are expected. These are typical properties of systems where the feeding and feedback cycle is most relevant, thus aligning with the motivation of this work.
\par We fully report in Appendix \ref{app:data} the data reduction techniques we adopted for the VLBA, MUSE, and {\it Chandra} data. These steps allowed us to obtain final maps of the radio emission on parsec scales, of the hot gas and of the warm gas distributions. From these maps, we identified the position of the SMBH, of the H$\alpha$ peak, and of the X-ray peak, respectively (see the corresponding relevant subsections in Appendix \ref{app:data}). For the VLBA-undetected systems, the position of the SMBH was measured from high-resolution VLA (radio) or HST (optical) imaging (see Appendix \ref{appsub:vlba} and Tab.~\ref{tab:completeinfo} for details). We report in Tab.~\ref{tab:completeinfo} the 5~GHz VLBA radio power $P_{5\text{GHz}}$ of the sources in our sample, measured from the peak flux in the VLBA maps and assuming a typical spectral index of $\alpha = -0.7$ (e.g., \citealt{condon2002}). For undetected sources, we report the upper limit on $P_{5\text{GHz}}$ given by 5$\sigma_{\text{rms}}$. We also report the RA, DEC J2000 coordinates of the SMBH, of the X-ray peak, and of the H$\alpha$ peak. 
\par We measured the projected physical distance in kpc between the position of the SMBH traced by the 5~GHz VLBA data, the peak of the hot gas traced by the X-ray {\it Chandra} maps, and the peak of the warm gas traced by the H$\alpha$ MUSE maps. The results are reported in the last two columns of Tab.~\ref{tab:completeinfo}. The uncertainties in the offsets $\Delta$ were computed as:
\begin{equation}
    \delta \Delta = \sqrt{\delta_1^{2} + \delta_2^{2}},
\end{equation}
where $\delta_1$ and $\delta_2$ are the positional accuracies of the peak in map 1 and 2 (see the relevant subsections in Appendix \ref{app:data}). The sum of squares is usually dominated by the MUSE or \textit{Chandra} uncertainty ($0.5\arcsec - 1.0\arcsec$), as the positional accuracy in VLBA maps (a fraction of milliarcsec) is orders of magnitude smaller.

\section{Results}\label{sec:results}
Our primary aim is to relate the multiphase gas offsets to the SMBH quiescent or active state based on the radio power measured in 5~GHz VLBA data (i.e., an undetected radio core provides an upper limit on radio power). To this end, we show in Fig.~\ref{fig:offset1} the radio power at 5~GHz of the 25 BCGs in our sample versus the distance between the position of the SMBH and of the hot (X-ray, left panel) and warm (H$\alpha$, right panel) gas peaks. We show in Fig. \ref{fig:multiexample} and Fig. \ref{fig:multi1} the distribution of the hot and warm gas phases in the 25 systems of the sample, as well as the location of the SMBH, the X-ray peak, and the H$\alpha$ peak. 
\subsection{SMBH - Hot gas (X-ray) peak offset}\label{subsec:Xray}
As shown in Fig.~\ref{fig:offset1} (left), most systems in our sample (20/25) exhibit significant spatial offsets (that is, for which the distance $\Delta$ between the SMBH and the gas peak is larger than the uncertainty $\delta\Delta$) between the position of the SMBH and the X-ray peak. Quantitatively, the maximum measured offset is $\Delta_{\text{X-ray}}^{\text{SMBH}} = 15\pm3$~kpc, with the majority of the systems having their X-ray peak at less than 10~kpc from the SMBH. The average SMBH -- X-ray peak offset is $4.7$~kpc, with a dispersion of $3.8$~kpc. These results are in agreement with literature studies showing that cool core systems can show $\leq$10-20~kpc-scale offsets between the core of the BCG and the X-ray peak of the hot gas in cool cores \citep{sanderson2009,2010A&A...513A..37H}. Moreover, the systems in our sample with a significant X-ray peak offset correspond to about 80\% of the total number. This fraction matches the fraction of cool core clusters where sloshing is taking place \citep{ueda2020}. As noted in Section \ref{sec:intro}, X-ray peak offsets are typically attributed to bulk motions of the hot gas (e.g., \citealt{hamer2012}), although AGN feedback processes such as uplift might also contribute \citep{rosignoli2024}. In this context, we note that in some systems (NGC~5846, NGC~5044, Abell~3581, and Abell~2052; see Fig.~\ref{fig:multi1}) the X-ray peak is located along the rims of X-ray cavities, suggesting that AGN-driven uplift can be able of influencing the hot gas distribution as much as sloshing, either by pushing the density peak outwards or by compressing the gas and locally increasing its density. Finally, we note that there is no evident correlation between the magnitude of the offset and the radio power on pc scales of the SMBH, and that there appears to be no clear connection between the magnitude of the offset and the radio detection of the SMBH on parsec scales.

\subsection{SMBH - Warm gas (H$\alpha$) peak offset}\label{subsec:Halpha}
The right panel of Fig. \ref{fig:offset1} shows the comparison between the radio power on parsec scales and the H$\alpha$ peak - SMBH offset. We observe that in most systems (21/25), the H$\alpha$ peak coincides spatially with the SMBH, while only four systems show significant offsets. The average $\Delta^{\text{SMBH}}_{\text{H}\alpha}$ for the 25 systems is $0.6$~kpc, with a standard deviation of $1.4$~kpc. The offsets are quite small compared to the X-ray case, up to $\sim$5~kpc at most, being $\Delta^{\text{SMBH}}_{H\alpha} = 1.4\pm0.5$~kpc for Abell~133, $\Delta^{\text{SMBH}}_{H\alpha} = 1.3\pm0.7$ Abell~2495, $\Delta^{\text{SMBH}}_{H\alpha} = 2.9\pm1.1$~kpc for RX~J0821.0+0752, and $\Delta^{\text{SMBH}}_{H\alpha} = 5.3\pm0.9$~kpc for Abell~2566. Most strikingly, all four systems showing an H$\alpha$ peak - SMBH offset of $\geq$1~kpc also lack detected radio cores in VLBA observations, with upper limits on $P_{\text{5~GHz}}$ of $\leq10^{21 - 22}$~W/Hz. This strong correlation suggests that it is the warm gas peaking at the galaxy's center -- rather than the hot X-ray gas -- that plays a decisive role in triggering SMBH activity. We further discuss this scenario in Section \ref{sec:disc}.
\par As a note of caution, we considered a potential vicious circle: if the AGN in our sample were highly efficient at photo-ionizing the warm gas in their proximity (as in e.g., Seyferts, \citealt{kauffmann2003}), then the observed spatial coincidence between the H$\alpha$ peak and the radio AGN might reflect stronger ionization than in the rest of the nebula rather than a real association with the gas distribution. To test this possibility, we performed a Baldwin, Phillips \& Terlevich (BPT) analysis (e.g., \citealt{baldwin1981,kewley2001,kauffmann2003,kewley2006,cidfernandes2010}) of the warm gas emission lines (the details and plots are presented in Appendix~\ref{app:bpt}). Although a detailed analysis of the BPT diagrams is outside the scope of the paper, the main findings relevant to this work are summarized as follows: none of the systems exhibit emission consistent with AGN photoionization; instead, the ionization is dominated by a combination of LINER-like and star formation processes, as is commonly seen in BCGs (e.g., \citealt{fogarty2015,hamer2016,polles2021}). Moreover, there are no significant differences between the ionization conditions at the H$\alpha$ peak and those across the rest of the nebula, suggesting that the warm gas near the peak shares the same ionization mechanism as the extended filamentary structure.

\section{Discussion}\label{sec:disc}
\begin{figure*}[ht!]
    \centering
    \includegraphics[width=0.50\linewidth]{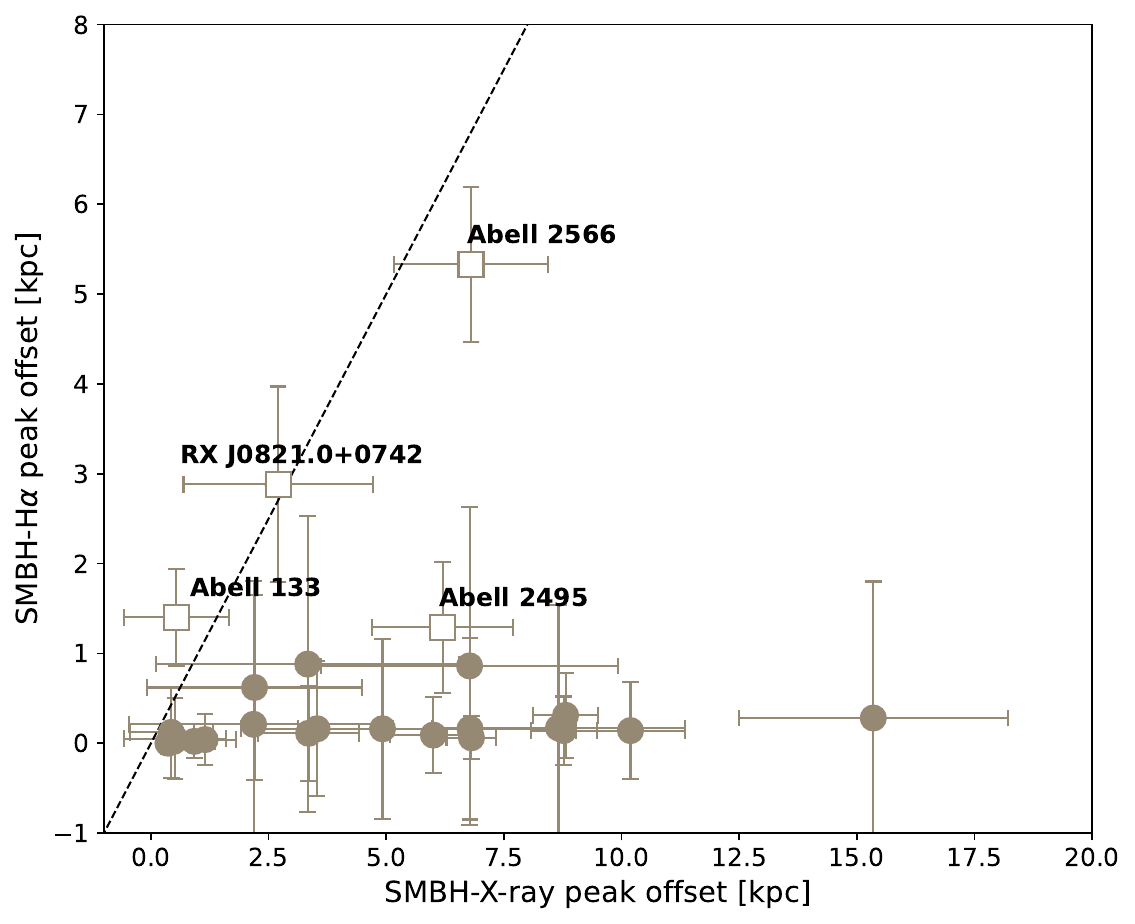}
    \includegraphics[width=0.49\linewidth]{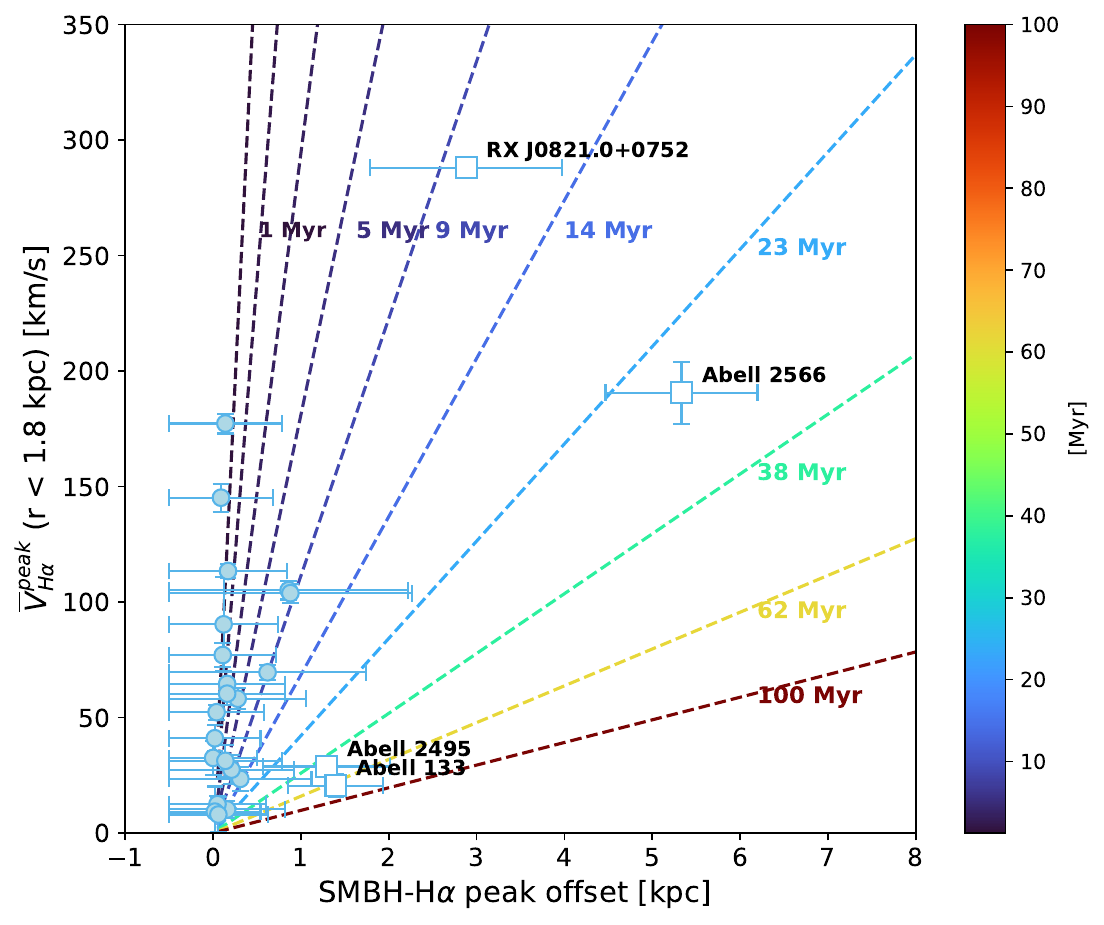}
    \caption{{\it Left}: Comparison between the SMBH - H$\alpha$ peak offset with the SMBH - X-ray peak offset; the dashed line marks the line of 1:1 scaling between the two distances. {\it Right:} Velocity of the warm gas, measured from the MUSE data within an extraction region centered on the H$\alpha$ peak with radius 1.8~kpc, versus the SMBH–H$\alpha$ peak offset. Dashed lines show the time required to traverse the distance on the x-axis at various (constant) velocities on the y-axis. For example, the brown line indicates that for a velocity of 50 km/s, the time needed to travel a distance of 6 kpc is approximately 100 Myr. Thus, depending on the warm gas velocity and the offset of the H$\alpha$ peak from the SMBH, each system lies along a different crossing time line.} In both panels, open squares represent systems with a significant H$\alpha$ offset, while filled circles represent the remaining objects in the sample.
    \label{fig:compareoffset}
\end{figure*}
\subsection{The fragmentation of multi-phase gas in cooling cores}\label{subsec:fragm}
To highlight the complexity of the spatial relationship between the X-ray peak, the H$\alpha$ peak, and the SMBH, we show in Fig~\ref{fig:compareoffset} the comparison between $\Delta^{\text{SMBH}}_{H\alpha}$ and $\Delta^{\text{SMBH}}_{\text{X-ray}}$. Within uncertainties, many systems exhibit a noticeable displacement of the X-ray peak relative to the BCG even when the H$\alpha$ emission remains coincident with the galaxy center. Three H$\alpha$-offset systems lie along the 1:1 line -- in these cases, this plot would suggest an overall alignment of the multi-phase gas peaks. However, the picture is more complex. In these systems -- RX~J0821.0+0752, Abell~2566, and Abell~133 -- the H$\alpha$ and X-ray peaks are both offset from the BCG by similar distances, but as evident in Fig. \ref{fig:multiexample}, they are not spatially coincident with each other. In contrast, Abell~2495 has a 5 times larger X-ray peak offset than the H$\alpha$ one (see also \citealt{rosignoli2024}). Clearly, these variations demonstrate the diversity in the spatial configuration of cool cores. We stress here that we are not suggesting a complete decoupling between the X-ray and H$\alpha$ phases. In 22 out of 25 systems, the X-ray peak lies within the H$\alpha$ nebula (Fig.~\ref{fig:multiexample} and \ref{fig:multi1}), indicating that an overall spatial association persists. Furthermore, we find that also the H$\alpha$ peak or the SMBH usually lie within a region of enhanced X-ray emission: Fig.~\ref{fig:contours1} shows that in 17 out of the 20 systems with a significant X-ray peak - SMBH/H$\alpha$ peak offset, the SMBH or the H$\alpha$ peak still lie within the region where the X-ray emissivity ($\appropto n_{e}^{2}$) is statistically the highest.
\par It is difficult to identify the cause of the above diversity. The dynamics of gas in the systems of our sample is likely influenced by the same physical mechanisms (sloshing due to minor mergers, jet-driven uplift, e.g., \citealt{million2010,hamer2012,mcnamara2016}). The observed variation among the 25 systems therefore suggests that they are likely being observed at different stages of the same mechanisms driving the offset.
For example (see Sec.~\ref{subsec:timescales}), in some systems, the X-ray and H$\alpha$ peaks may still be moving away from the SMBH, while in others, they may be falling back toward it.
\par Additionally, the fact that the X-ray and H$\alpha$ peaks are not spatially coincident is difficult to reconcile with a simple, yet well-supported, cooling scenario -- where different phases of the cooling gas are expected to be co-spatial (e.g., \citealt{mcnamara2016,gaspari2017,voit2017}). The fact that the X-ray offsets are, on average, larger than the H$\alpha$ ones is informative, possibly indicating that one phase has decoupled more rapidly than the other from the central potential (see also \citealt{hamer2012}).
Due to the typically higher densities and lower temperature of H$\alpha$-emitting filaments compared to those of the hot gas (e.g., \citealt{olivares2025}), the warm gas may be less easily displaced by hydrodynamic effects (e.g., ram pressure) than the diffuse hot ICM. Clearly, if the warm gas originates from ICM cooling, then the separation must have occurred \textit{after} the warm phase had cooled out of the hot phase -- otherwise, the warm gas peak would also appear offset and coincident with the X-ray peak.
\subsection{Timescales of the warm gas offsets}\label{subsec:timescales}
Verifying whether the warm gas cooled out of the ICM before or after the separation occurred would require reconstruction of the sequence of events, likely through tailored numerical simulations. However, we can still provide a rough estimate of the timescales over which the observed H$\alpha$ offsets develop and compare these with the hot gas cooling time. To do so, we use the kinematics of the H$\alpha$ line, by measuring the average velocity of the warm gas peak with respect to the redshift of the central galaxy (see Tab.~\ref{tab:completeinfo}) within a circular aperture of radius 1.8 kpc\footnote{The smallest physical scale consistently probed across our sample is set by the largest uncertainty in the measured distance between the H$\alpha$ peak and the SMBH, approximately 1.8 kpc.} for each system and plotting it against the projected H$\alpha$–SMBH offset in Fig. \ref{fig:compareoffset} (right). Overlaid on this plot are lines representing the time required to traverse the distance on the x-axis at various velocities. The plot shows that the typical velocities in systems with coincident H$\alpha$ peak and SMBH range between 0 -- 200 km/s. The four H$\alpha$-offset systems cover a wide parameter space in this plot, reflecting once again the diversity in cooling cores. Notably, the two systems with the smallest offsets ($\sim$1.3 kpc; Abell~2495 and Abell~133) also exhibit the lowest velocities ($\sim20-30$~km/s), while the two with the largest offsets ($\sim$3–5 kpc; Abell~2566 and RX J0821.0+0752) show the highest velocities ($\sim200-300$~km/s). This trend may supports a scenario in which, when the H$\alpha$ peak is displaced from the SMBH, stronger bulk motions produce greater separations between gas components. However, there are at least two caveats. The first is that any motion (whether due to sloshing or AGN-driven uplift) is unlikely to occur at constant velocity. In sloshing scenarios, for example, the gas likely follows an oscillatory trajectory, with velocities varying along the path (e.g., \citealt{johnson2012}). Statistically, one would expect the velocity to be smallest at the largest offsets (i.e., near apocenter) and highest near the smallest offsets (pericenter), which would produce the opposite trend to what is observed for the four H$\alpha$ peak offset systems. The second and possibly most relevant one is that projection effects are likely significant (see Fig. \ref{fig:multi1}). Abell~133 and Abell~2495 show narrow, elongated H$\alpha$ morphologies, suggesting motion predominantly in the plane of the sky (e.g., \citealt{hatch2006}). In these cases, the observed offset may closely approximate the 3D displacement, while the line-of-sight velocity might underestimate the 3D velocity -- meaning these systems would shift upward in Fig. \ref{fig:compareoffset} (right), implying shorter timescales. Conversely, Abell~2566 and RX J0821.0+0752 exhibit rounder H$\alpha$ morphologies, and high $\overline{V}^{peak}_{H\alpha}$, consistent with non-negligible motion along the line of sight. For these systems, the true 3D spatial offsets may exceed the projected values, implying longer timescales (i.e., a shift to the right in the plot). Taken together, these geometrical considerations suggest that all four systems likely lie along similar time tracks ($t_{\text{offset}} \sim 20-40$  Myr). Nevertheless, because the true 3D geometry is not directly measurable, we adopt the projected quantities shown in Fig.~\ref{fig:compareoffset} (right) and conclude that the characteristic timescales over which these offsets develop range from approximately 10 to 70 Myr. We note that the observed timescales are comparable to the typical free-fall time of gas at $r \leq 6$~kpc within a $10^{10}$–$10^{11}$~M$\odot$ galaxy ($t{\text{ff}} \propto r^{3/2} \times M^{-1/2} \lesssim 25$~Myr), suggesting that the warm gas may be moving ballistically. At the same time, these timescales are longer than the typical sound-crossing time at the same radius in gas with $0.5 \leq kT,[\text{keV}] \leq 5$ ($t_{\text{sc}} \propto r \times kT^{-1/2} \lesssim 15$~Myr), which is consistent with subsonic motions of the hot gas (e.g., \citealt{su2017}).  Moreover, the timescales are shorter than the central cooling time $t_{cool}\appropto T^{1/2} n_e^{-1}$ of the X-ray gas ($\sim$400~Myr for Abell~133, \citealt{2009ApJS..182...12C}; $\sim$350~Myr for Abell~2495, \citealt{rosignoli2024}; $\sim$400~Myr for Abell~2566, from the temperature and density values in \citealt{kadam2024}; $\sim$350~Myr for RX~J0821.0+0752, \citealt{2019ApJ...870...57V}). Therefore, the fact that the X-ray peak is offset from the warm gas peak is in not in contrast with a cooling origin -- the warm and hot gas phases may have detached from one another after condensation.

\subsection{Connection to the SMBH activation}\label{subsec:connectsmbh}
Our analysis of Section \ref{sec:results} showed that all four systems exhibiting a spatial offset between the H$\alpha$ peak and the SMBH also show no detectable VLBA radio cores, with upper limits on their radio power of $P_{\text{5~GHz}}\leq10^{21 - 22}$~W/Hz. The absence of radio-bright jets in H$\alpha$-offset systems suggests that the central concentration of warm gas plays a key role in sustaining powerful SMBH activity. This observational coincidence may support the idea that the mode of accretion in central galaxies is not dominated by accretion of hot gas at the Bondi rate (\citealt{bondi1952}; see e.g., \citealt{allen2006,fujita2014,prasad2024}), but instead by the intermittent infall of cooled gas clouds condensed out of the ICM (e.g., \citealt{pizzolato2005,gaspari2013,voit2017}; see also \citealt{russell2013,bambic2023} for detailed discussions). Our results seem to suggest that warm gas, when peaking near the SMBH, provides a sufficient fuel reservoir to enhance the accretion rate and trigger jet formation, which results in a relatively high parsec-scale radio power of the SMBH ($P_{\text{5~GHz}}$ up to $10^{24 - 25}$~W/Hz, see Fig. \ref{fig:offset1}). This aligns with theoretical models of cold accretion, where cold molecular gas (originating from the warm phase) primarily fuels the SMBH \citep{pizzolato2005,gaspari2013,gaspari2017,voit2017}.
\par Our conclusion on the inactive state of the SMBHs in the four H$\alpha$ offset systems is additionally supported  by existing studies of AGN feedback (i.e., the presence of X-ray cavities) in these objects. In Abell 2495, the detected cavities seem associated with past outbursts that are now fading \citep[see][]{2019ApJ...885..111P,rosignoli2024}; the cavities are also asymmetric with respect to the X-ray peak. In Abell 133, cavities are detected on one side of the X-ray peak, at large radii ($\sim$60 kpc), and co-spatial with remnant, aged radio emission, consistent with diffuse bubbles that likely detached long ago \citep[see][]{Randall_2010}. In Abell 2566, no clear cavities are visible in the {\it Chandra} data, and the dedicated analysis by \citet{kadam2024} does not report any. Finally, in RXJ 0821+0752, one depression is found on one side of the core \citep[see][]{bayer-kim_2002,2019ApJ...870...57V}. If interpreted as a cavity, this feature would most likely trace a fading past outburst, given the lack of associated extended radio emission. Taken together, these results indicate that jet feedback has indeed been ongoing in these systems, but primarily in the past, since the observed cavities are all linked to fading outbursts. This supports our conclusion that no major AGN fueling episode has occurred in recent times (within a few Myr, i.e. the typical age of young X-ray cavities).
\par For clarity, we stress that we are not suggesting that all H$\alpha$-offset systems have an  inactive SMBH, but only that any core radio emission is too faint to be detected at current VLBA sensitivities. If future studies identify cases of VLBI-detected SMBH in central galaxies with offset H$\alpha$ peaks, our results suggest that such systems will be characterized by very low radio powers. However, this prediction is sensitive to the timescales. As shown in Fig. \ref{fig:compareoffset}, depending on the combination of spatial distance and warm gas velocity, the timescales might be short enough to allow the periodic passages of the H$\alpha$ peak near the SMBH, the deposition of fuel, and the maintenance of AGN activity (see also \citealt{rosignoli2024}). The relevant timescale to match is the synchrotron cooling timescale of relativistic electrons, which depends on frequency and magnetic field strength as $t_{syn}\propto\nu^{-1/2}B^{-3/2}$ (e.g., \citealt{longair2011}). For magnetic field strengths typical of parsec-scale radio jets (i.e., $B\geq0.1$~mG, e.g., \citealt{xu2000,dallacasa2021}), the cooling time at GHz frequencies is $t_{syn}\leq5$~Myr. Therefore, if an offset event occurred within the past few Myr, the SMBH could still show radio emission on pc scales. 
\par Finally, we return to our four H$\alpha$-offset, VLBA undetected systems, and consider the case in which future, deeper VLBI observations reveal emission with radio power just below our upper limits. First, we note that in all four systems with offset H$\alpha$ peaks, the SMBH is still embedded in the extent of the warm gas nebula (see Fig.~\ref{fig:multi1}; in A2566, a secondary H$\alpha$ blob is also present at the SMBH location), suggesting that some fuel might still be available -- though seemingly not enough to support radio powers exceeding $10^{22}$~W/Hz. Second, we note that even a VLBI detection may not imply that these systems host jets. At low radio powers, other mechanisms can become the dominant origin of synchrotron emission (e.g., \citealt{panessa2019} for a review). For the case of VLBI scales, and considering the early type galaxies at the center of clusters and groups, the most likely alternative synchrotron source is the advection-dominated accretion flow (ADAF) surrounding the SMBH (e.g., \citealt{doi2011,panessa2019}; see also \citealt{schellenberger2024} for an example of ADAF emission in a central group galaxy). The maximal ADAF radio power at a given frequency is a function of the SMBH mass (e.g., \citealt{wucao2005}), with the dependence at 5~GHz being as $P_{ADAF}\propto\,M_{BH}^{1.25}$. Based on the K-band magnitudes of their host galaxies \citep{2007MNRAS.379..711G}, the SMBHs in the four H$\alpha$-offset systems have masses of $1-2\times10^{9}$~M$_{\odot}$, implying maximum ADAF radio power at 5~GHz of about $1-3\times10^{21}$~W/Hz. These values are just below our current 5$\sigma$ upper limits (see Tab. \ref{tab:completeinfo}). Thus, if future high-sensitivity VLBI observations reveal faint (and unresolved) radio emission in these systems, it is still possible that this could arise from an ADAF rather than a jet. 

\section{Summary}\label{sec:summary}
In this work, we explored spatial offsets between the multi-phase gas around central galaxies, and their connection to the SMBH activation, in a sample of 25 cool core galaxy groups and clusters. Our strategy combined observations from some of the highest angular resolution instruments available in their respective bands: {\it Chandra} for the X-ray-emitting hot gas, VLT/MUSE for optical emission lines tracing the warm gas, and VLBA for pc-scale radio emission from the AGN. Our results can be summarized as follows:
\begin{itemize}
   \item On average, the offsets between the X-ray peak and the SMBH ($\langle\Delta^{\text{SMBH}}{\text{X-ray}}\rangle = 4.8$~kpc) are significantly larger than those between the H$\alpha$ peak and the SMBH ($\langle\Delta^{\text{SMBH}}{\text{H}\alpha}\rangle = 0.6$~kpc). There is also a higher incidence of X-ray offsets ($\sim$80\%) than of H$\alpha$ offsets ($\sim$15\%). This evidence supports a scenario in which gas sloshing primarily drives these displacements, with the denser warm phase being more difficult to separate from the central galaxy than the diffuse hot halo. 
   \item The spatial configuration of multi-phase gas in cooling cores shows a large degree of variation. While the extent of the warm ionized  nebula often overlaps with the location of the X-ray peak, the peak of the H$\alpha$ emission is rarely coincident with it. This evidence can still be reconciled with a cooling scenario -- where the warm gas cools out of the hot gas -- by considering the relevant timescales. Our data indicate that offsets between the H$\alpha$ peak and the SMBH develop on timescales of 10 -- 70 Myr, significantly shorter than the typical cooling time of the central hot gas (300 -- 400 Myr). This suggests that the warm and hot gas phases may have detached from one another after condensation.
   \item A striking result emerges from our study of how the X-ray and H$\alpha$ offsets relate to SMBH activation. While there is no apparent correlation between the magnitude of the X-ray offset and the radio power of the SMBH, all 4 systems showing an H$\alpha$ peak - SMBH offset of $\geq$1~kpc also lack detected radio cores, with upper limits on $P_{\text{5~GHz}}$ of $\leq10^{21 - 22}$~W/Hz. In the remaining 21 systems, the pc-scale radio powers can reach $P_{5~\text{GHz}} \sim 10^{24-25}$~W/Hz. These results suggest that a central concentration of warm gas plays an important role in sustaining powerful SMBH activity. This, in turn, supports the idea that cold-mode accretion contributes to fueling AGN with relatively high radio power.
\end{itemize}
In an upcoming paper (paper II, Ubertosi et al. in preparation), we will expand this analysis to the cold molecular phase, which represents the end product of the multiphase cooling cascade and, according to cold accretion models, is the most closely related to AGN fueling. We also highlight that this analysis would not have been possible without the high spatial resolution provided by the {\it Chandra} X-ray telescope, matching that of VLT/MUSE. The future of multi-phase gas studies relies on maintaining such capabilities, either through the continued operation of {\it Chandra} or through next-generation observatories. In this context, the Advanced X-ray Imaging Satellite (AXIS; see \citealt{axis2023}), with its proposed 1.5" angular resolution, represents a promising successor for high-resolution X-ray imaging of galaxy cluster cores.

\begin{acknowledgments}
The authors sincerely thank the anonymous referee for providing a clear and constructive report on our manuscript. F. Ubertosi thanks the Smithsonian Astrophysical Observatory for the hospitality and support during his visits in 2024 and 2025, during which this project was developed and finalized, respectively. F. Ubertosi thanks E. Giunchi for the useful discussion on optical data analysis. FU and MG acknowledge support from the research project PRIN 2022 ``AGN-sCAN: zooming-in on the AGN-galaxy connection since the cosmic noon", contract 2022JZJBHM\_002 -- CUP J53D23001610006. EOS acknowledges support from the Smithsonian Combined Support for Life on a Sustainable Planet, Science, and Research administered by the Office of the Under Secretary for Science and Research. PT acknowledges support from NASA’s NNH22ZDA001N Astrophysics Data and Analysis Program under award 24-ADAP24-0011. The National Radio Astronomy Observatory is a facility of the National Science Foundation operated under cooperative agreement by Associated Universities, Inc. Based on data obtained from the ESO Science Archive Facility with DOI(s): \citet{ESO_DOI}. This research has made use of a list of \textit{Chandra} datasets, obtained by the \textit{Chandra} X-ray Observatory, contained in the \textit{Chandra} Data Collection (CDC)~\dataset[doi:10.25574/cdc.470]{https://doi.org/10.25574/cdc.470}. Basic research in radio astronomy at the Naval Research Laboratory is supported by 6.1 Base funding.

\end{acknowledgments}

\begin{contribution}

FU developed the research concept, analyzed the data, and was responsible for writing and submitting the manuscript. GS and EOS developed the idea of analyzing a large collection of VLBA observations of central galaxies. EOS provided funding to support the publication of the manuscript. FB provided extensive support on the data analysis and interpretation. All co-authors contributed equally to the review of this manuscript.


\end{contribution}

%
\facilities{NRAO (VLBA), CXO, VLT (MUSE)}

\software{AIPS \citep{aips1990,Greisen03}, CASA \citep{casa2007}, CIAO \citep{2006SPIE.6270E..1VF}, PLATEFIT \citep{tremonti2004}.
          }


\appendix

\begin{table}[ht!]
\centering
\caption{Measured properties for the 25 objects in our sample,  ordered by increasing redshift.} \label{tab:completeinfo}\renewcommand{\arraystretch}{0.8}
\begin{tabular}{l|c|c|c|c|c|c|c}
Name & z & $P_{5\text{GHz}}$& SMBH & X-ray peak & H$\alpha$ peak  & $\Delta X^{\text{SMBH}}_{\text{X-ray}}$ & $\Delta X^{\text{SMBH}}_{\text{H}\alpha}$  \\
 &  & [$10^{22}$~W/Hz] & & & & [kpc] & [kpc]  \\
\hline
\multirow{2}{*}{NGC~5846} & \multirow{2}{*}{0.00572} & \multirow{2}{*}{$0.040 \pm 0.004$} & 15:06:29.2917 & 15:06:29.0231& 15:06:29.2945  & \multirow{2}{*}{0.48$\pm$0.12} & \multirow{2}{*}{0.02$\pm$0.06} \\
& & & +1:36:20.342 & +1:36:20.828 & +1:36:20.473 & & \\
\hline

\multirow{2}{*}{NGC~5044} & \multirow{2}{*}{0.009} & \multirow{2}{*}{$0.28 \pm 0.03$} & 13:15:23.9613 & 13:15:24.0621 & 13:15:23.9627  & \multirow{2}{*}{0.36$\pm$0.18} & \multirow{2}{*}{0.00$\pm$0.12}\\
& & & -16:23:07.549 & -16:23:06.255 & -16:23:07.529 & & \\
\hline

\multirow{2}{*}{Abell~3581} & \multirow{2}{*}{0.02179} & \multirow{2}{*}{$16.9 \pm 1.7$} & 14:07:29.7622 & 14:07:29.8564 & 14:07:29.7639  & \multirow{2}{*}{0.92$\pm$0.44} & \multirow{2}{*}{0.02$\pm$0.19} \\
& & & -27:01:04.293 & -27:01:05.952 & -27:01:04.323 & & \\

\hline

\multirow{2}{*}{NGC~7237} & \multirow{2}{*}{0.02621} & \multirow{2}{*}{$2.7 \pm 0.3$} & 22:14:46.8819 & 22:14:46.0953 & 22:14:46.8739  & \multirow{2}{*}{6.81$\pm$0.53} & \multirow{2}{*}{0.06$\pm$0.24} \\
& & & +13:50:27.115 & +13:50:21.129 & +13:50:27.150 & & \\

\hline

\multirow{2}{*}{Abell~496} & \multirow{2}{*}{0.03273} & \multirow{2}{*}{$10.8 \pm 1.1$} & 04:33:37.8413 & 4:33:37.7749 & 04:33:37.8357 & \multirow{2}{*}{1.15$\pm$0.65} & \multirow{2}{*}{0.04$\pm$0.28} \\
& & & -13:15:42.989 & -13:15:41.566 & -13:15:43.016 & & \\

\hline

\multirow{2}{*}{Abell~2052} & \multirow{2}{*}{0.03453} & \multirow{2}{*}{$91.4 \pm 9.1$} & 15:16:44.5130 & 15:16:43.6611 & 15:16:44.4943  & \multirow{2}{*}{8.81$\pm$0.69} & \multirow{2}{*}{0.31$\pm$0.47} \\
& & & +7:01:18.100 & +7:01:19.779 & +7:01:17.731 & & \\

\hline

\multirow{2}{*}{ZwCl 0335+0956} & \multirow{2}{*}{0.03520} & \multirow{2}{*}{$0.67 \pm 0.07$} & 3:38:40.5505 & 3:38:41.0218 & 3:38:40.5557  & \multirow{2}{*}{8.77$\pm$0.70} & \multirow{2}{*}{0.14$\pm$0.38} \\
& & & +9:58:12.042 & +9:58:01.585 & +9:58:11.853 &  & \\

\hline

\multirow{2}{*}{ACT-CL J1521.8+0742$^{(a)}$} & \multirow{2}{*}{0.04419} & \multirow{2}{*}{$\leq0.08$} & 15:21:51.8509 & 15:21:51.8860 & 15:21:51.8545 & \multirow{2}{*}{0.43$\pm$0.88} & \multirow{2}{*}{0.12$\pm$0.51}\\
& & & +7:42:31.831 & +7:42:32.015 & +7:42:31.703 & & \\

\hline

\multirow{2}{*}{Abell~1644} & \multirow{2}{*}{0.047} & \multirow{2}{*}{$56.4 \pm 7.4$} & 12:57:11.5961 & 12:57:11.9829 & 12:57:11.6007 & \multirow{2}{*}{6.00$\pm$0.92} & \multirow{2}{*}{0.09$\pm$0.42} \\
& & & -17:24:34.135 & -17:24:30.740 & -17:24:34.166 & & \\

\hline

\multirow{2}{*}{Abell~2626} & \multirow{2}{*}{0.055} & \multirow{2}{*}{$3.0 \pm 0.4$} & 23:36:30.4916 & 23:36:30.5021 & 23:36:30.4939 & \multirow{2}{*}{3.35$\pm$1.07} & \multirow{2}{*}{0.11$\pm$0.53} \\
& & & +21:08:47.460 & +21:08:50.607 & +21:08:47.557 & & \\

\hline

\multirow{2}{*}{Abell~85} & \multirow{2}{*}{0.05536} & \multirow{2}{*}{$0.29 \pm 0.03$} & 0:41:50.4696 & 0:41:50.4974 & 0:41:50.4661 & \multirow{2}{*}{0.51$\pm$1.08} & \multirow{2}{*}{0.05$\pm$0.45} \\
& & & -9:18:11.394 & -9:18:11.656 & -9:18:11.399 & & \\

\hline

\multirow{2}{*}{Abell~133$^{(a)}$} & \multirow{2}{*}{0.057} & \multirow{2}{*}{$\leq0.20$} & 1:02:41.7460 & 1:02:41.7175 & 1:02:41.7328 & \multirow{2}{*}{0.54$\pm$1.11} & \multirow{2}{*}{1.40$\pm$0.54}\\
& & & -21:52:55.757 & -21:52:56.057 & -21:52:54.536 & & \\

\hline

\multirow{2}{*}{Abell~1991} & \multirow{2}{*}{0.05921} & \multirow{2}{*}{$6.5 \pm 0.7$} & 14:54:31.4879 & 14:54:31.4247 & 14:54:31.4871 & \multirow{2}{*}{10.19$\pm$1.15} & \multirow{2}{*}{0.14$\pm$0.54} \\
& & & +18:38:32.413 & +18:38:41.231 & +18:38:32.521 & & \\

\hline

\multirow{2}{*}{Abell~2495$^{(b)}$} & \multirow{2}{*}{0.0794} & \multirow{2}{*}{$\leq0.30$} & 22:50:19.700 & 22:50:19.4448 & 22:50:19.6567 & \multirow{2}{*}{6.20$\pm$1.50} & \multirow{2}{*}{1.29$\pm$0.73} \\
& & & +10:54:12.526 & +10:54:14.254 & +10:54:12.075 & & \\

\hline

\multirow{2}{*}{Abell~478} & \multirow{2}{*}{0.086} & \multirow{2}{*}{$5.9 \pm 0.6$} & 4:13:25.2778 & 4:13:25.1296 & 4:13:25.2677 & \multirow{2}{*}{3.53$\pm$1.61} & \multirow{2}{*}{0.16$\pm$0.75} \\
& & & +10:27:54.777 & +10:27:54.930 & +10:27:54.786 & & \\
\hline

\multirow{2}{*}{Abell~2566$^{(c)}$} & \multirow{2}{*}{0.0871} & \multirow{2}{*}{$\leq0.37$} & 23:16:05.0230 & 23:16:04.7542 & 23:16:04.9838 & \multirow{2}{*}{6.80$\pm$1.63} & \multirow{2}{*}{5.33$\pm$0.86} \\
& & & -20:27:48.400 & -20:27:46.701 & -20:27:45.169 & & \\

\hline

\multirow{2}{*}{MCXC J1558.3-1410} & \multirow{2}{*}{0.097} & \multirow{2}{*}{$793.5 \pm 79.4$} & 15:58:21.9481 & 15:58:21.8052 & 15:58:21.9482 & \multirow{2}{*}{4.92$\pm$1.80} & \multirow{2}{*}{0.16$\pm$1.00}\\
& & & -14:09:59.052 & -14:09:57.276 & -14:09:58.969 & & \\

\hline
\multirow{2}{*}{PKS 0745-191} & \multirow{2}{*}{0.1028} & \multirow{2}{*}{$16.9 \pm 1.7$} & 7:47:31.3264 & 7:47:31.0735 & 7:47:31.3230 & \multirow{2}{*}{6.78$\pm$1.89} & \multirow{2}{*}{0.16$\pm$1.01} \\
& & & -19:17:39.986 & -19:17:40.013 & -19:17:39.910 & & \\

\hline

\multirow{2}{*}{RX J0821.0+0752$^{(d)}$} & \multirow{2}{*}{0.11007} & \multirow{2}{*}{$\leq0.78$} & 8:21:02.2550 & 8:21:02.3319 & 8:21:02.2555  & \multirow{2}{*}{2.70$\pm$2.01} & \multirow{2}{*}{2.88$\pm$1.09} \\
& & & +7:51:47.222 & +7:51:47.907 & +7:51:48.672 & & \\

\hline

\multirow{2}{*}{Abell~1664} & \multirow{2}{*}{0.12798} & \multirow{2}{*}{$31.4 \pm 5.8$} & 13:03:42.5652 & 13:03:42.5658 & 13:03:42.5658 & \multirow{2}{*}{2.20$\pm$2.29} & \multirow{2}{*}{0.62$\pm$1.03} \\
& & & -24:14:42.216 & -24:14:41.312 & -24:14:42.484 & & \\

\hline

\multirow{2}{*}{RX J1350.3+0940} & \multirow{2}{*}{0.13255} & \multirow{2}{*}{$1345 \pm 140$} & 13:50:22.1360 & 13:50:21.6860 & 13:50:22.1428 & \multirow{2}{*}{15.35$\pm$2.86} & \multirow{2}{*}{0.28$\pm$1.52} \\
& & & +9:40:10.656 & +9:40:09.878 & +9:40:10.724 & & \\

\hline

\multirow{2}{*}{Abell~2204} & \multirow{2}{*}{0.152} & \multirow{2}{*}{$54.7 \pm 5.5$} & 16:32:46.9447 & 16:32:46.8972 & 16:32:46.9425 & \multirow{2}{*}{2.18$\pm$2.64} & \multirow{2}{*}{0.21$\pm$1.60} \\
& & & +5:34:32.747 & +5:34:32.364 & +5:34:32.679 & & \\

\hline

\multirow{2}{*}{MCXC J2014.8-2430} & \multirow{2}{*}{0.1555} & \multirow{2}{*}{$104.0 \pm 11.4$} & 20:14:51.7332 & 20:14:51.5104 & 20:14:51.7376 & \multirow{2}{*}{8.66$\pm$2.69} & \multirow{2}{*}{0.17$\pm$1.37} \\
& & & -24:30:22.723 & -24:30:23.007 & -24:30:22.522 & & \\

\hline

\multirow{2}{*}{Abell~383} & \multirow{2}{*}{0.18884} & \multirow{2}{*}{$38.0 \pm 4.6$} & 2:48:03.3955 & 2:48:03.3037 & 2:48:03.3904 & \multirow{2}{*}{6.77$\pm$3.15} & \multirow{2}{*}{0.86$\pm$1.77} \\
& & & -3:31:44.856 & -3:31:46.521 & -3:31:44.963 & & \\

\hline

\multirow{2}{*}{MCXC J0331.1-2100} & \multirow{2}{*}{0.19276} & \multirow{2}{*}{$61.7 \pm 9.9$} & 3:31:06.0316 & 3:31:05.9913 & 3:31:06.0394 & \multirow{2}{*}{3.33$\pm$3.22} & \multirow{2}{*}{0.88$\pm$1.65} \\
& & & -21:00:33.018 & -21:00:32.467 & -21:00:33.006 & & \\

\hline
\end{tabular}
\tablecomments{(1) Name; (2) redshift; (3) 5~GHz radio power from the VLBA data, measured from the peak flux in the VLBA maps; upper limits are given at 5$\sigma_{rms}$ (see Sec~\ref{sec:sampledata}); (4) coordinates of the SMBH from VLBA data (see notes for non-detections; see also App.~\ref{appsub:vlba}); (5) Coordinates of the X-ray peak from {\it Chandra} data (see App.~\ref{appsub:chandra}); (6) Coordinates of the H$\alpha$ peak from MUSE data (see App.~\ref{appsub:muse}); (7) Spatial offset between the SMBH and the X-ray peak; (8) Spatial offset between the SMBH and the H$\alpha$ peak. $^{(a)}$: SMBH located in archival VLA images at 1.4~GHz (see the \href{http://www.vla.nrao.edu/astro/nvas/}{NVAS project}), with positional accuracy of 0.12" (10\% of the beam FWHM). $^{(b)}$: SMBH located in archival VLA images at 4.8~GHz \citep{2019ApJ...885..111P} with positional accuracy of 0.1" (10\% of the beam FWHM). $^{(c)}$: SMBH located in VLASS images at 3.0~GHz with positional accuracy of 0.2" (see the \href{https://science.nrao.edu/vlass/vlass-se-continuum-users-guide}{VLASS continuum user guide}). $^{(d)}$: SMBH located in HST images \citep{2019ApJ...870...57V}, with positional accuracy of 0.1".
}
\end{table}

\section{The sample}\label{app:sample}
We report in Tab.~\ref{tab:completeinfo} the list of 25 systems selected for our study (ordered by increasing redshift), along with the positions of the SMBH, of the X-ray peak, and of the H$\alpha$ peak. We also report the spatial offsets between the different components. 

\section{Data reduction and analysis}\label{app:data}
\subsection{VLBA data}\label{appsub:vlba}
We reduced the VLBA data using standard reduction techniques in \texttt{AIPS}\footnote{\url{https://www.aips.nrao.edu/}.}, including the correction for the Earth orientation parameters and ionospheric delays, the application of digital sampling corrections, the removal of instrumental delays in phases, the bandpass calibration, and the correction for time-dependent delays in phases. Ultimately, we applied the calibration and averaged channels in the different spectral windows. No self-calibration was performed on the target visibilities, to avoid losing the information on the absolute position. 
Imaging was performed in AIPS adopting briggs \citep{briggs1995} weighting (with \texttt{robust = 0}). Since the phase-center of the data did not always match the exact position of the radio core, we imaged the whole field of view (a few arcseconds) to secure the identification of sources far from the phase center. 
\par For the 19 detections, the peak in the maps was used as a tracer for the position of the SMBH\footnote{Possible core-shift effects or self-absorption masking the true core (e.g., \citealt{sokolovsky2011}) introduce systematic errors of $\sim$0.01$\arcsec$ at most, far negligible with respect to the {\it Chandra} and MUSE resolution.}. For the 5 non-detections, we considered the position of the radio core in high-resolution VLA imaging (for ACT-CL~J1521.8+0742, Abell~133, Abell~2495, and Abell~2566) or of the BCG optical peak in HST imaging (for RX~J0821.0+0752), both of which have positional accuracies of $\sim$0.1'', as tracer of the position of the SMBH (see details in Tab. \ref{tab:completeinfo}). To check for potential biases in using these alternative methods to locate the SMBH in VLBA-undetected systems, we proceeded with the following validation procedures:
\begin{itemize}
    \item To validate the high-resolution VLA imaging method, we considered the 20 VLBA detections in our sample and compared the position of the radio core in VLBA images with the position of the radio core in 3~GHz images from the Very Large Array Sky Survey (VLASS\footnote{See \url{https://science.nrao.edu/vlass/data-access/vlass-epoch-1-quick-look-users-guide}.}), which has a typical positional uncertainty of 0.2", comparable to that of the VLA images we employed (see the notes in Tab.~\ref{tab:completeinfo}). The comparison shown in Fig. \ref{fig:validatepeak} (left panel) demonstrates that the lower-resolution radio images provide good tracers of the position of the SMBH, as the distribution is centered around zero with a standard deviation in $\delta \text{RA}, \delta \text{DEC} = 0.16"$ (consistent with the VLASS positional accuracy, $\sim0.2"$). 
    \item To validate the HST imaging method, we considered the 11 VLBA detections in our sample with archival HST images\footnote{See \url{https://hla.stsci.edu/}.} (NGC~5846, NGC~5044, Abell~3581, NGC~7237, Abell~496, Abell~2052, ZwCl~0335+0956, Abell~2626, Abell~478, PKS~0745-191, and Abell~2204), and compared the position of the radio core in VLBA images with the position of the BCG optical peak. The comparison shown in Fig. \ref{fig:validatepeak} (middle panel) demonstrates that the HST images provide good tracers of the position of the SMBH, as the distribution is centered around zero with a standard deviation in $\delta \text{RA}, \delta \text{DEC} = 0.05"$ (comparable with the positional accuracy of HST, $\leq0.1"$). 
\end{itemize}
Based on these tests, we are confident that using VLA images (for ACT-CLJ1521.8+0742, Abell~133, Abell~2495, and Abell~2566) or HST images (for RX~J0821.0+0752) to estimate SMBH positions in VLBA-undetected systems does not introduce systematic biases in our results.

\subsection{VLT/MUSE data}\label{appsub:muse}
We present optical nebular emission line maps for the 25 sources in our sample from an analysis of data obtained with the Multi-Unit Spectroscopic Explorer (MUSE). 
The MUSE data were reduced using the standard recipes provided by the MUSE pipeline \citep{weilbacher2014}. Then, we checked the astrometry of the cubes by checking the coordinates of detected sources against external catalogues (PanSTARRS and GAIA). When necessary, we corrected the cubes for any astrometric shift, which we found to be always smaller (below 0.3$\arcsec$) than the average seeing ($\sim$0.8$\arcsec$). 
We fitted the data following the same method described by \citet{olivares2019}, that is using \texttt{PLATEFIT}\footnote{\url{https://pyplatefit.readthedocs.io/en/latest/}.} \citep{tremonti2004} to simultaneously fit the stellar continuum and emission lines.  
We imposed a threshold in signal-to-noise ratio (SNR) of 3 for the detection of emission lines, 
and we produced maps of H$\alpha$ intensity (see Figure \ref{fig:multiexample} and \ref{fig:multi1}). The position of the warm gas peak corresponds to the pixel with the largest H$\alpha$ flux. The uncertainty on the location of the H$\alpha$ peak is given by the seeing of the observation. The velocities of the warm gas peak, traced by the H$\alpha$ line and shown in Fig.~\ref{fig:compareoffset}, are measured relative to the redshift of the central galaxy (see Tab.~\ref{tab:completeinfo}).

\subsection{Chandra data}\label{appsub:chandra}
We retrieved ACIS data from the \textit{Chandra} archive (\url{cda.harvard.edu/chaser}) and reprocessed the observations using \texttt{CIAO-4.16}\footnote{\url{https://cxc.cfa.harvard.edu/ciao/}.}; for each object, we restricted the analysis to all the available observations obtained with ACIS-S and ACIS-I, without gratings. 
The data were reprocessed with the \texttt{chandra\_repro} script; for each object with multiple ObsIDs, the astrometry of each ObsID was matched to that of the longest ObsID. We filtered the data from periods contaminated by background flares. Blank-sky event files were selected as background files, and normalized by the 9-12 keV count-rate of the observation. Then, we merged the event files and the background images obtaining mosaiced images for the target source and the background in the 0.5 - 7 keV band. For each target, we verified if the astrometry of the \textit{Chandra} mosaic needed further corrections beyond its nominal pointing accuracy (0.4$''$), and proceeded to update the coordinates when necessary.
As the peak in the X-rays can be dominated by non-thermal emission from point sources in the image (the BCG itself, in some cases; \citealt{russell2013}), we first identified point sources in the mosaics with \texttt{wavdetect}, and masked them by interpolating the counts from an annular region surrounding the point source and extending to 1.5$\times$ its extent. Then, to mitigate potential pixels to pixels variations (especially in the low-counts regime), we convolved the maps with a 1.0$\arcsec$ FWHM Gaussian (see also \citealt{2010A&A...513A..37H}). Ultimately, we identified the X-ray brightest pixel in the map and considered a 0.5$\arcsec$ radius as the uncertainty on the location of the X-ray peak. 
\par
\begin{figure*}[h!]
    \centering
    \includegraphics[width=0.33\linewidth]{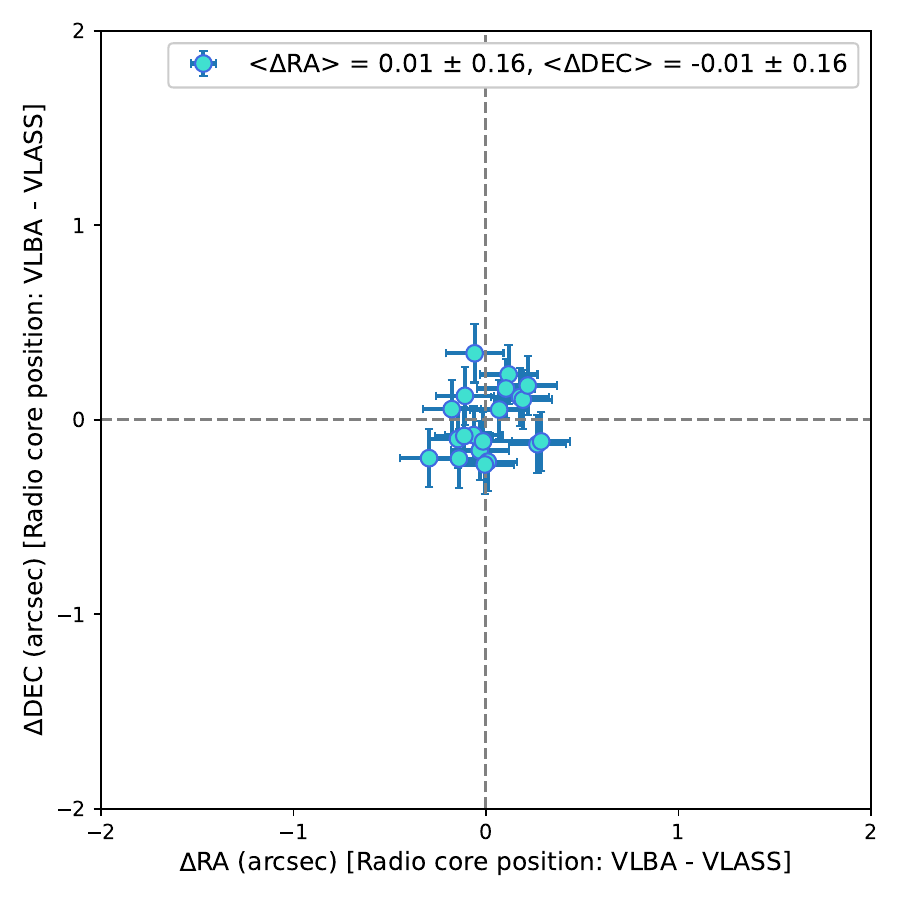}
    \includegraphics[width=0.33\linewidth]{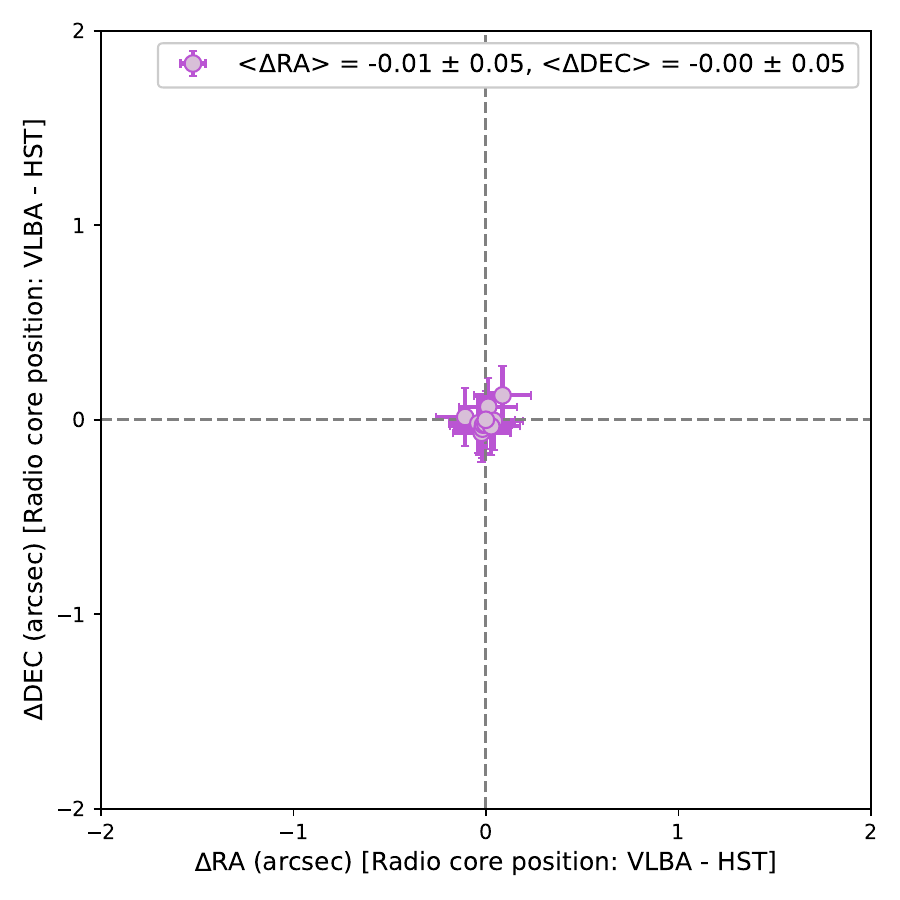}
    \includegraphics[width=0.33\linewidth]{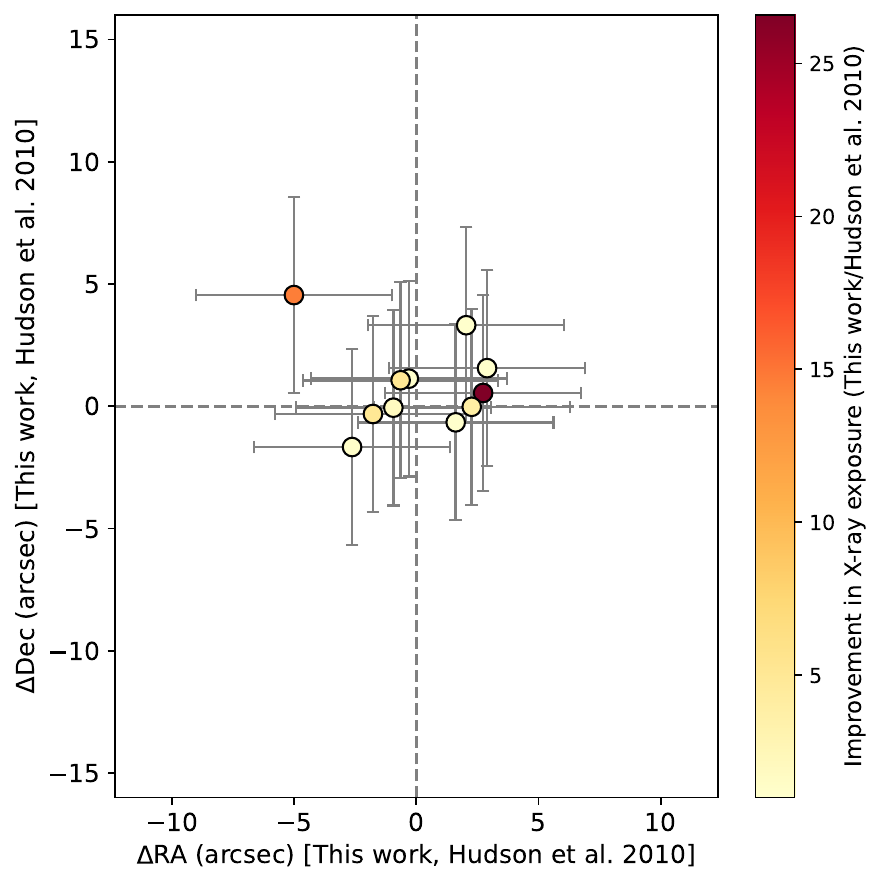}
    \caption{{\it Left}: Comparison between the position of the radio core in VLBA images and that in VLASS 3~GHz images for the 20 systems with VLBA detections (see Appendix \ref{appsub:vlba}). The distribution is centered near zero, with a dispersion of $\sim0.16"$, consistent with the VLASS positional accuracy ($\sim0.2"$). As noted in Appendix \ref{appsub:vlba}, this supports using VLA observations to estimate SMBH positions in VLBA-undetected systems (i.e., ACT-CL~J1521.8+0742, Abell~133, Abell~2495, and Abell~2566). {\it Center}: Comparison between the position of the radio core in VLBA images and the optical peak of the galaxy in HST images for the 11 systems with both VLBA detections and available HST data (see Appendix \ref{appsub:vlba} for details). The distribution is centered around zero, with a dispersion of approximately $0.05"$ -- comparable with the positional accuracy of HST ($\leq0.1"$). As discussed in Appendix \ref{appsub:vlba}, this validates the use of HST imaging to locate the SMBH in VLBA-undetected systems (i.e., RX~J0821.0+0752). {\it Right}: Comparison between the X-ray peak positions reported in this work (Table~\ref{tab:completeinfo}) and those from \citet{2010A&A...513A..37H} for 11 overlapping systems (see Appendix \ref{appsub:chandra}). Points are color-coded by the ratio of {\it Chandra} exposure times between this work and that of \citet{2010A&A...513A..37H}. As detailed in Appendix \ref{appsub:chandra}, the agreement validates our X-ray peak identification method. The only outlier is Abell~3581, with an offset of $\Delta$RA$=5.2"\pm4.1"$ and $\Delta$DEC~$=4.9"\pm4.1"$, likely explained by the 15$\times$ deeper exposure used in our analysis.}
    \label{fig:validatepeak}
\end{figure*}

 To validate our method of identifying the X-ray peak, we compare our results with those of \citet{2010A&A...513A..37H}, focusing on the 11 systems common to both samples. For each system, we considered the X-ray ray peak position from our work (Tab.~\ref{tab:completeinfo}) and from table 2 in \citet{2010A&A...513A..37H}, and then computed the angular offset in right ascension and declination between the two sets. We note that \citet{2010A&A...513A..37H} convolved the {\it Chandra} maps with a 8.0$\arcsec$ FWHM Gaussian, therefore we considered a 4$\arcsec$ radius as the uncertainty on their location of the X-ray peak. We show the comparison in Fig. \ref{fig:validatepeak} (right panel), where each point is color-coded according to the ratio of the {\it Chandra} exposure time used in our analysis to that used in the earlier study. This is meant to provide context for interpreting the agreement (longer exposures yield better SNR and potentially more accurate peak localization). We find a good agreement between our study and that of \citet{2010A&A...513A..37H}, since the distribution is centered at zero and 10/11 systems are consistent with a $0"$ difference. The only outlier is Abell~3581, with an offset of $\Delta$RA$=5.2"\pm4.1"$ and $\Delta$DEC~$=4.9"\pm4.1"$. This is explained by the 15$\times$ deeper {\it Chandra} exposure we employed ($\sim$90~ks here vs $\sim$6~ks in \citealt{2010A&A...513A..37H}), which likely provides a more accurate localization of the X-ray peak.
\par
\begin{figure}[ht!]
    \centering
    \includegraphics[width=0.19\linewidth, trim={0.2cm 0 1cm 1.6cm}]{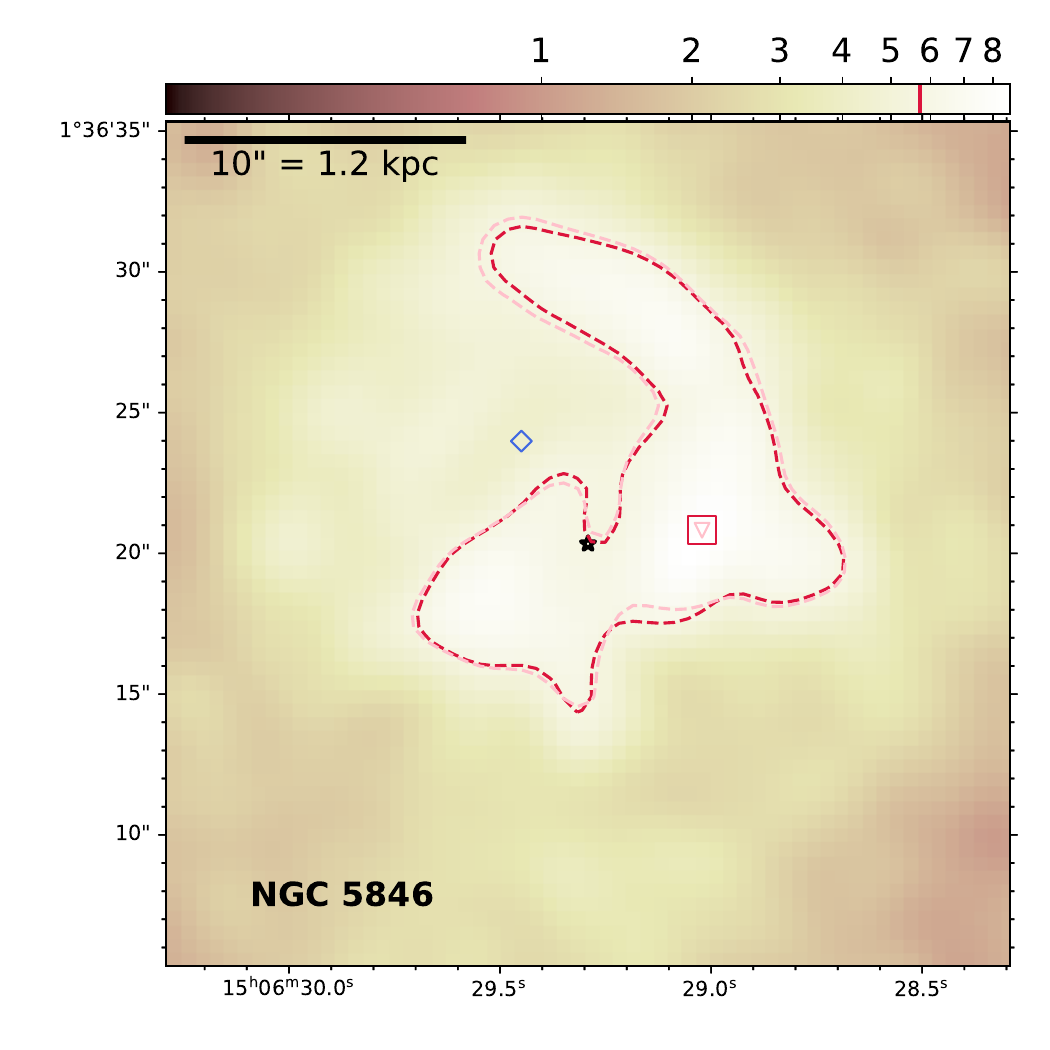}
    \includegraphics[width=0.19\linewidth, trim={0.2cm 0 1cm 1.6cm}]{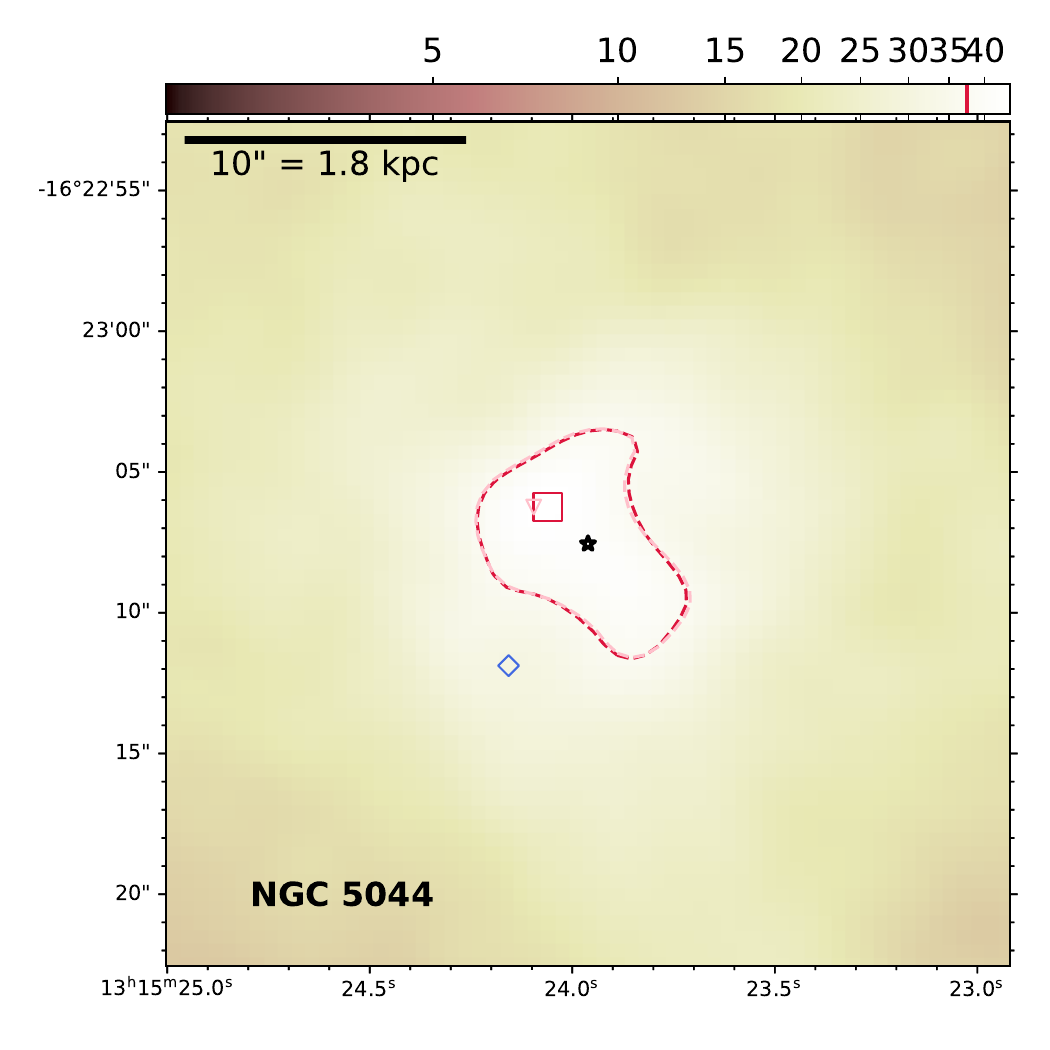}
    \includegraphics[width=0.19\linewidth, trim={0.2cm 0 1cm 1.6cm}]{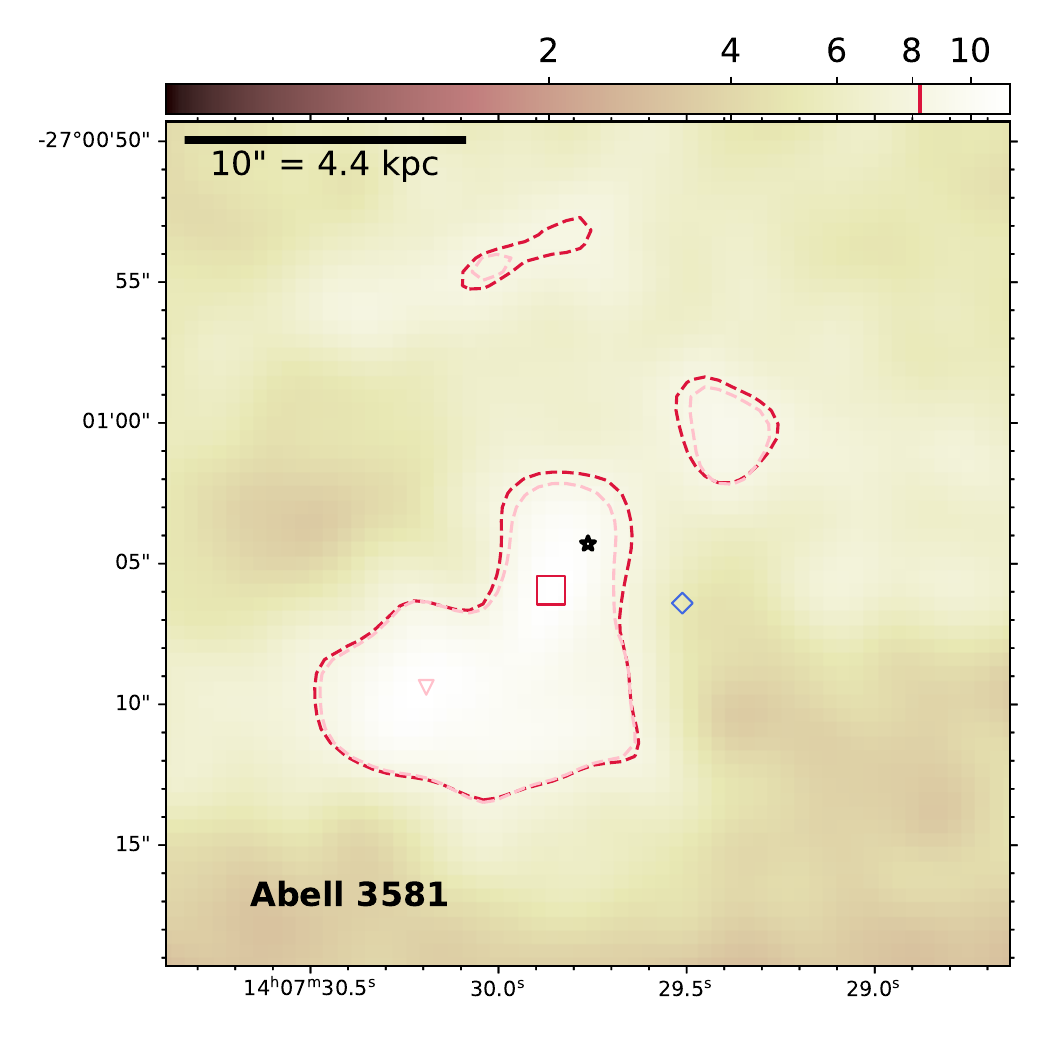}
    \includegraphics[width=0.19\linewidth, trim={0.2cm 0 1cm 1.6cm}]{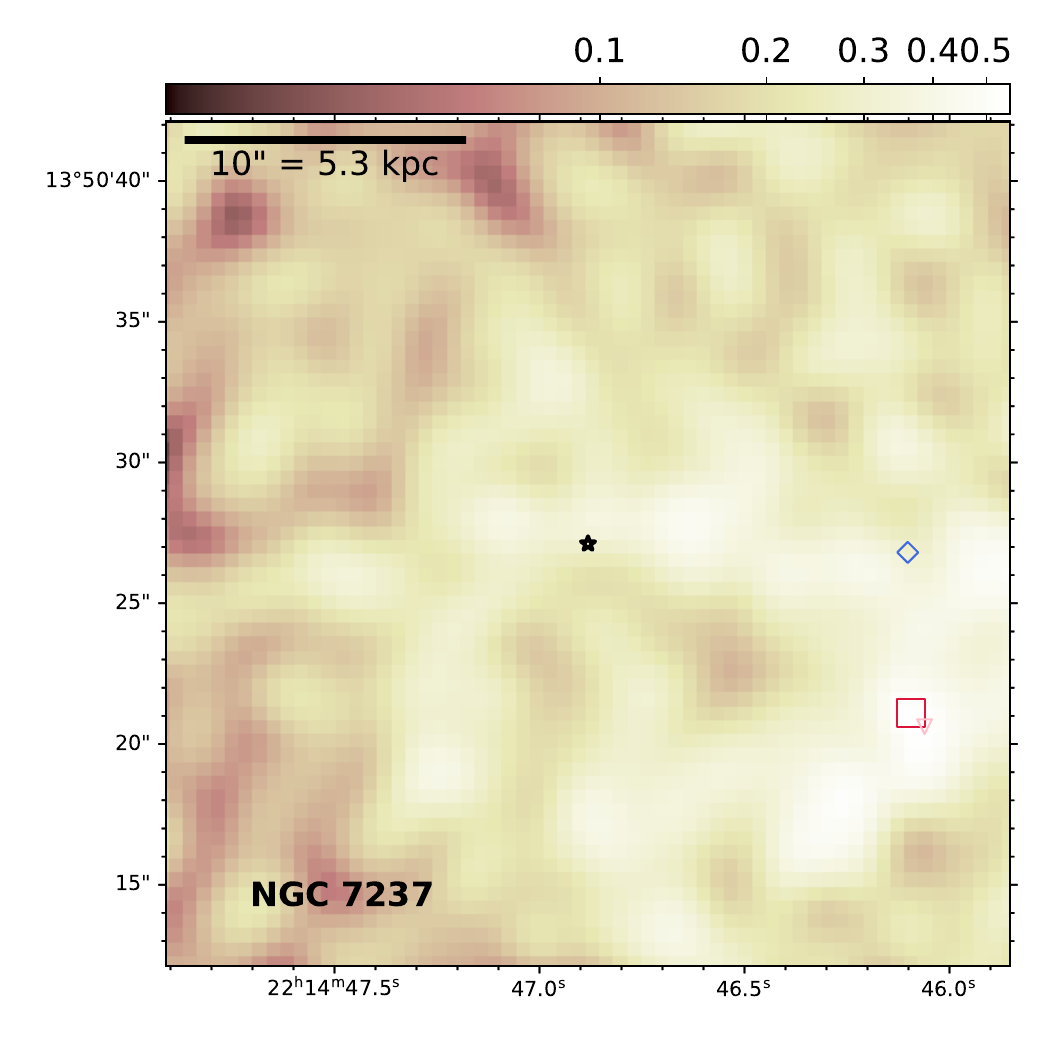}
    \includegraphics[width=0.19\linewidth, trim={0.2cm 0 1cm 1.6cm}]{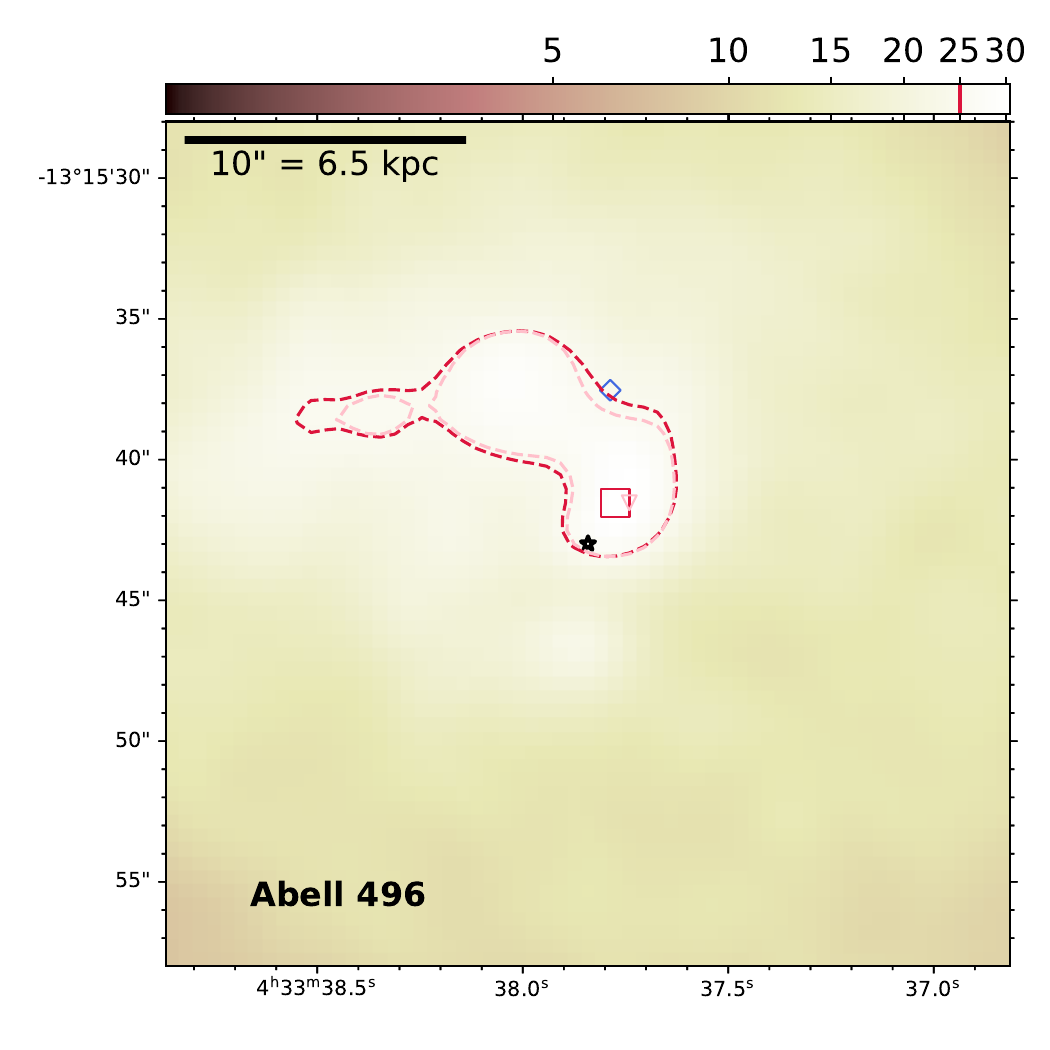}\\
    \includegraphics[width=0.19\linewidth, trim={0.2cm 0 1cm 1.6cm}]{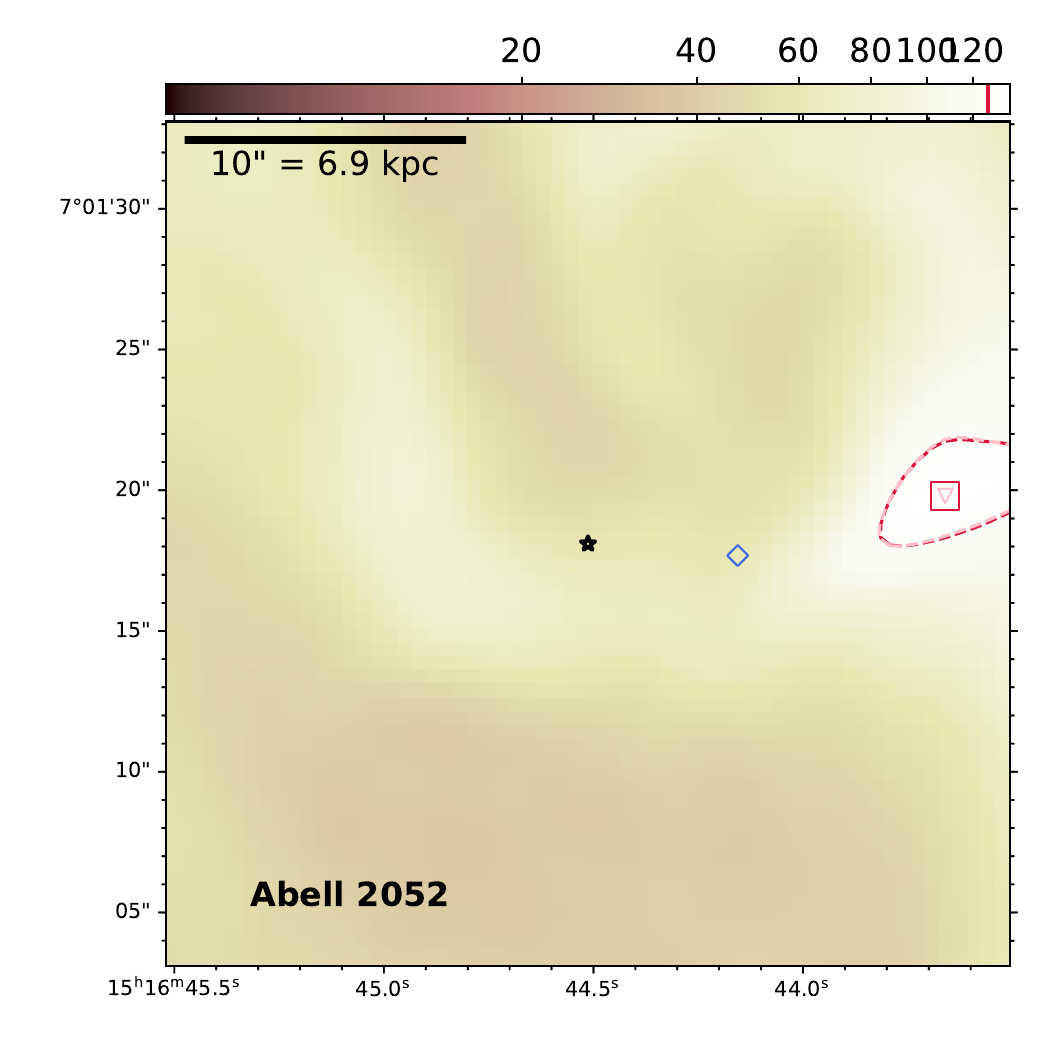}
    \includegraphics[width=0.19\linewidth, trim={0.2cm 0 1cm 1.6cm}]{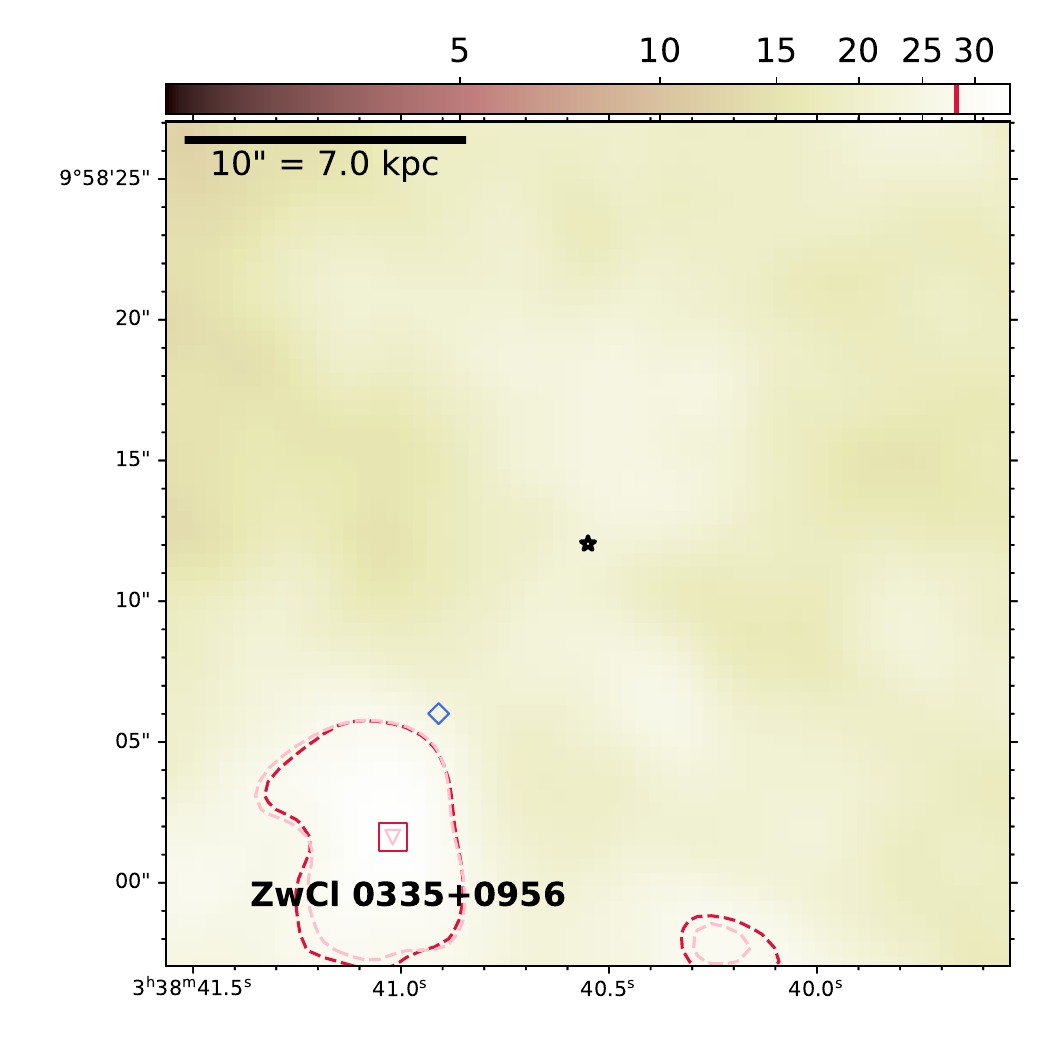}
    \includegraphics[width=0.19\linewidth, trim={0.2cm 0 1cm 1.6cm}]{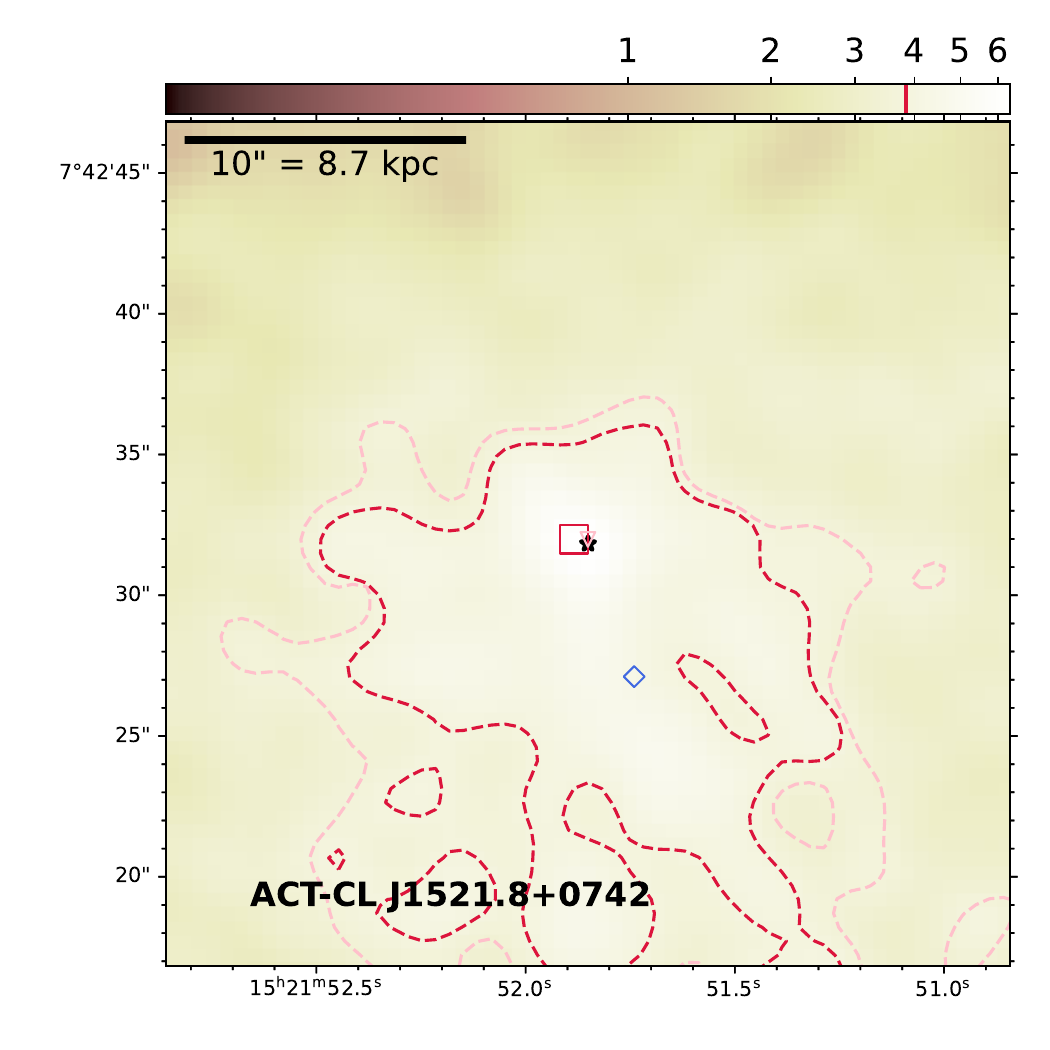}
    \includegraphics[width=0.19\linewidth, trim={0.2cm 0 1cm 1.6cm}]{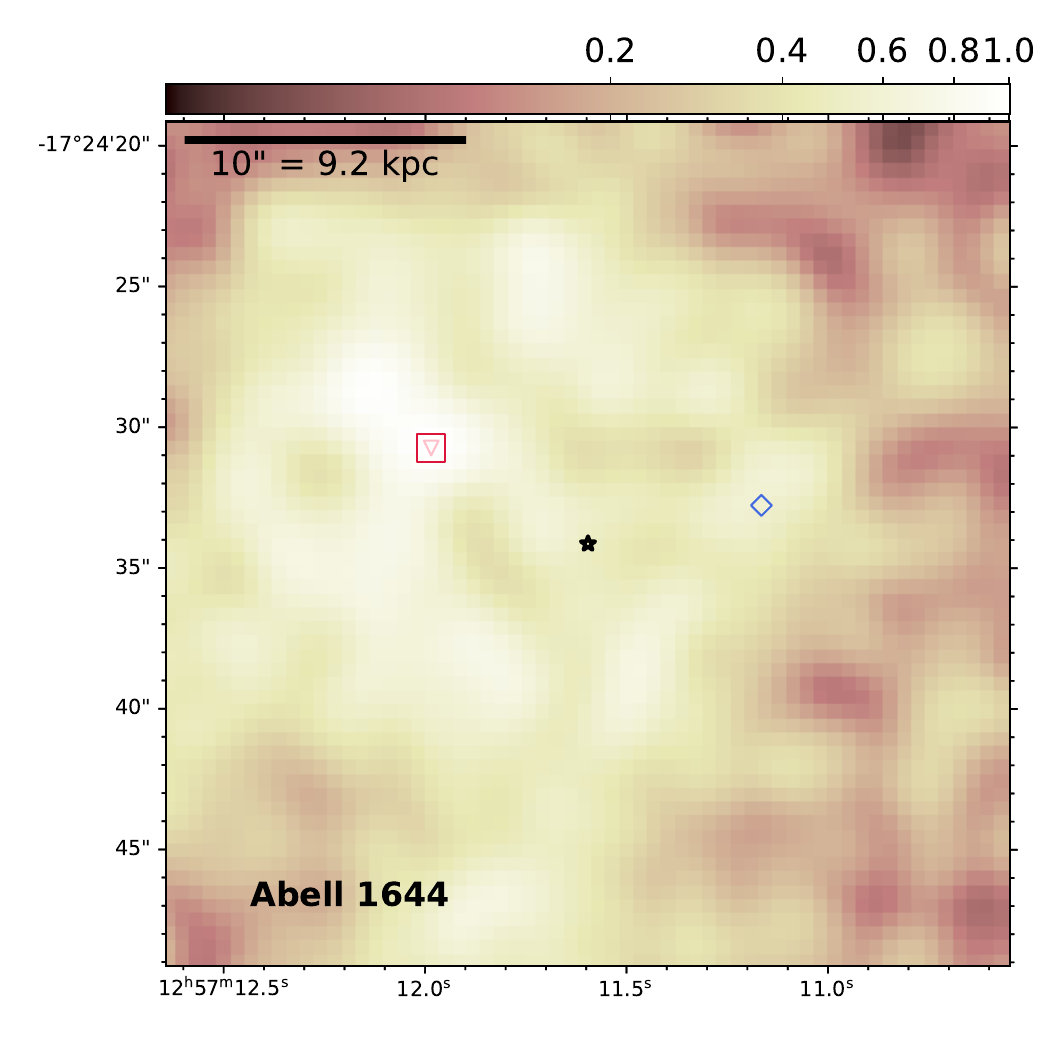}
    \includegraphics[width=0.19\linewidth, trim={0.2cm 0 1cm 1.6cm}]{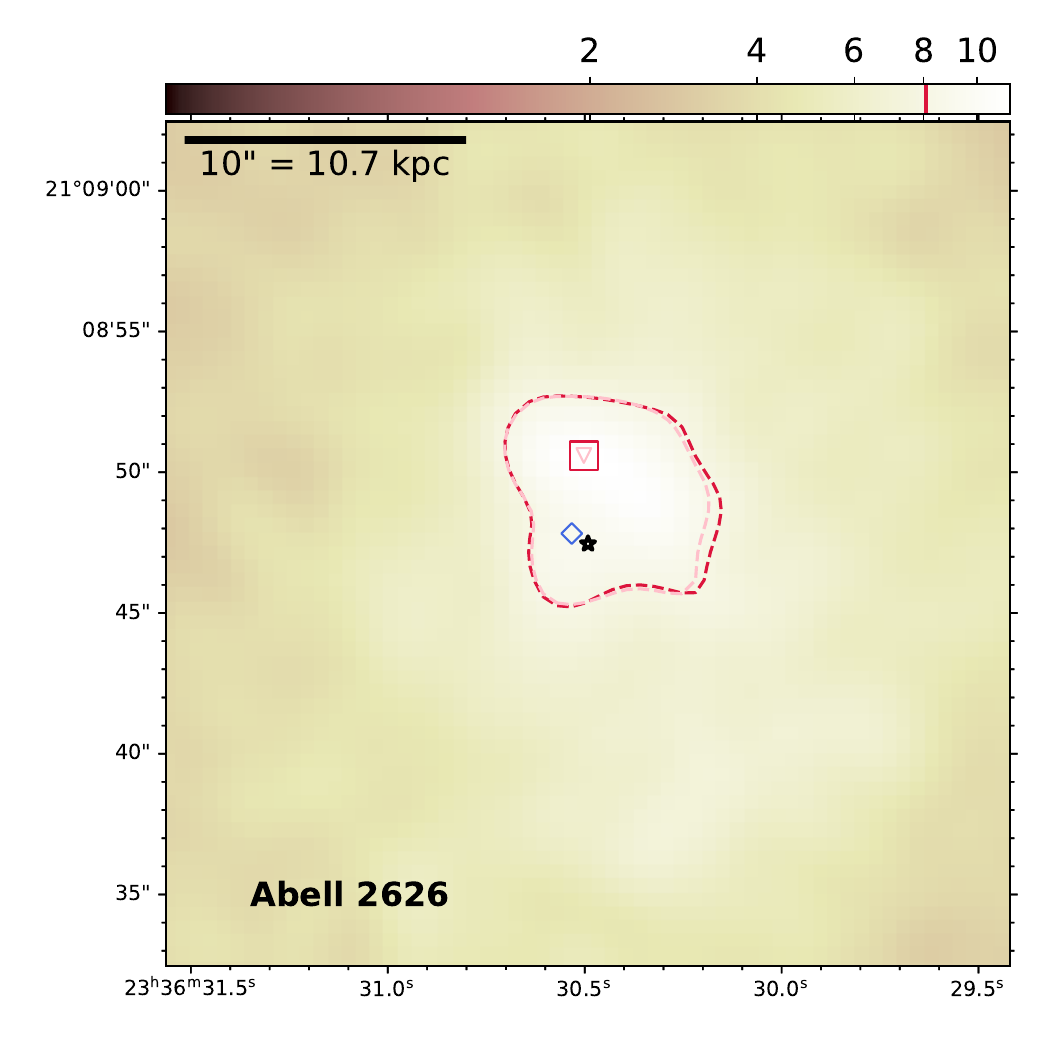}\\
    \includegraphics[width=0.19\linewidth, trim={0.2cm 0 1cm 1.6cm}]{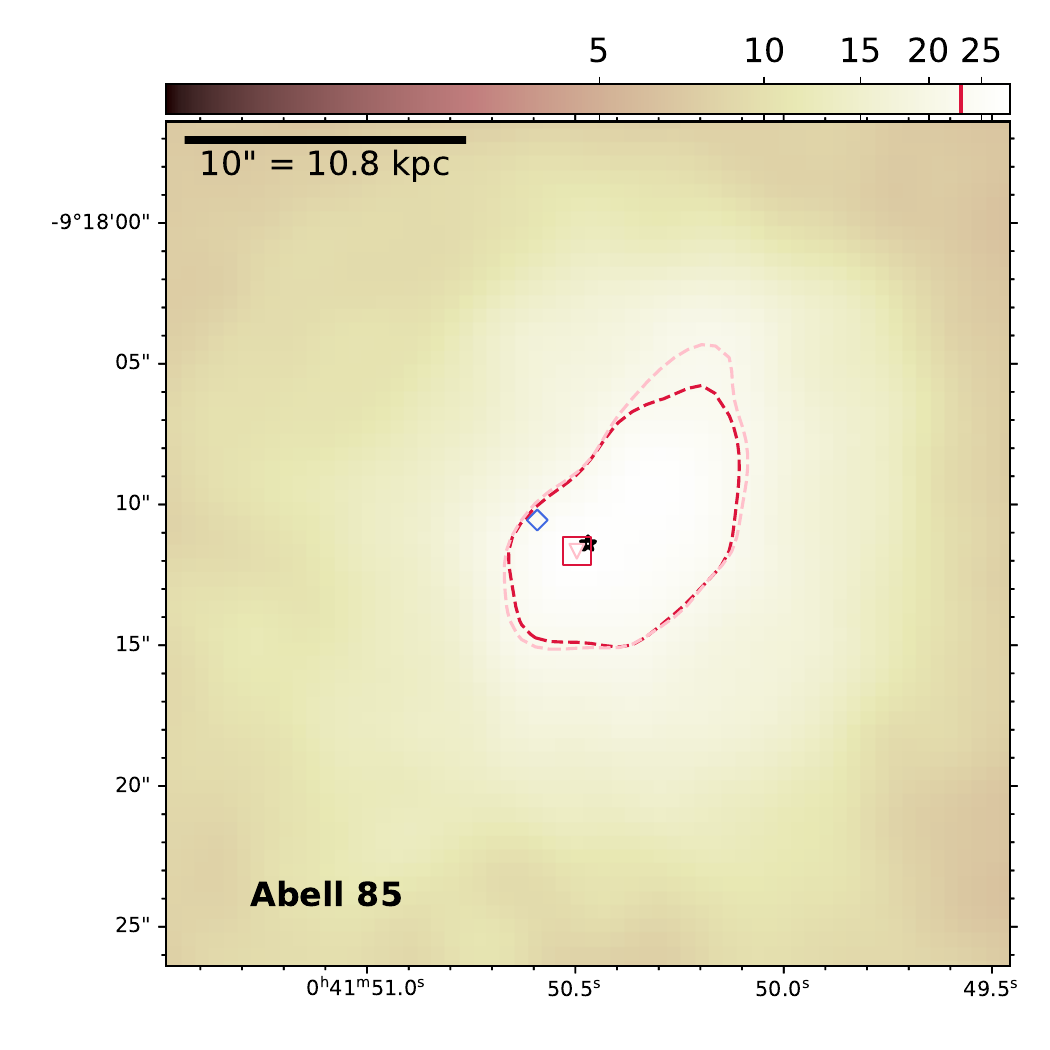}
    \includegraphics[width=0.19\linewidth, trim={0.2cm 0 1cm 1.6cm}]{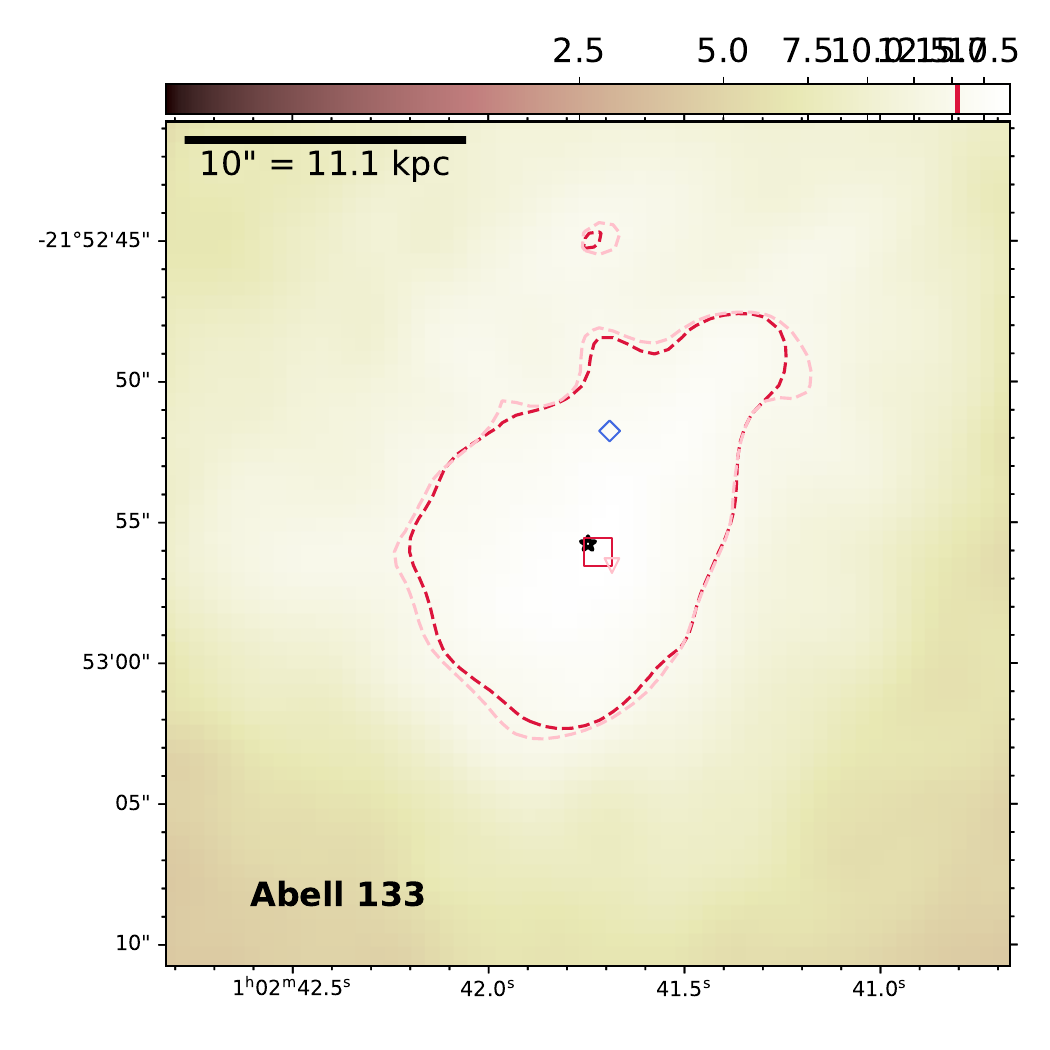}
    \includegraphics[width=0.19\linewidth, trim={0.2cm 0 1cm 1.6cm}]{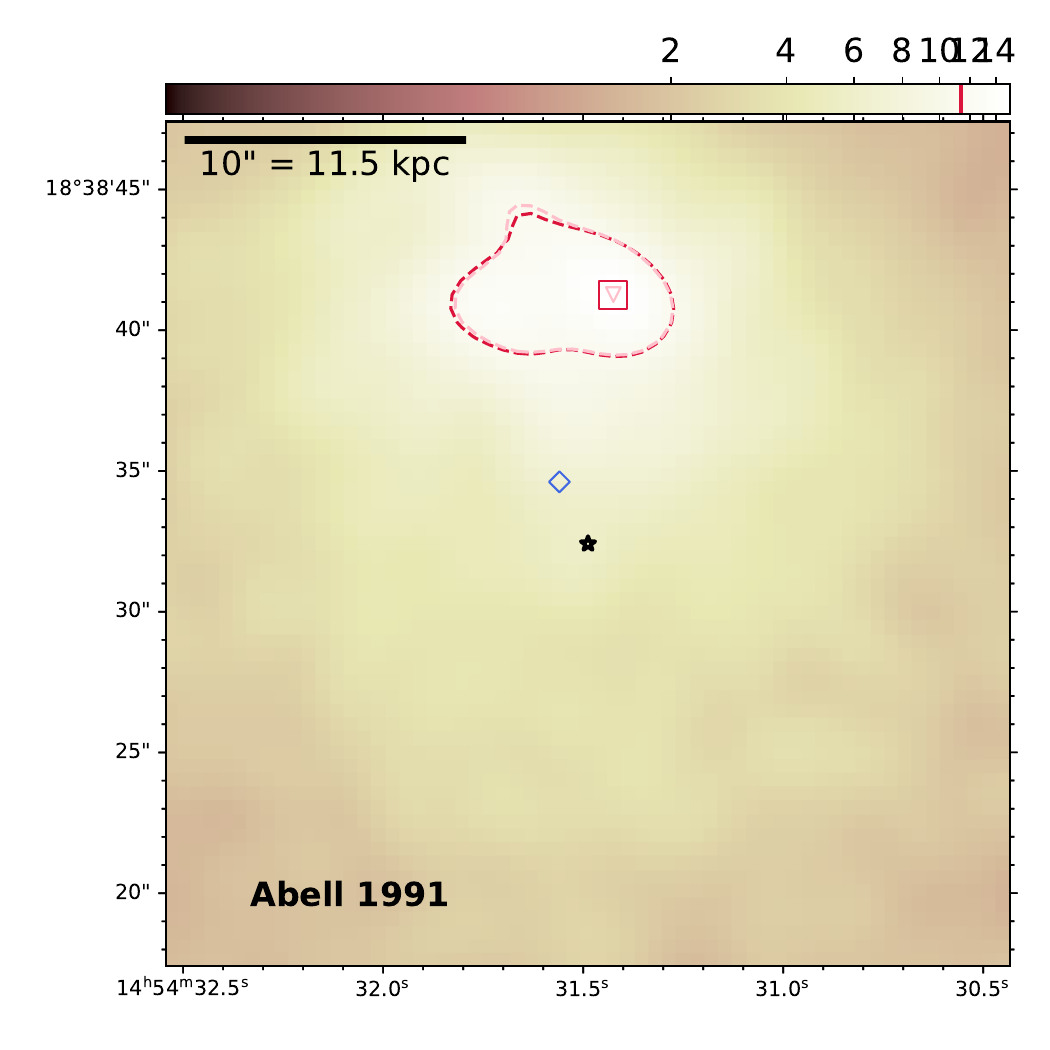}
    \includegraphics[width=0.19\linewidth, trim={0.2cm 0 1cm 1.6cm}]{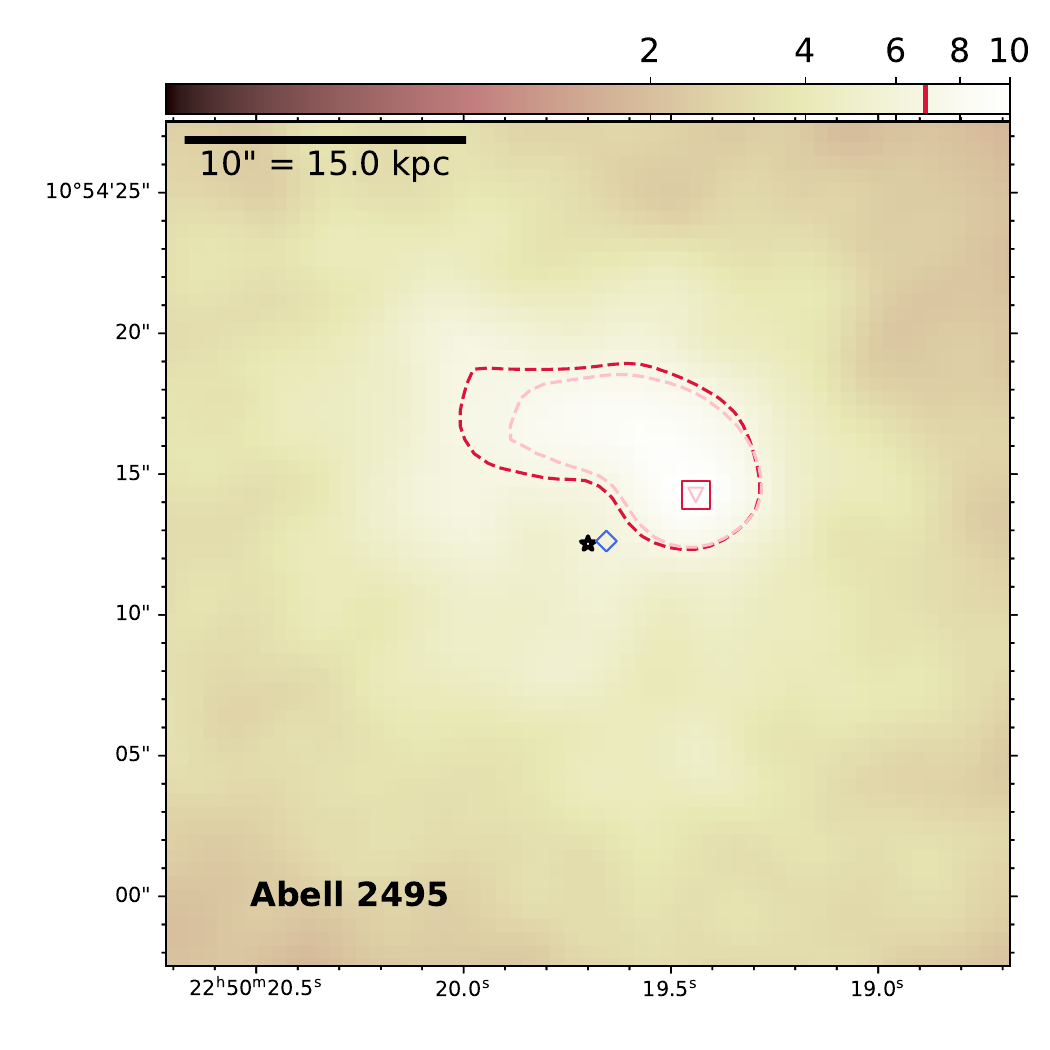}
    \includegraphics[width=0.19\linewidth, trim={0.2cm 0 1cm 1.6cm}]{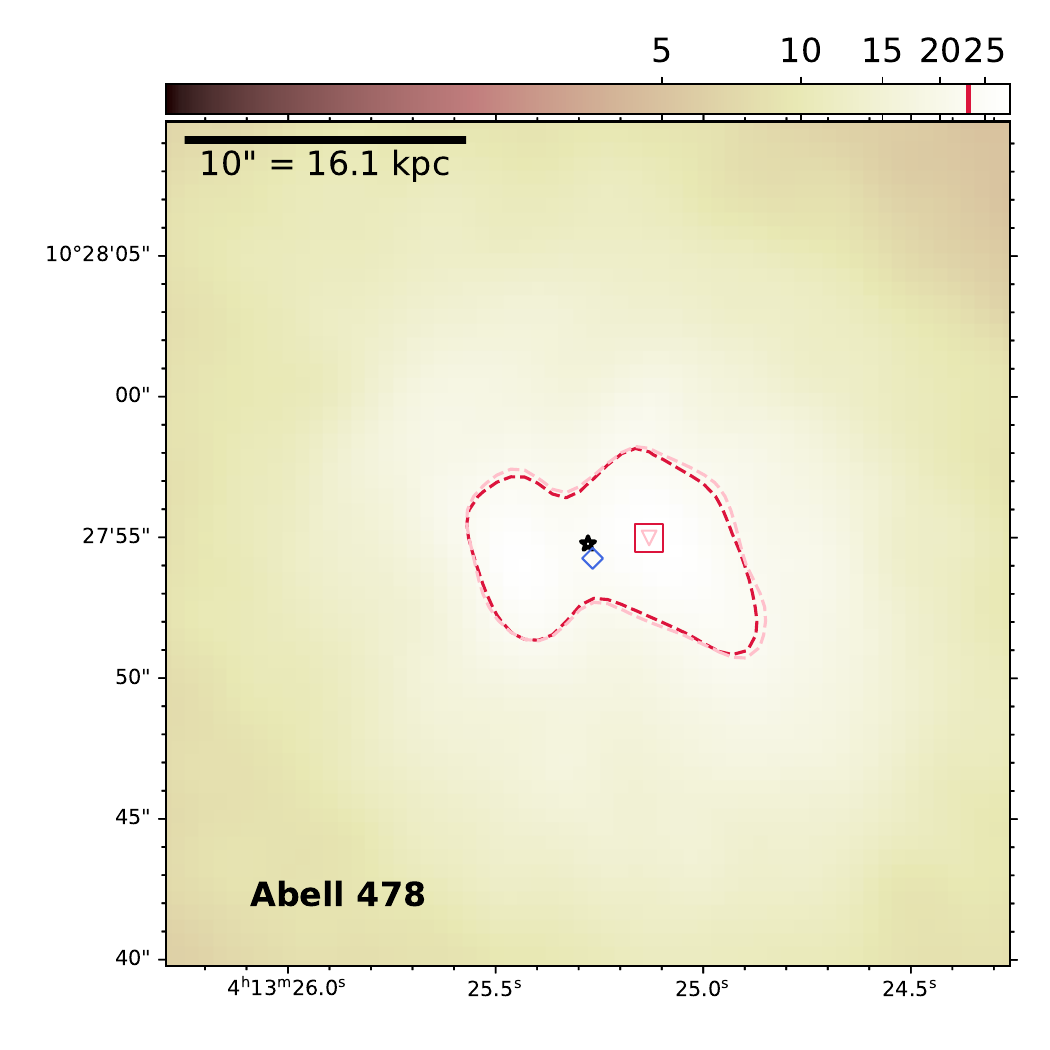}\\
    \includegraphics[width=0.19\linewidth, trim={0.2cm 0 1cm 1.6cm}]{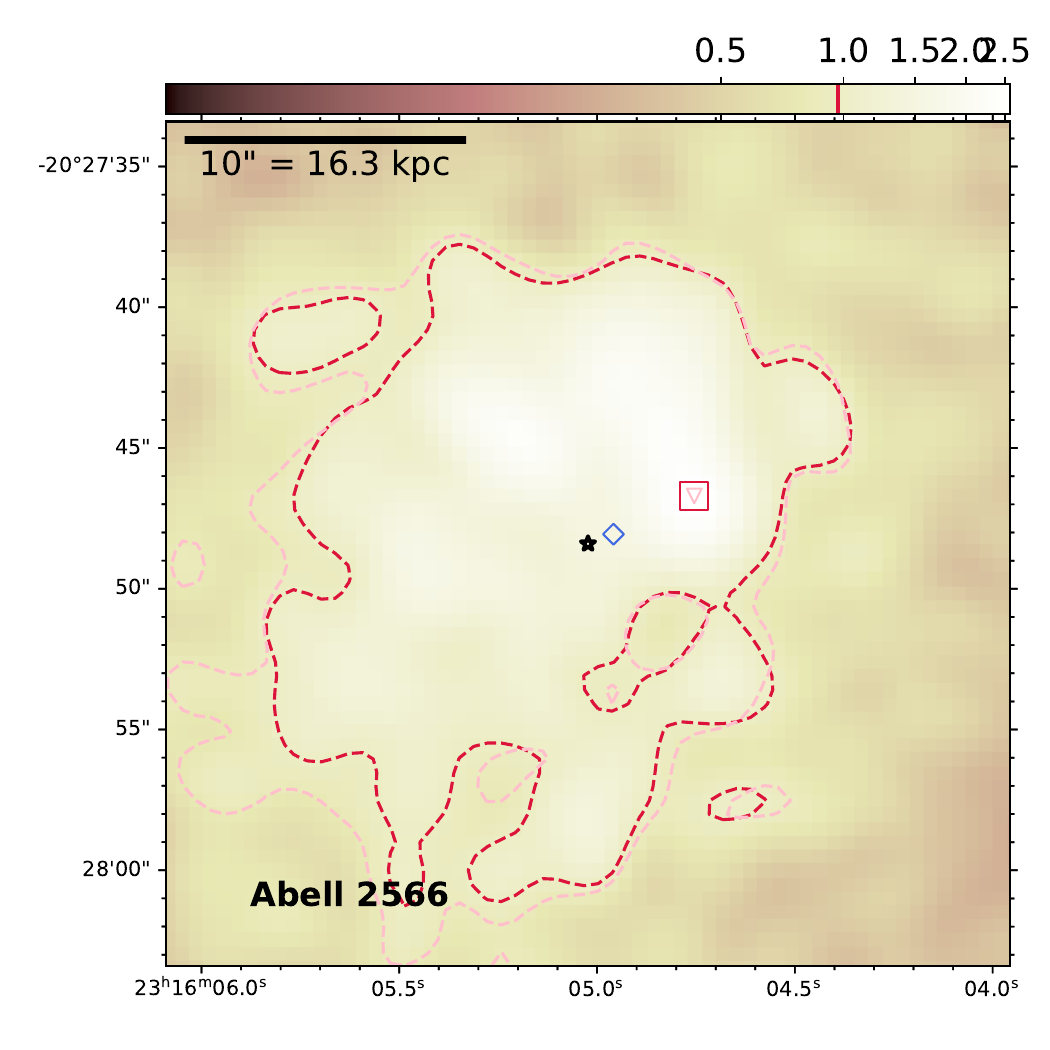}
    \includegraphics[width=0.19\linewidth, trim={0.2cm 0 1cm 1.6cm}]{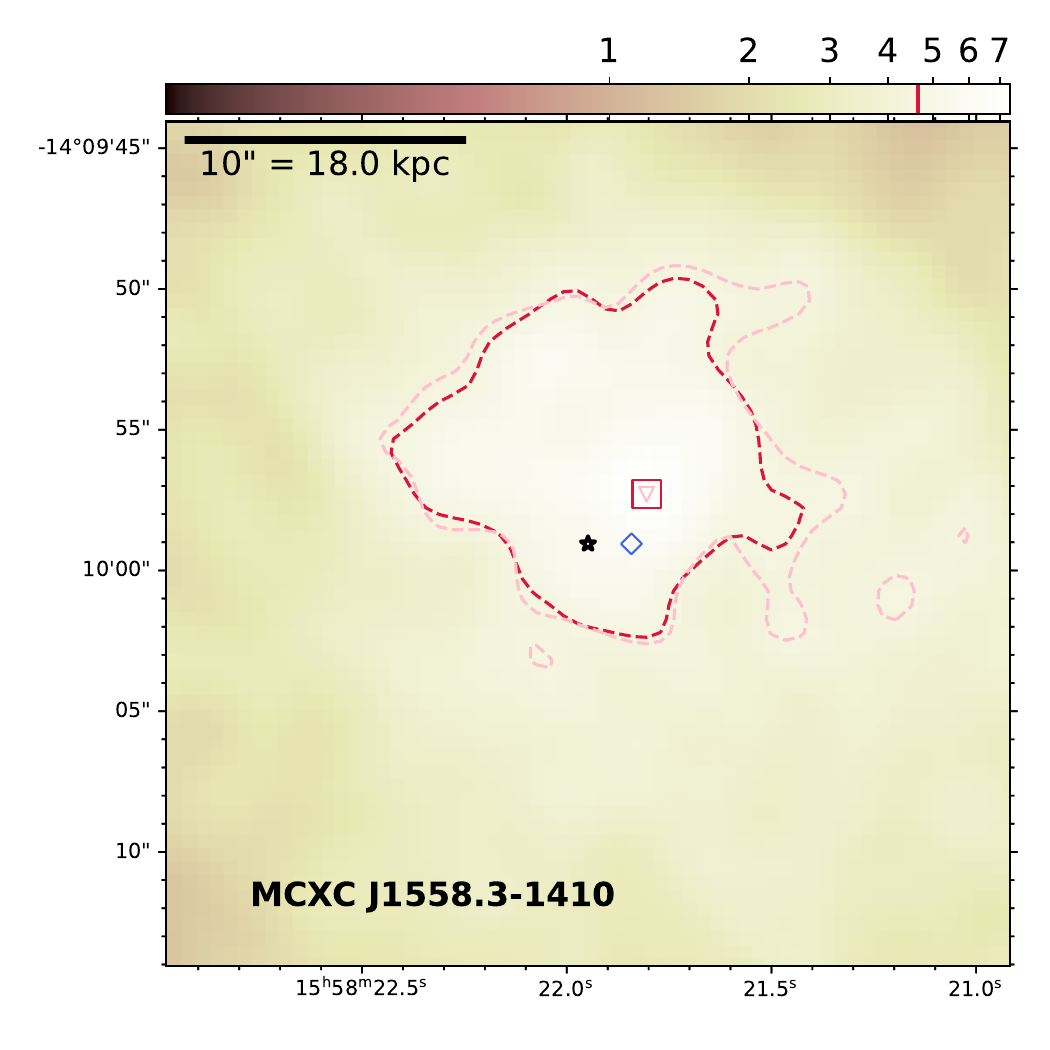}
    \includegraphics[width=0.19\linewidth, trim={0.2cm 0 1cm 1.6cm}]{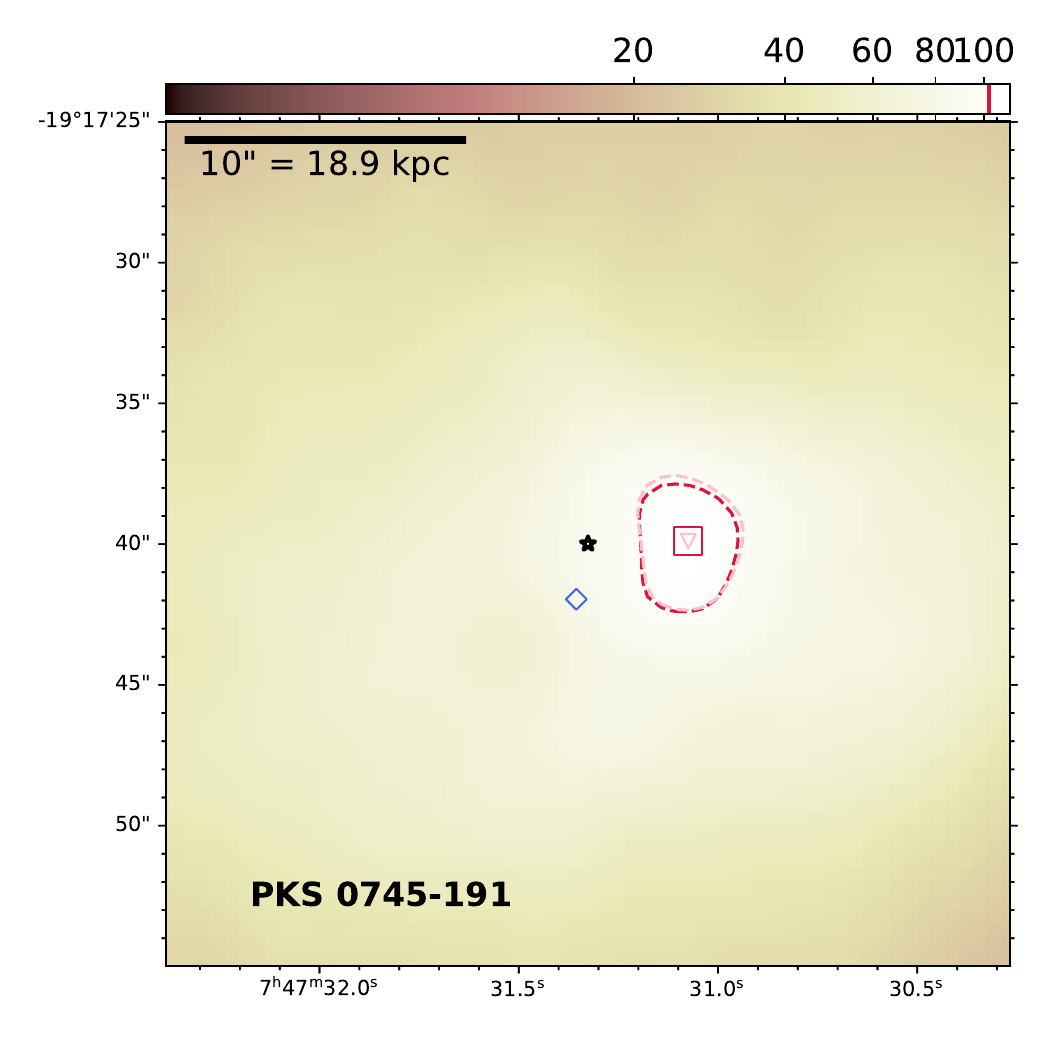}
    \includegraphics[width=0.19\linewidth, trim={0.2cm 0 1cm 1.6cm}]{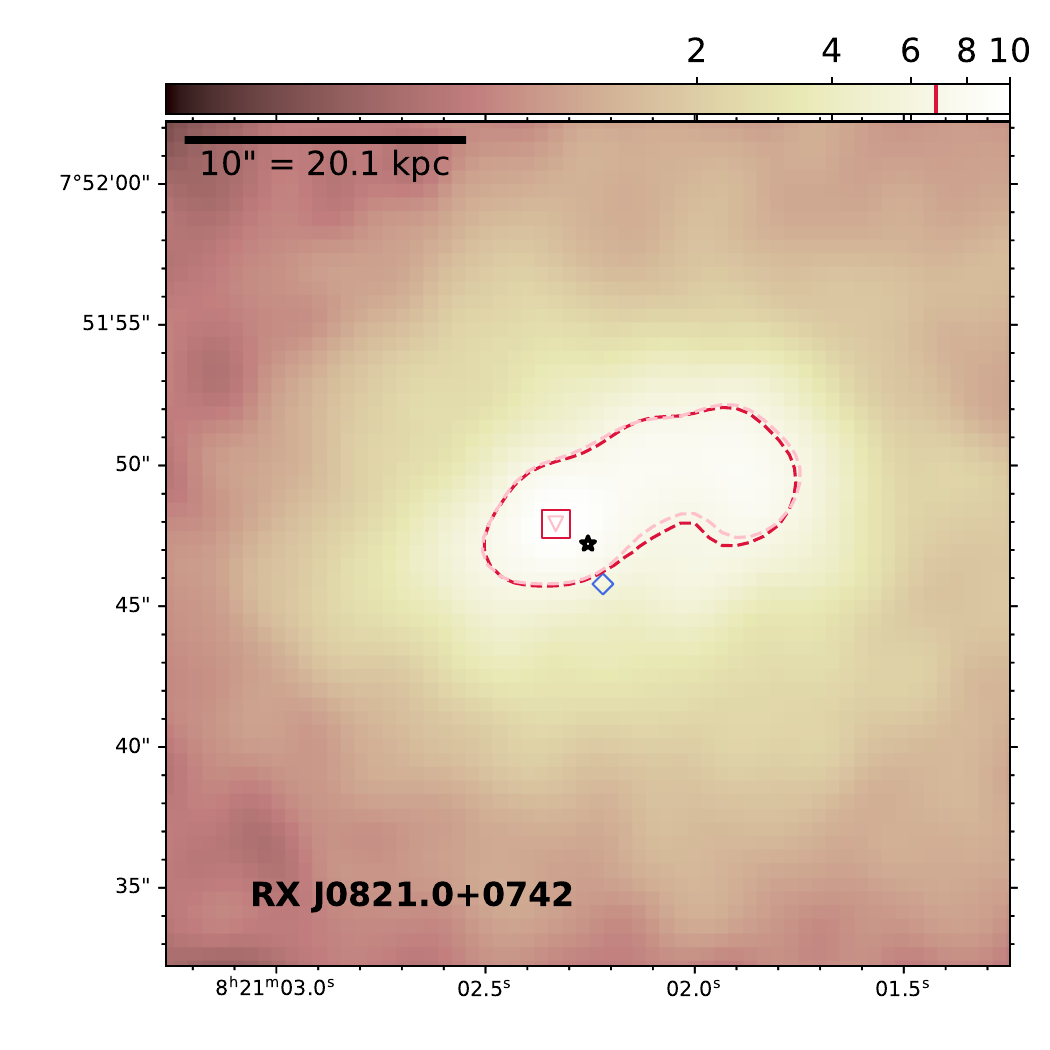}
    \includegraphics[width=0.19\linewidth, trim={0.2cm 0 1cm 1.6cm}]{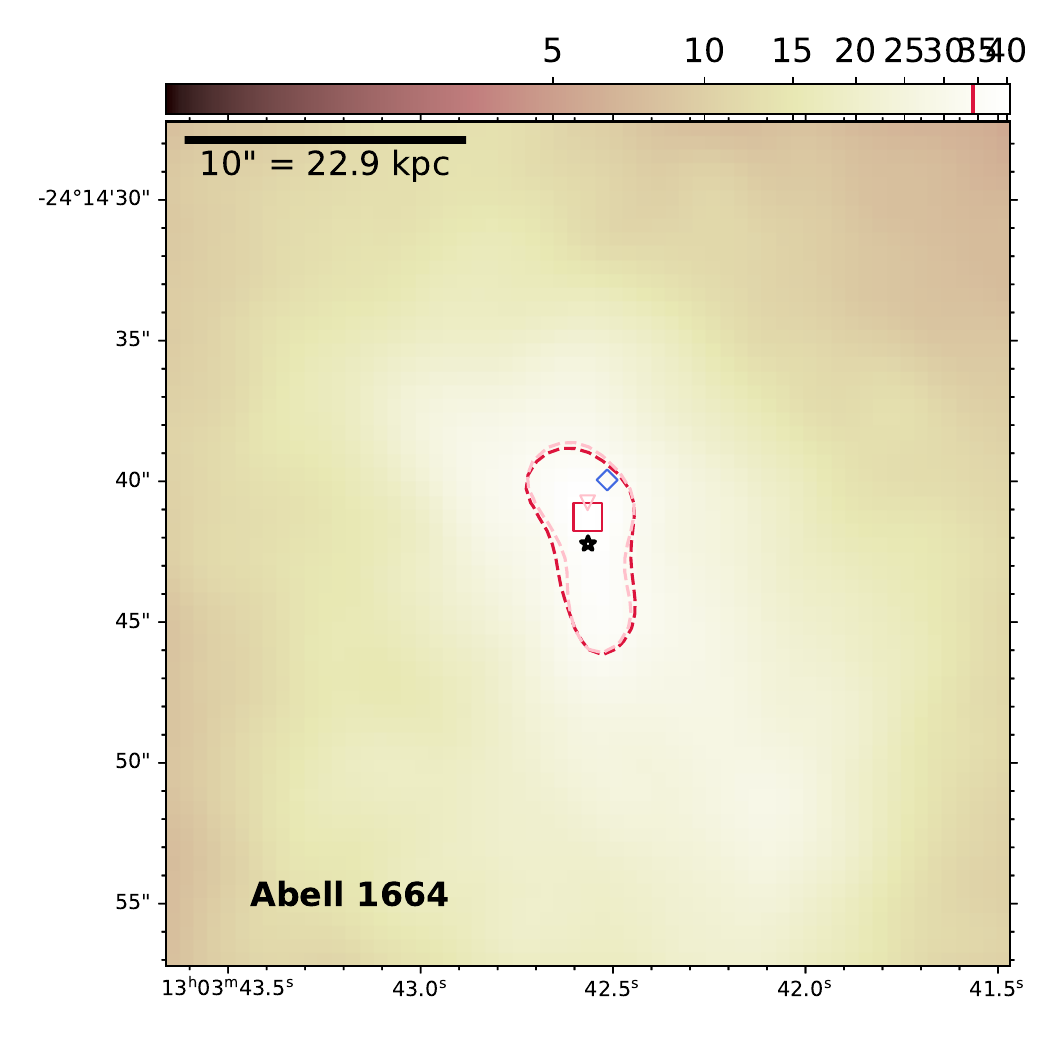}\\
    \includegraphics[width=0.19\linewidth, trim={0.2cm 0 1cm 1.6cm}]{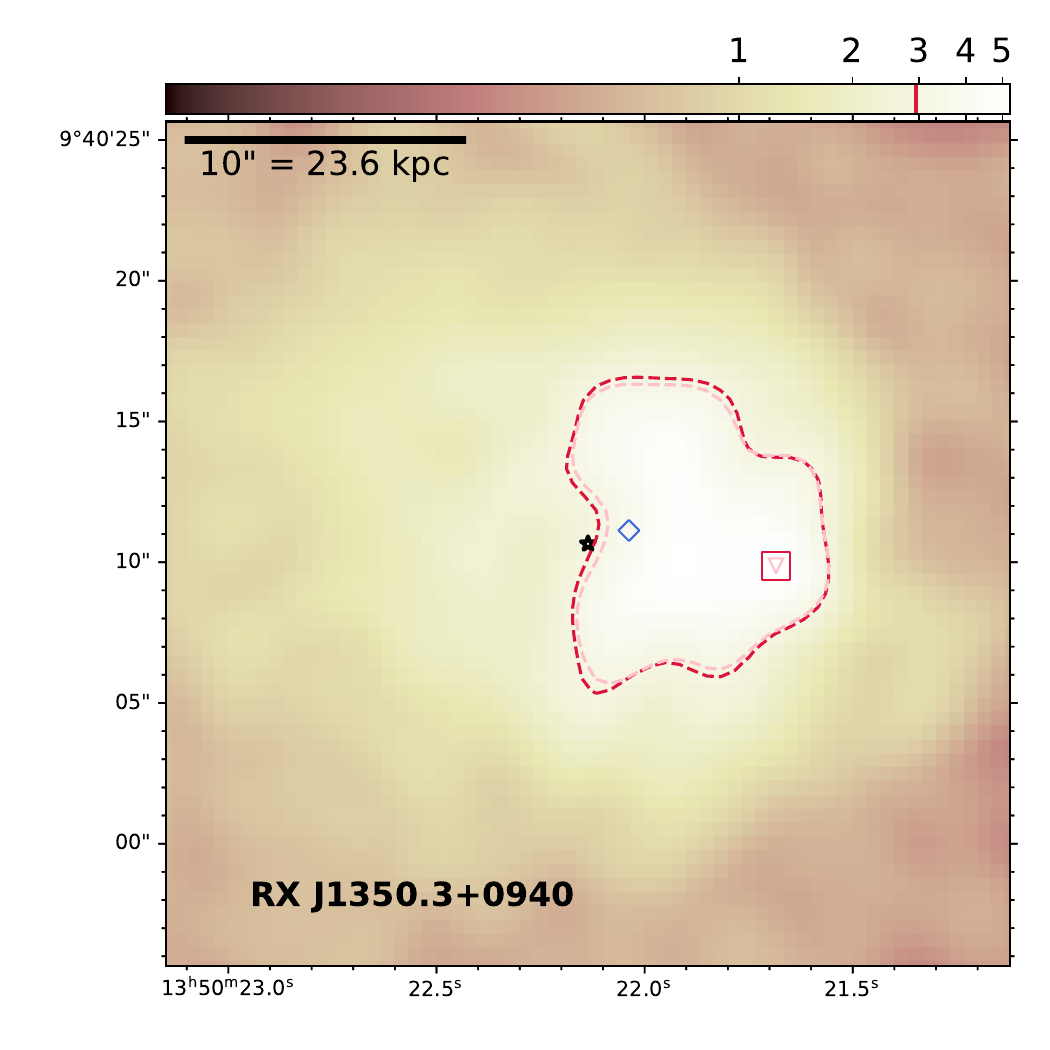}
    \includegraphics[width=0.19\linewidth, trim={0.2cm 0 1cm 1.6cm}]{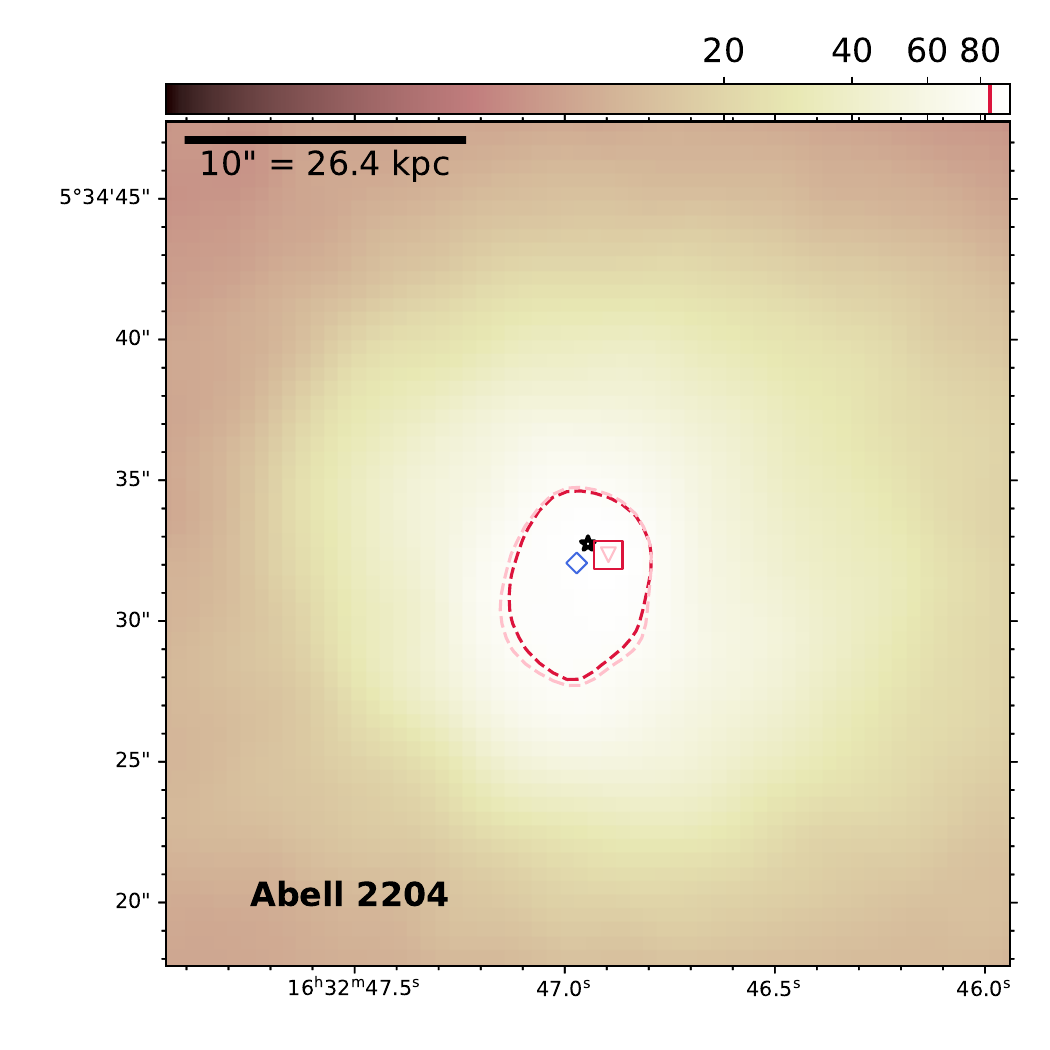}
    \includegraphics[width=0.19\linewidth, trim={0.2cm 0 1cm 1.6cm}]{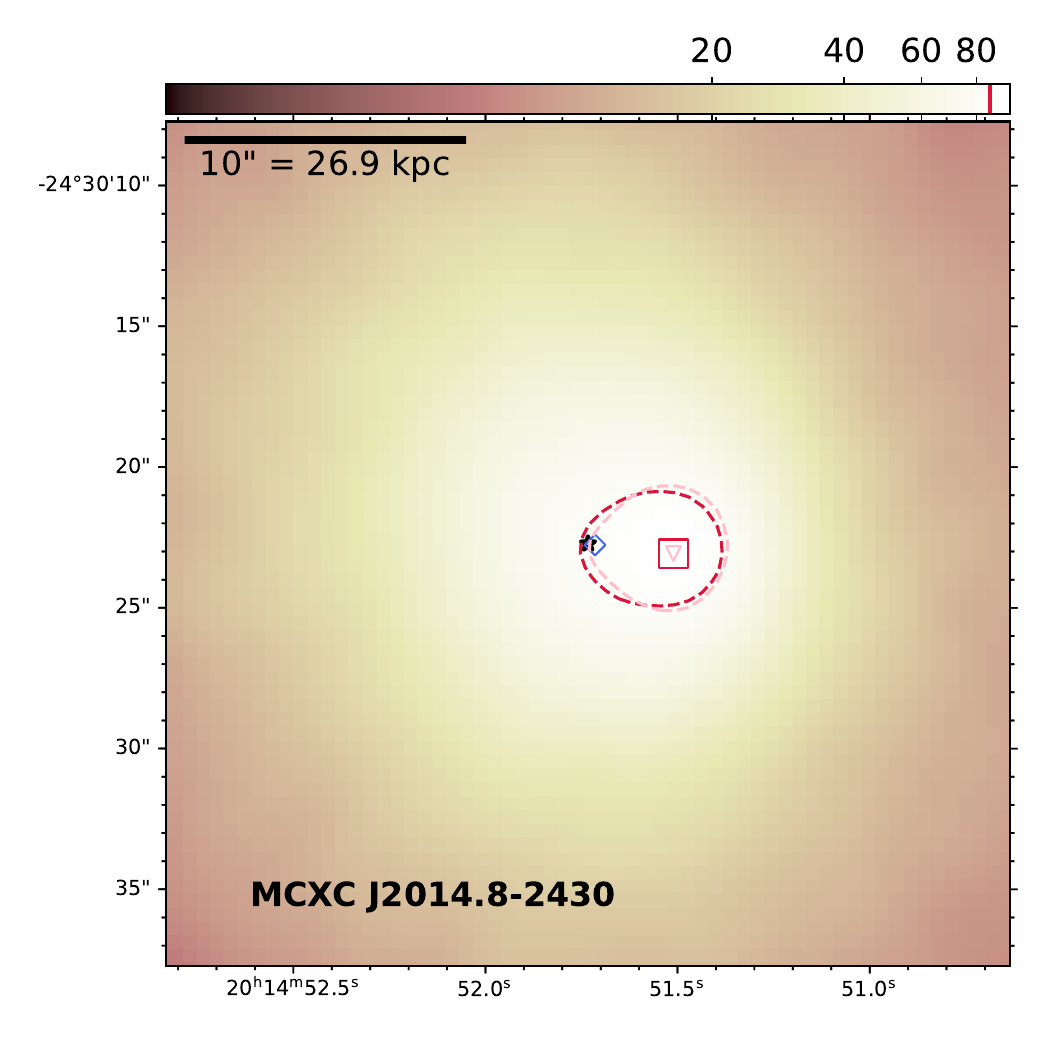}
    \includegraphics[width=0.19\linewidth, trim={0.2cm 0 1cm 1.6cm}]{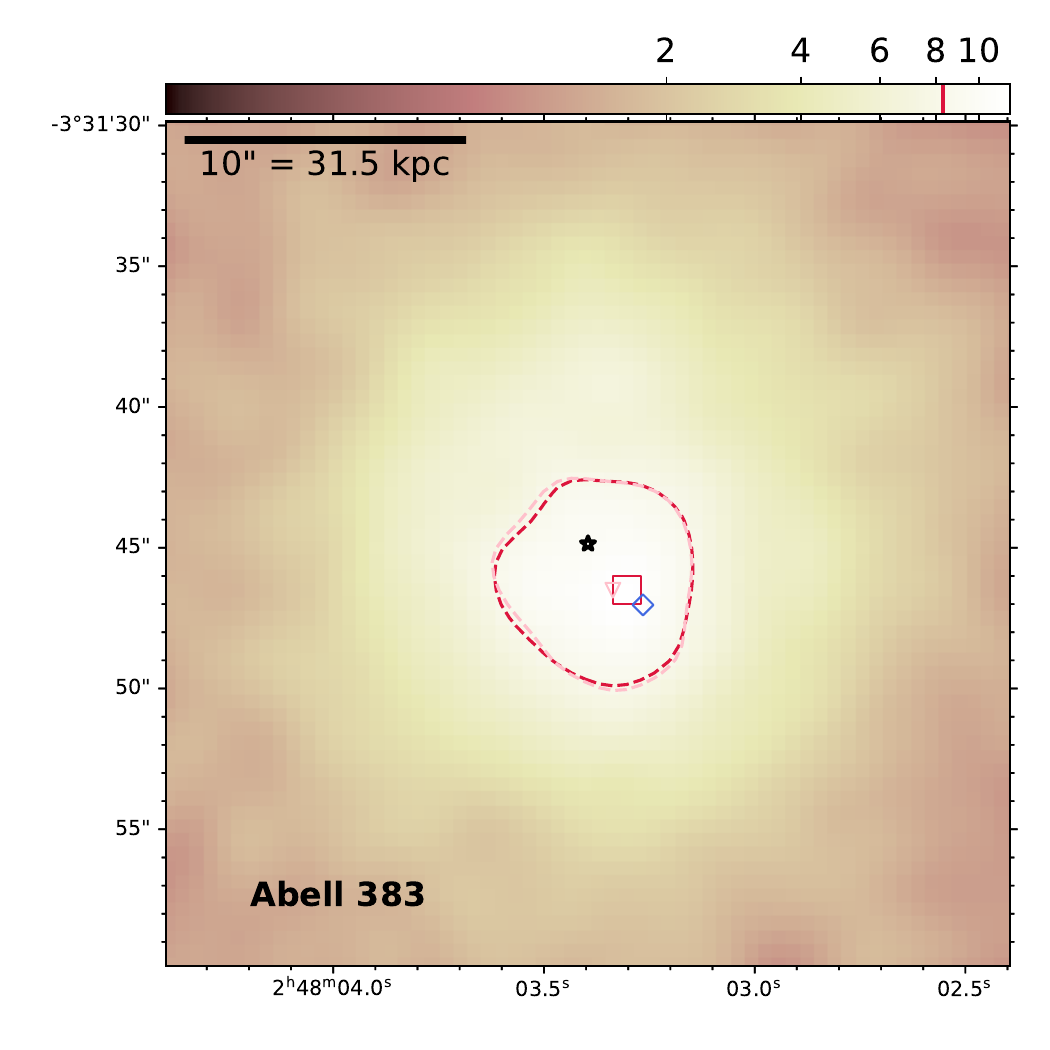}
    \includegraphics[width=0.19\linewidth, trim={0.2cm 0 1cm 1.6cm}]{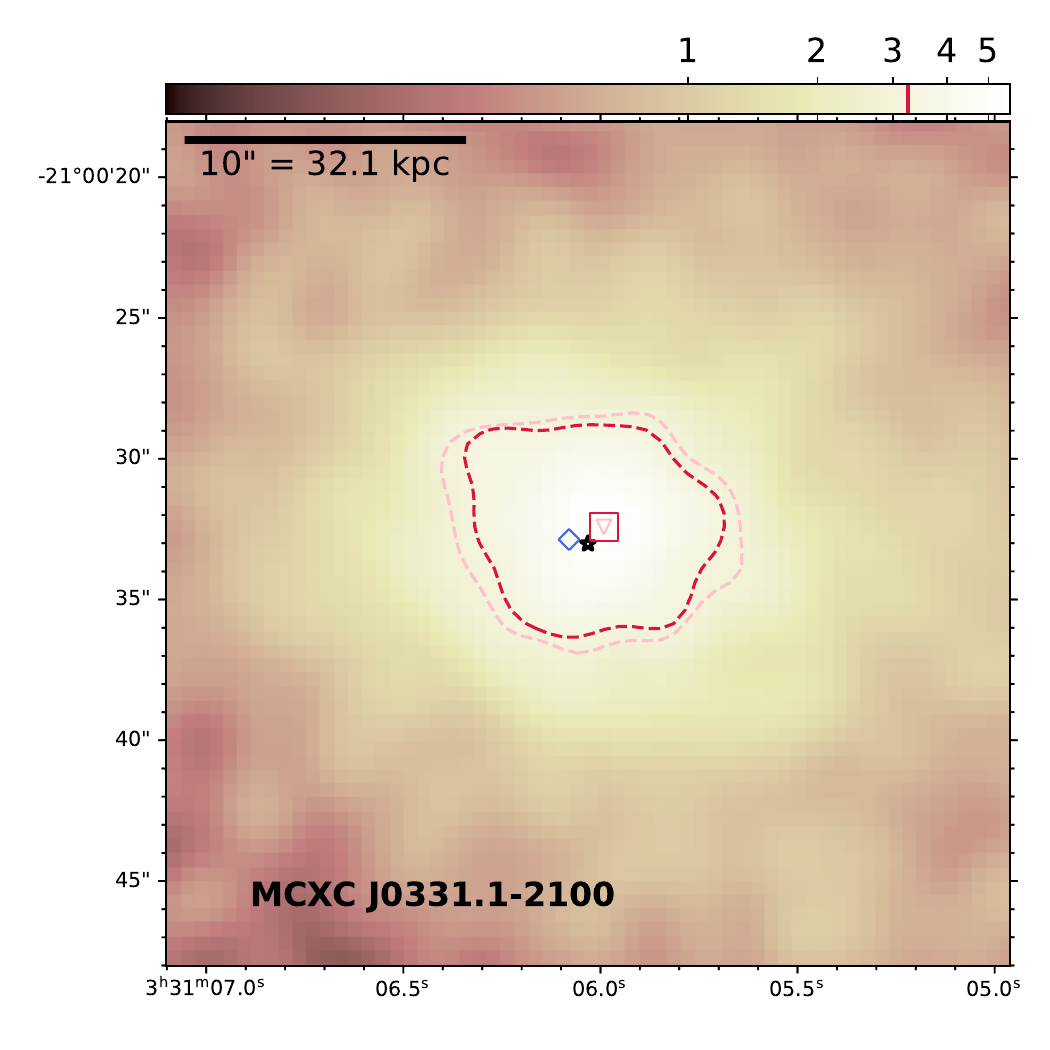}\\
    \caption{Same as Fig. \ref{fig:multi1} but without H$\alpha$ contours. The red square and pink triangle mark the position of the X-ray peak in the broad (0.5 - 7 keV) and soft (0.5 - 1.2 keV) band X-ray images, respectively (see Appendix \ref{appsub:chandra} for details). The dashed red and pink contours outline a "statistical confidence region" for the position of the X-ray peak (in the broad and soft band X-ray images, respectively), defined as the area containing pixels with X-ray counts $N_C > N_C^P - \sqrt{N_C^P}$, where $N_C^P$ is the number of counts in the pixel corresponding to the X-ray peak. The vertical red line on the colorbar indicates the count level that defines this contour (for the broad band). The black star marks the SMBH position (see Tab.~\ref{tab:completeinfo}). No contours are shown for ACT-CL~J1521.8+0742 and Abell~1644, as their X-ray surface brightness is below 1 count per pixel. In these cases, the statistical confidence region for the brightest pixel would formally extend to the edge of the \textit{Chandra} image. As noted in Appendix \ref{appsub:chandra}, these contours would represent the confidence region for the position of the X-ray peak in terms of X-ray counts, but we consider in our results the more conservative astrometric uncertainty of the X-ray data. We also show the location of the centroid of the X-ray surface brightness distribution with a blue diamond (see Appendix \ref{appsub:chandra} for details).
    }
    \label{fig:contours1}
\end{figure}

 For completeness, we also explored a statistical approach based on Poisson fluctuations in the count rate to estimate the positional uncertainty of the X-ray peak. For each system, we defined a “statistical confidence region” centered on the X-ray peak by selecting all pixels with counts greater than $N_C^P - \sqrt{N_C^P}$, where $N_C^P$ is the number of counts in the brightest pixel of the map. This threshold encompasses all pixels statistically consistent with the brightest one, under the assumption of Poisson statistics. We show in Fig. \ref{fig:contours1} the X-ray images where we overlay contours corresponding to this threshold. In systems with low surface brightness, such as ACT-CL~J1521.8+0742 and Abell~1644, where the peak pixel count is below 1 count/pixel, the resulting confidence region formally extends to the edge of the image. In the other systems, the extent of this region depends on the total number of counts within the X-ray peak and on how peaked the surface brightness distribution is. This alternative method allows us to visualize the extent of the region where the X-ray peak might be statistically located with the current X-ray exposure. However, in our analysis we keep the 0.5" radius uncertainty (see above) because of the following reasons: (1) the “statistical confidence region” does not provide a more physically-motivated indication of where the density peak (and thus the cooling peak) might be located; (2) the use of a fixed positional uncertainty based on the angular resolution allows us to maintain consistency with previous works (e.g., \citealt{2010A&A...513A..37H,hamer2012,2019ApJ...885..111P,pasini2021,rosignoli2024}); and (3) adopting the broader statistical regions as uncertainties would not alter our conclusion of Sect. \ref{subsec:fragm} and \ref{subsec:connectsmbh} that the position of the X-ray peak has a limited effect on the activation of the SMBH, but rather support it. 
\par We point out that while our analysis was performed on broad band (0.5 - 7 keV) {\it Chandra} images (as in \citealt{2010A&A...513A..37H}), previous observational and numerical studies point to a stronger correlation between warm ionized gas (traced by H$\alpha$) and soft ($\lessapprox$1 keV) X-rays \citep[e.g.,][]{Fabian_2006,Li_2014}. To evaluate this possibility, we repeated the analysis using soft band {\it Chandra} images, limited to the 0.5 – 1.2 keV band. From these, we identified the soft X-ray peak and compared it with that obtained from the broad-band images. The results are presented in Fig. \ref{fig:contours1}. We find that the soft X-ray peak position (pink triangle) is generally very similar to that derived from the broad-band image (red square), and the corresponding confidence regions (pink and red dashed contours) are nearly identical. The only noticeable difference occurs in Abell~3581, where the soft X-ray peak is offset slightly further from the SMBH compared to the broad-band peak, though it still falls within the same confidence region. For these reasons, we restrict the analysis to the results obtained from the broad band X-ray images.
\par Ultimately, we investigate how the X-ray peak position compares with the center of the potential well of the host group or cluster, taken as the centroid of the X-ray surface brightness distribution \citep[e.g.,][]{2019MNRAS.483.3545G}. As shown in Fig. \ref{fig:contours1}, the X-ray centroid (blue diamond) is usually not coincident with the X-ray peak, although they are always at $\leq$10~kpc from each other. Relative to the X-ray peak, the centroid lies closer to the SMBH in 11/25 systems, at a similar distance in 8/25 systems, and farther away in 6/25 systems. This suggests that the center of the potential well (the centroid) correlates slightly better than the density peak (the X-ray peak) with the SMBH position. We note, however, that in several cases where the centroid is formally closer to the SMBH than the X-ray peak, it does not coincide with the brightest X-ray emission region (e.g., Abell 2052, ZwCl 0335+096, Abell 1991, Abell 2495, Abell 2566; see Fig.~\ref{fig:contours1}). This is consistent with the centroid tracing the long-term center of the potential well rather than the present location of the cooling peak.

\section{BPT diagrams}\label{app:bpt}
While the X-ray emissivity is mostly proportional to the gas density, $\epsilon_{X}\propto n_e^{2}(kT)^{1/2}$, 
the H$\alpha$ luminosity is related both to the mass of warm gas and to the ionizing flux. In this sense, a potential vicious circle for systems where the H$\alpha$ peak lies on top of the radio AGN is represented by the possibility that the AGN itself is a dominant source of ionization. 
To verify this point, we made use of the wide wavelength range of MUSE and produced maps of H$\beta$4861$\AA$, [OIII]5007$\AA$, and [NII]6584$\AA$ intensity, in addition to the H$\alpha$, in order to use the well-known BPT diagram to separate between star-formation, AGN photoionization, LINER, and composite mechanisms as the dominant ionization source \citep{baldwin1981,kewley2001,kauffmann2003,kewley2006,cidfernandes2010}. In producing these maps, we considered only the pixels with a SNR greater than 3 in all the four lines (e.g., \citealt{poggianti2019}). We can thus assess the ionization source only for a fraction of the whole H$\alpha$ nebula. However, this is not an issue, since we are interested in the peak which typically has the highest SNR. For three systems (ACT-CL~J1521.8+0742, Abell~2566, Abell~2204), no spaxels met the SNR$>$3 threshold in all four lines -- primarily due to the weakness of [OIII]5007$\AA$ -- so BPT plots are not shown for these cases. We stress here that a detailed analysis of the BPT diagrams (such as the spatial variation in metallicity or density across the nebula) is outside the scope of this paper. Instead, we focus on the relative position of the H$\alpha$ peak and the surrounding nebula within the BPT plots.
\par Based on our inspection of the BPT diagrams (Fig. \ref{fig:bpt1}), we find that none of the spaxels with SNR$>$3 lie in the AGN photoionization region. The warm gas of the systems in our sample is predominatly ionized by star formation or LINER-like processes (e.g., shocks), as previously found in other works \citep{fogarty2015,hamer2016,polles2021}. Moreover, there are no significant differences between the ionization conditions at the H$\alpha$ peak and those across the surrounding warm gas, suggesting that the warm gas near the peak shares the same ionization mechanisms as the extended filamentary structure. This indicates that our analysis is not flawed by AGN photo-ionization producing a peak exactly where the AGN lies. We note that additional analysis using alternative BPT diagrams based on different line ratios  ([OI]6300$\AA$/H$\alpha$, [SII]6717,6731$\AA$/H$\alpha$, not shown here; e.g., \citealt{poggianti2019}) yielded consistent results.

\begin{figure}[ht!]
    \centering
    \includegraphics[width=0.245\linewidth, trim={0.2cm 0 1cm 1.7cm}, clip]{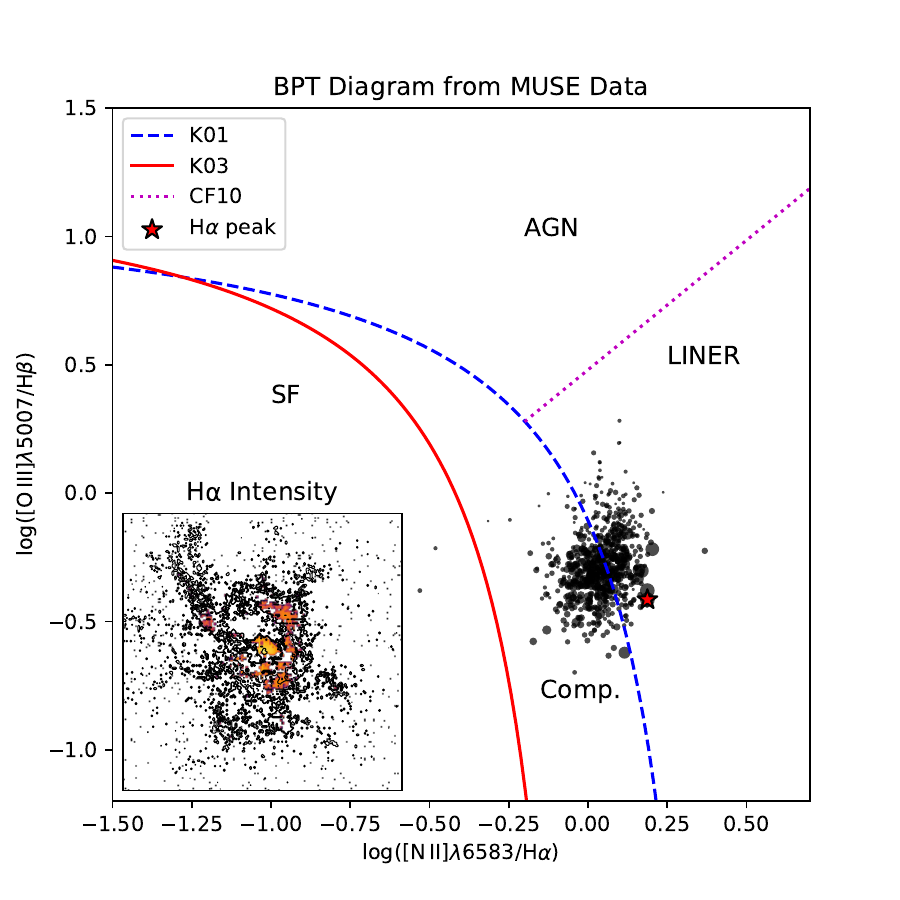}
    \includegraphics[width=0.245\linewidth, trim={0.2cm 0 1cm 1.7cm}, clip]{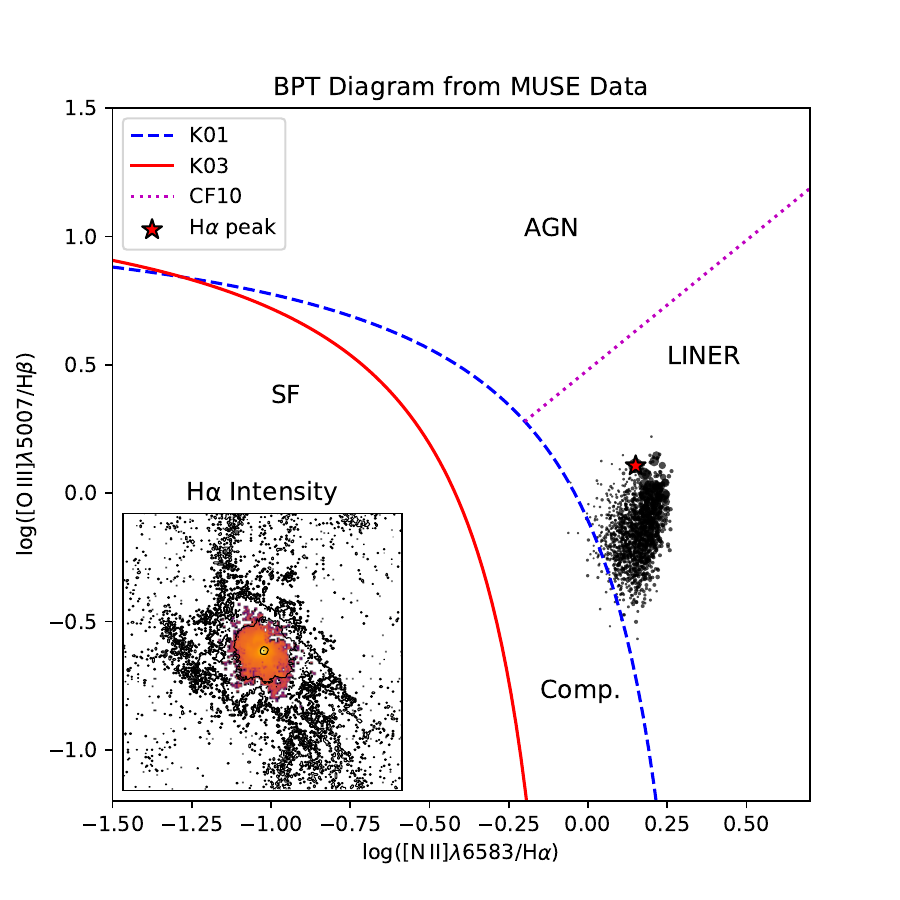}
    \includegraphics[width=0.245\linewidth, trim={0.2cm 0 1cm 1.7cm}, clip]{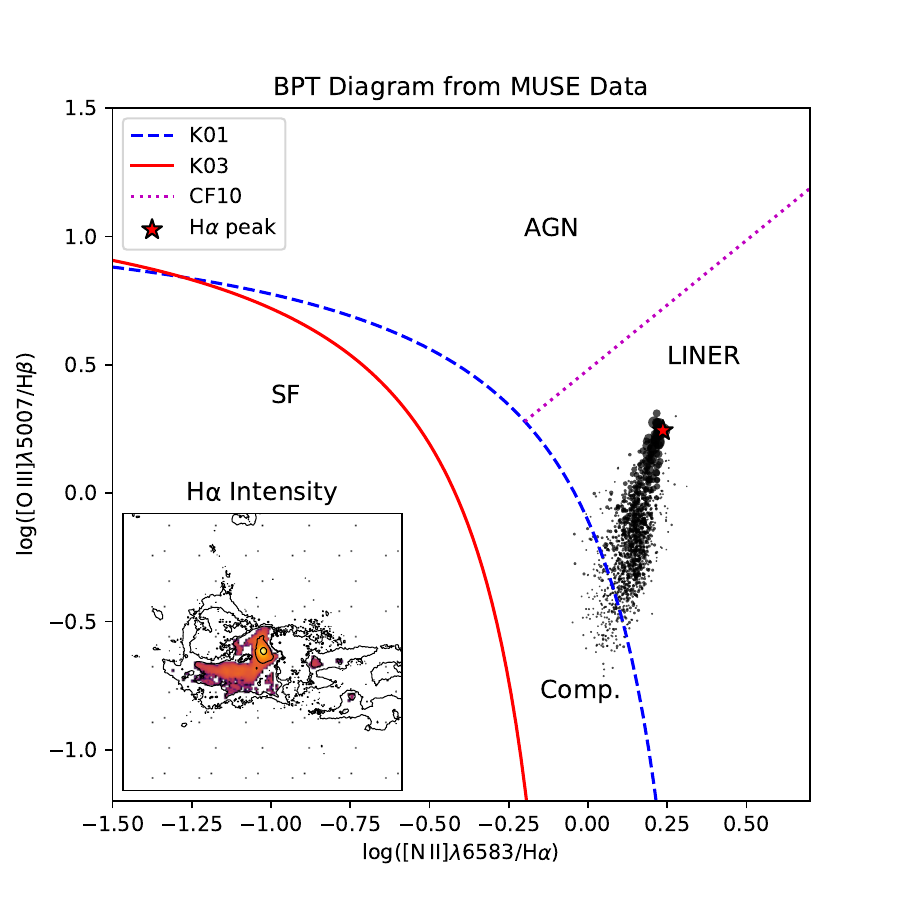}
    \includegraphics[width=0.245\linewidth, trim={0.2cm 0 1cm 1.7cm}, clip]{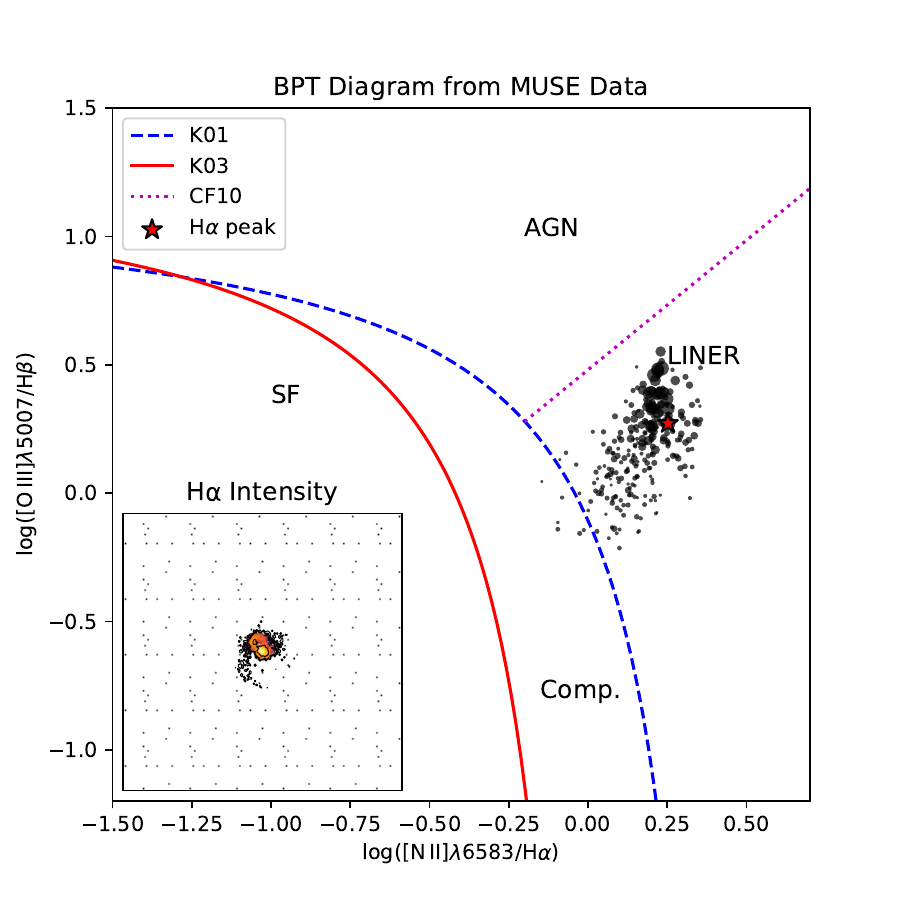}\\
    \includegraphics[width=0.245\linewidth, trim={0.2cm 0 1cm 1.7cm}, clip]{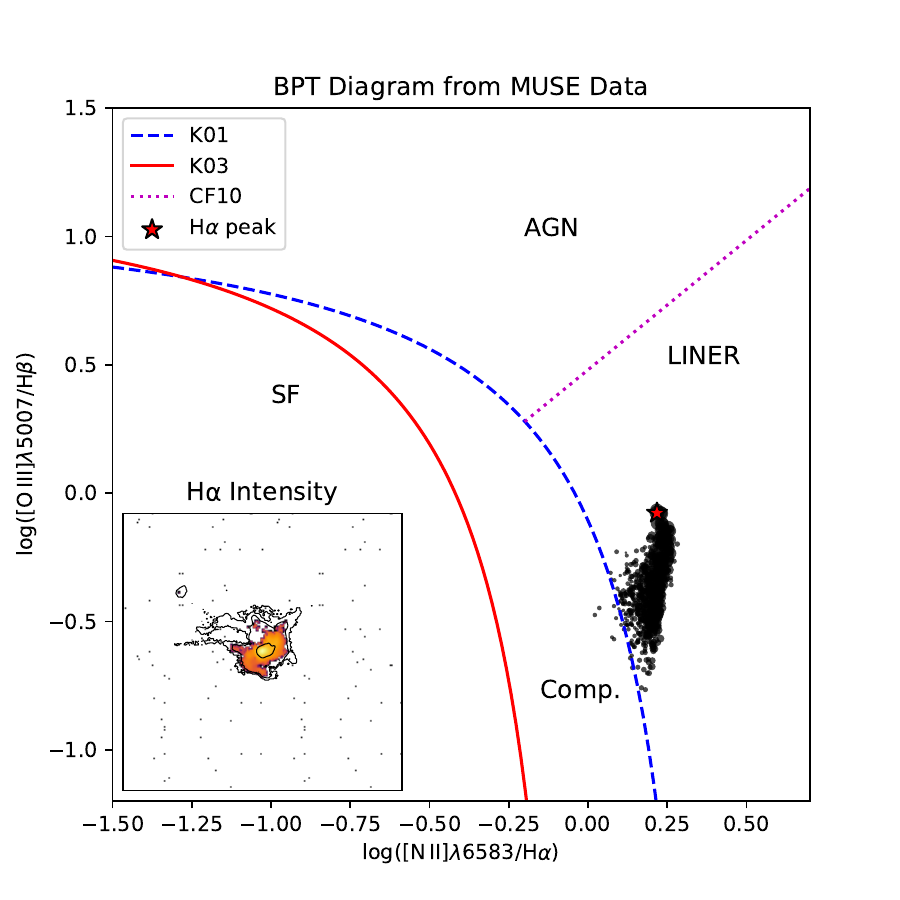}
    \includegraphics[width=0.245\linewidth, trim={0.2cm 0 1cm 1.7cm}, clip]{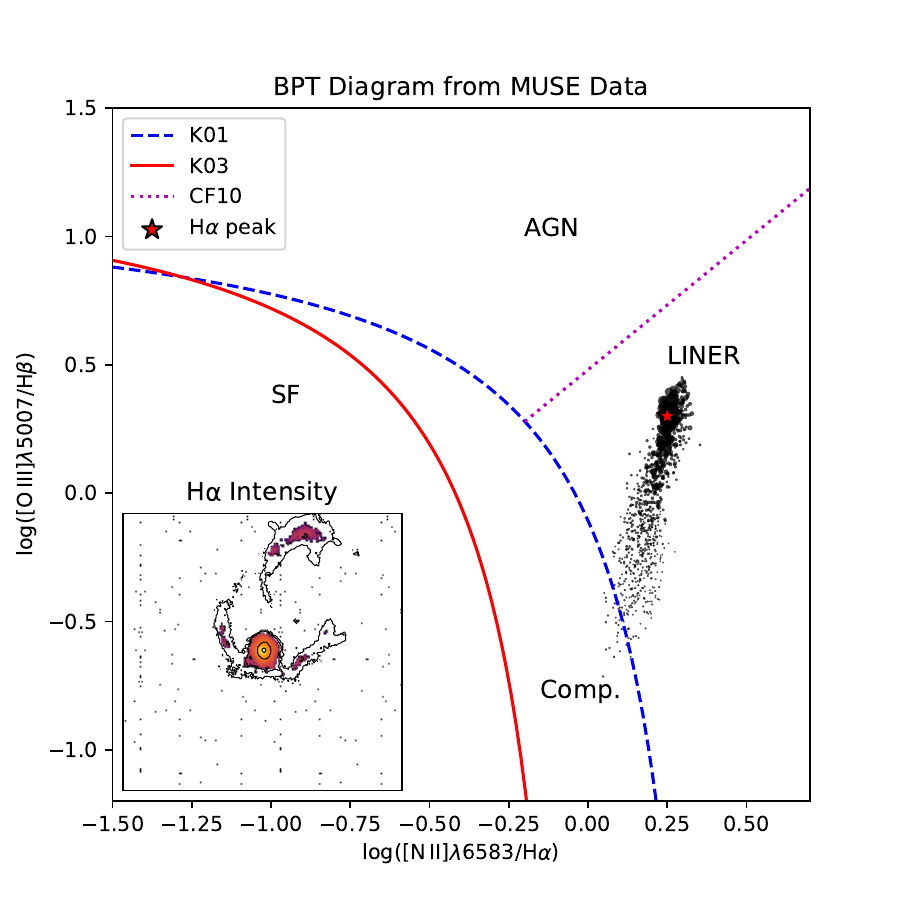}
    \includegraphics[width=0.245\linewidth, trim={0.2cm 0 1cm 1.7cm}, clip]{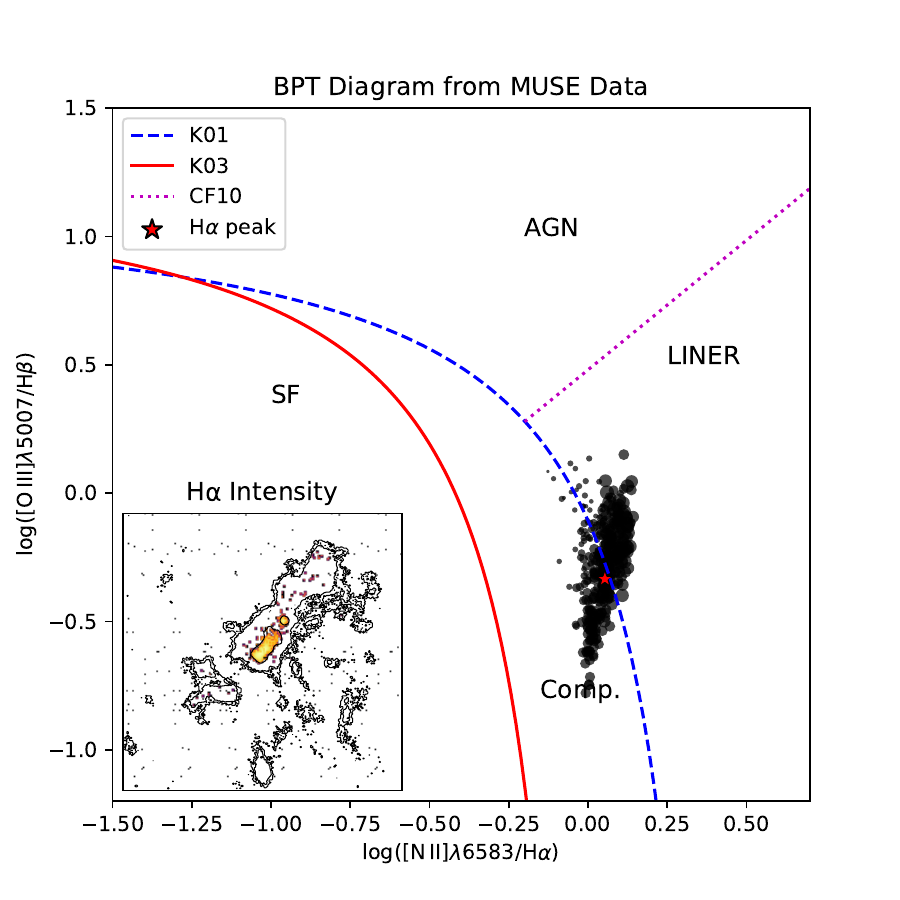}
    \includegraphics[width=0.245\linewidth, trim={0.2cm 0cm 1cm 1.7cm}, clip]{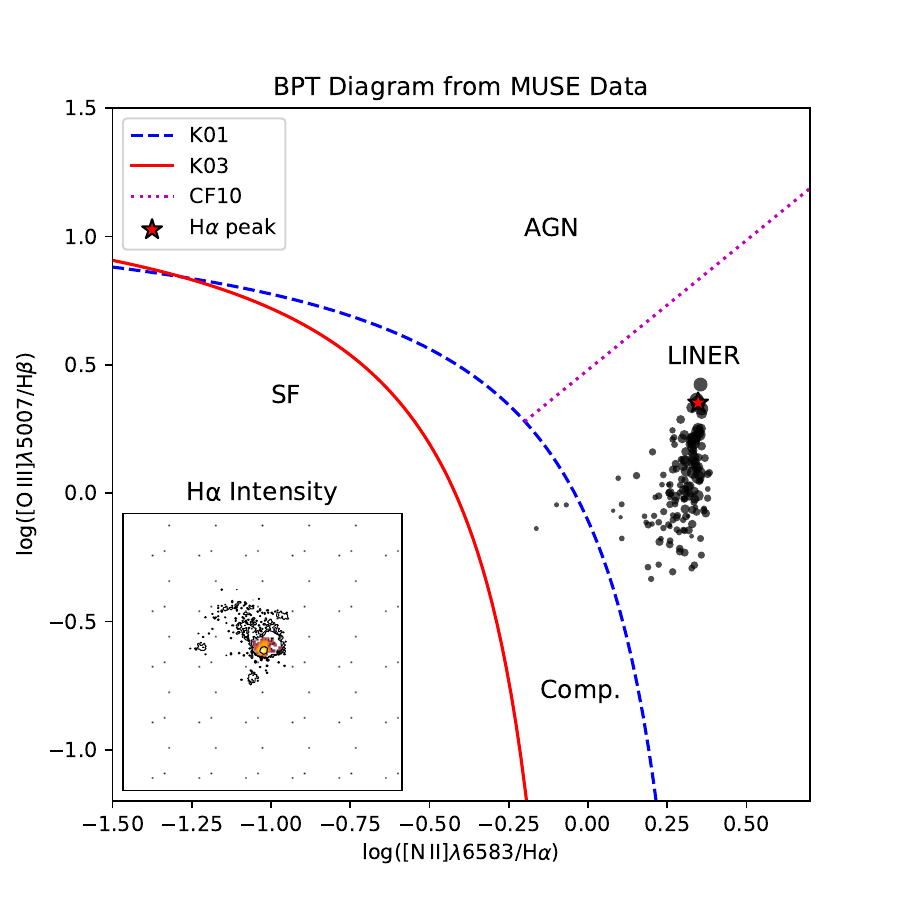}\\
        \includegraphics[width=0.245\linewidth, trim={0.2cm 0 1cm 1.7cm}, clip]{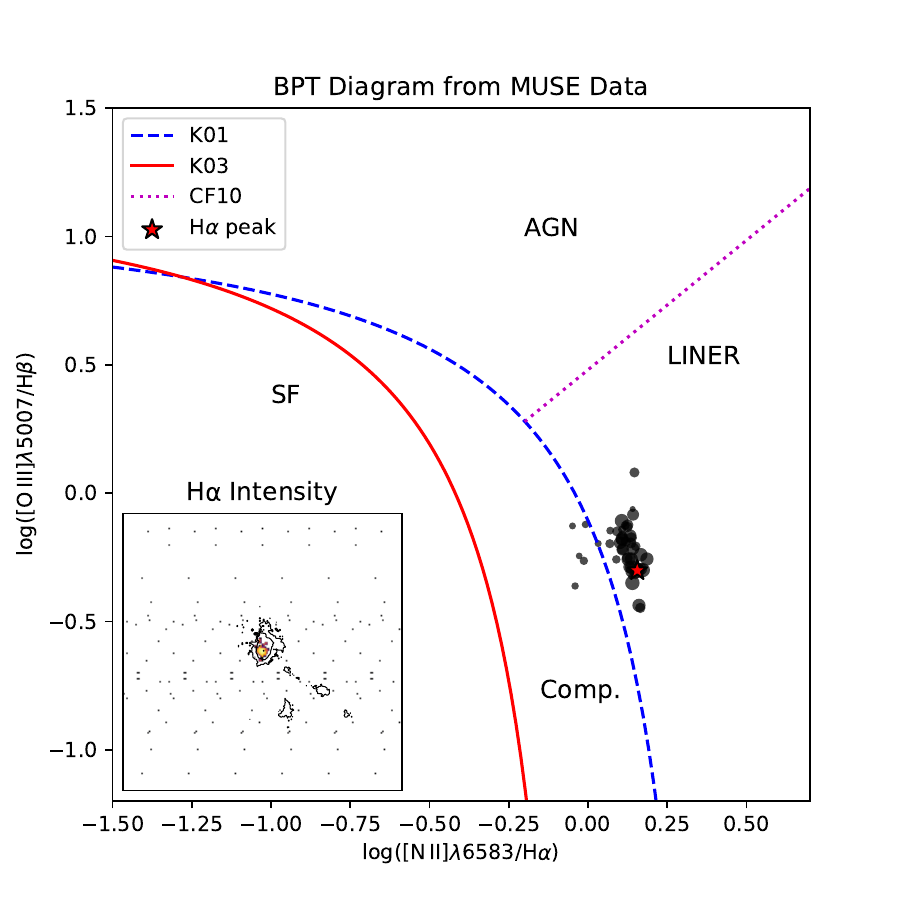}
    \includegraphics[width=0.245\linewidth, trim={0.2cm 0 1cm 1.7cm}, clip]{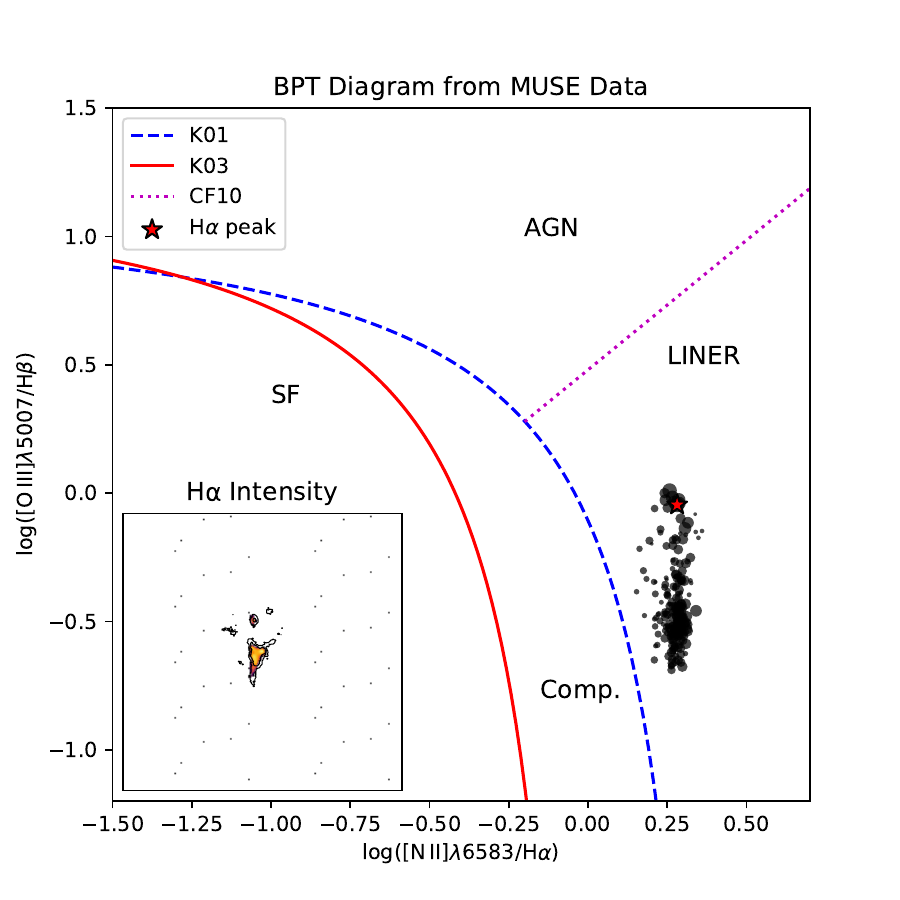}
    \includegraphics[width=0.245\linewidth, trim={0.2cm 0 1cm 1.7cm}, clip]{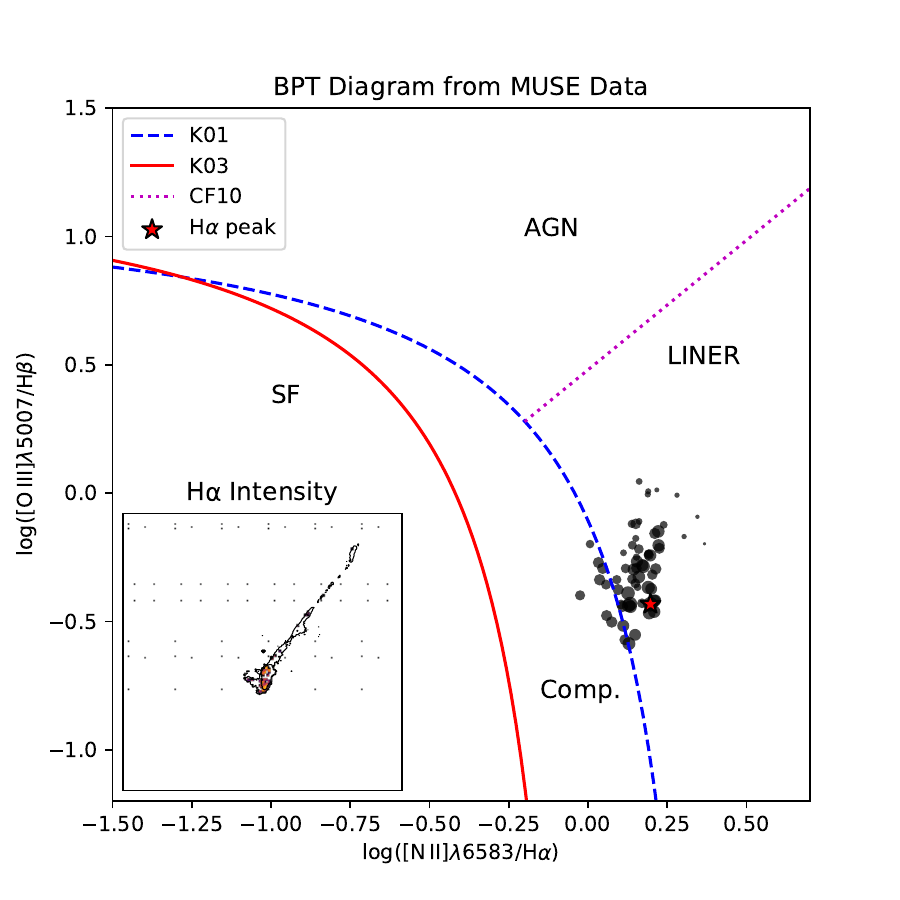}
    \includegraphics[width=0.245\linewidth, trim={0.2cm 0 1cm 1.7cm}, clip]{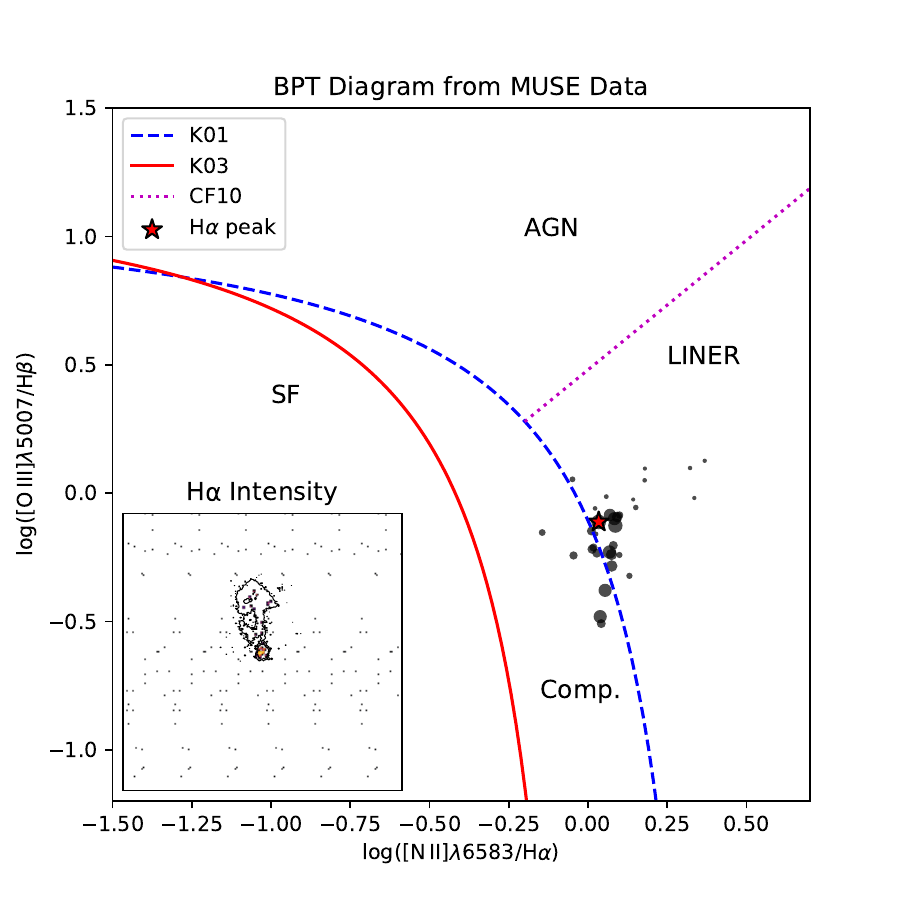}\\
    \includegraphics[width=0.245\linewidth, trim={0.2cm 0 1cm 1.7cm}, clip]{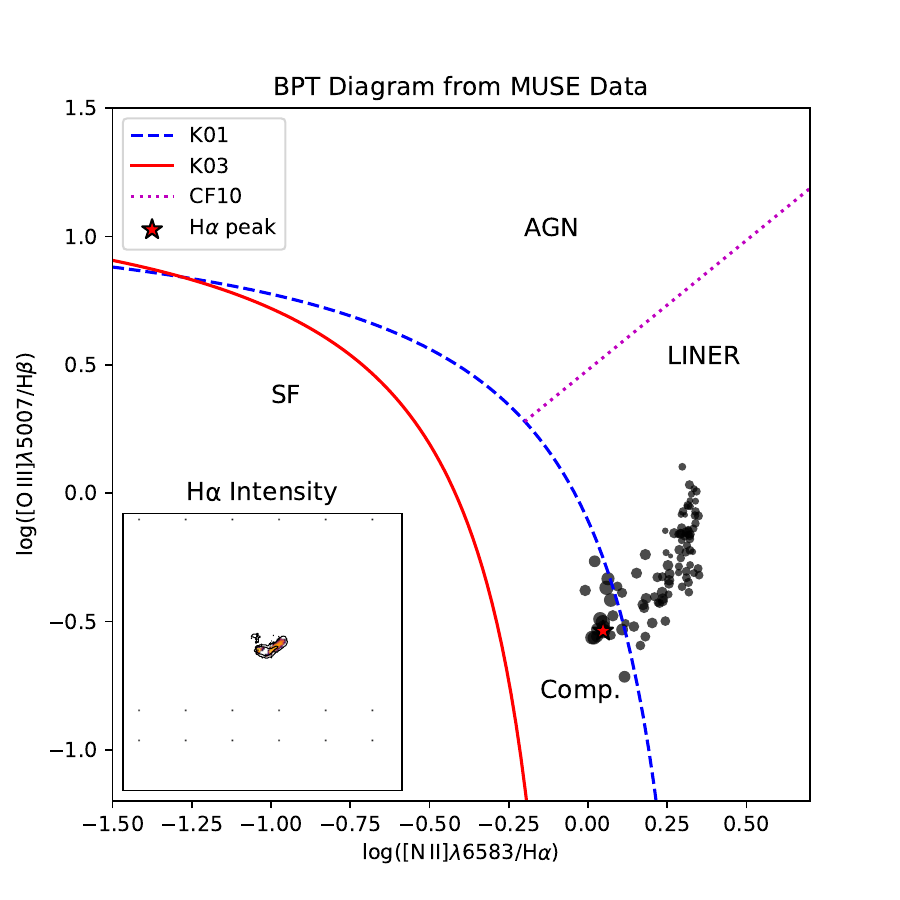}
    \includegraphics[width=0.245\linewidth, trim={0.2cm 0 1cm 1.7cm}, clip]{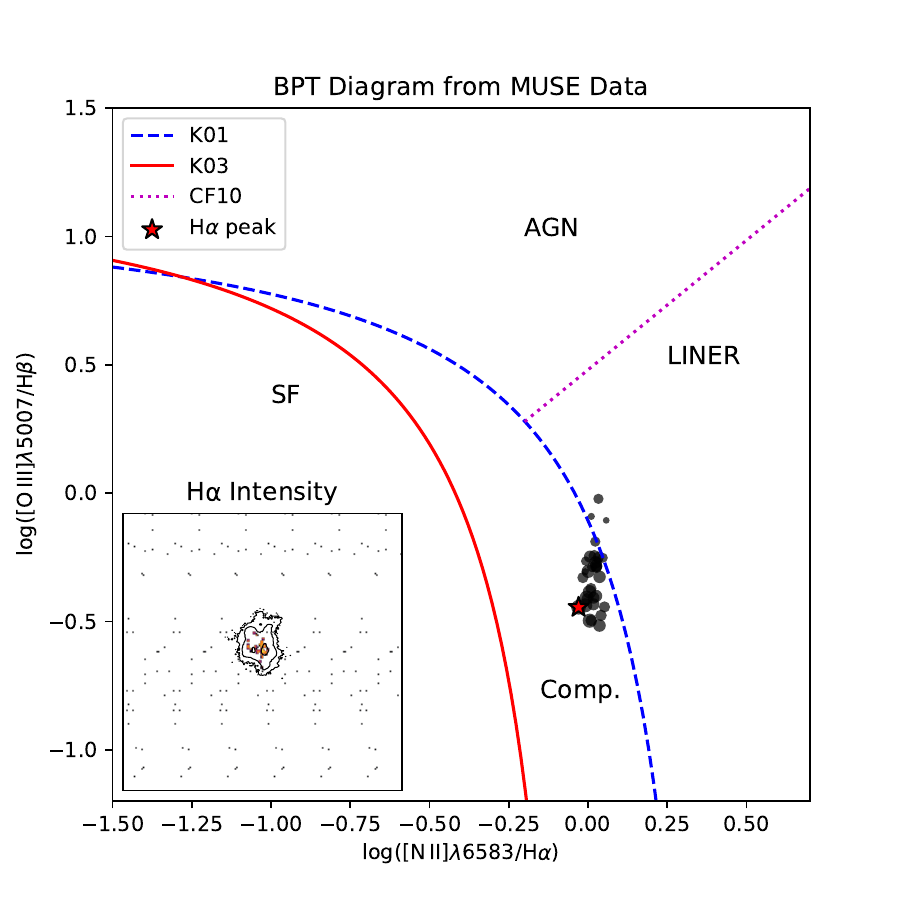}
    \includegraphics[width=0.245\linewidth, trim={0.2cm 0 1cm 1.7cm}, clip]{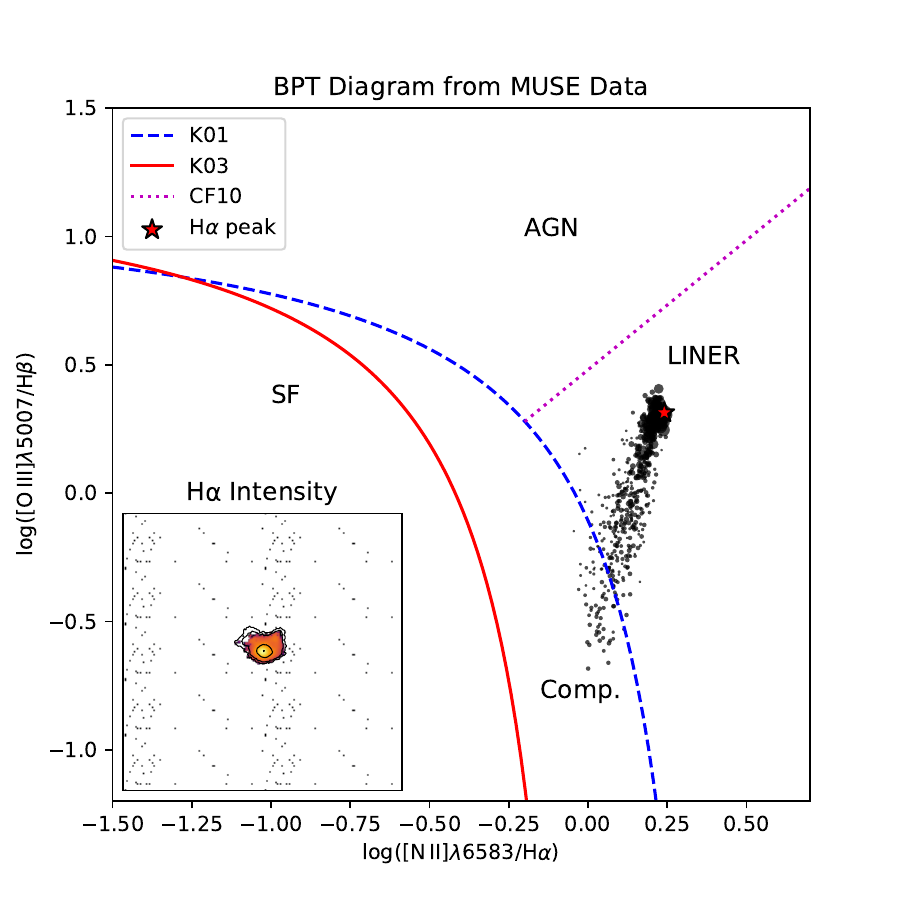}
    \includegraphics[width=0.245\linewidth, trim={0.2cm 0 1cm 1.7cm}, clip]{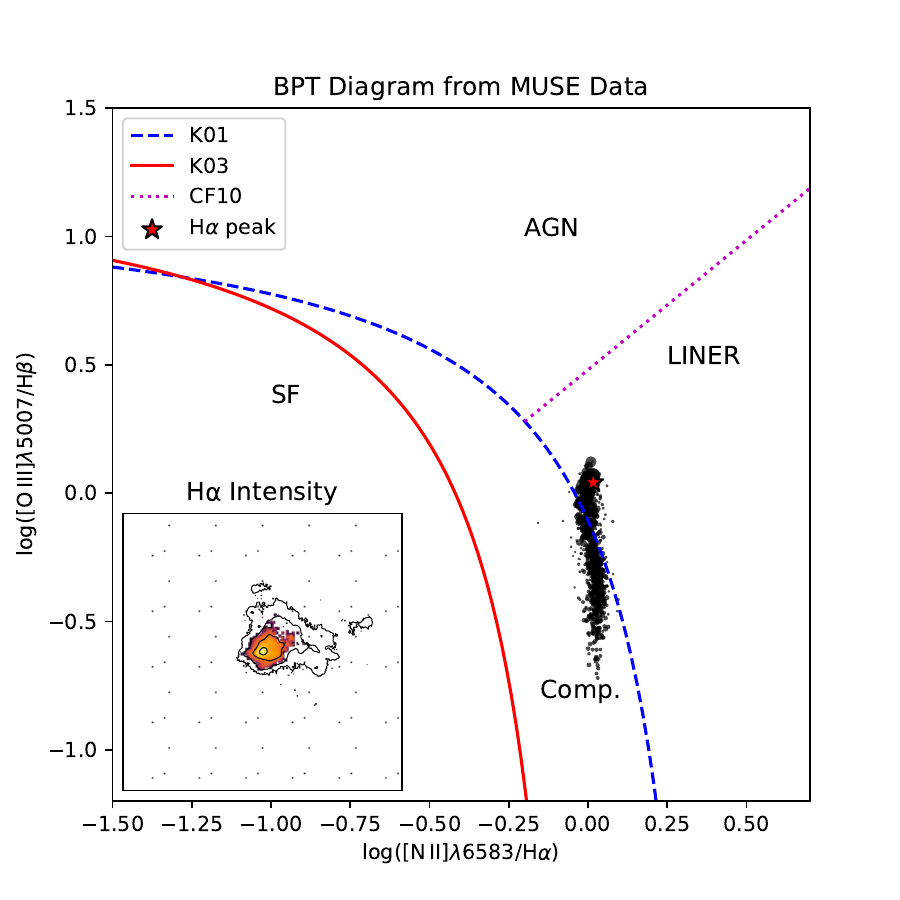}\\
    \caption{BPT diagram of the systems in our sample from MUSE data, showing the line ratio $\log([\mathrm{O\,III}]\,\lambda5007/\mathrm{H}\beta)$ versus $\log([\mathrm{N\,II}]\,\lambda6583/\mathrm{H}\alpha)$ for spaxels with S/N~$>3$ in all four lines. Systems that do not fulfill this requirement in any spaxel are marked. The standard demarcation curves from \citet{kewley2001} (K01), \citet{kauffmann2003} (K03), and \citet{cidfernandes2010} (CF10) are overlaid to distinguish between regions dominated by star formation, AGN photoionization, composite activity, and LINER-like excitation. The position of the H$\alpha$ peak is highlighted with a red star. The size of the black points is proportional to their H$\alpha$ intensity. The subplot in the left shows the H$\alpha$ intensity (black contours), and the subset of spaxels with S/N~$>3$ in all four lines (colors). The absence of points in the AGN region confirms that the warm gas emission is not dominated by AGN photoionization, supporting the interpretation of a non-AGN origin for the ionization near the H$\alpha$ peak. From left to right, top to bottom: NGC~5846, NGC~5044, Abell~3581, NGC~7237, Abell~496, Abell~2052, ZwCl~0335+096, Abell~1644, Abell~2626, Abell~85, Abell~133, Abell~1991, Abell~2495, Abell~478, MCXC~J1558.3-1410, PKS~0745-191.}
    \label{fig:bpt1}
\end{figure}
\begin{figure}[ht!]
    \centering
    \ContinuedFloat
    \includegraphics[width=0.245\linewidth, trim={0.2cm 0 1cm 1.7cm}, clip]{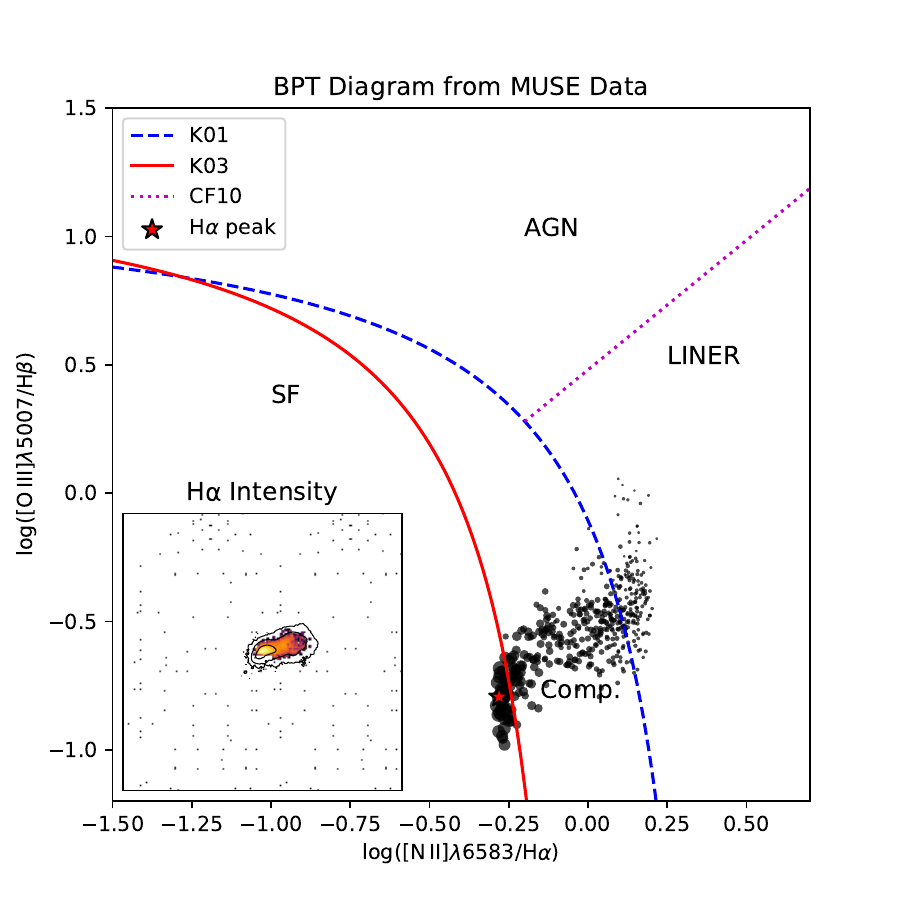}
    \includegraphics[width=0.245\linewidth, trim={0.2cm 0 1cm 1.7cm}, clip]{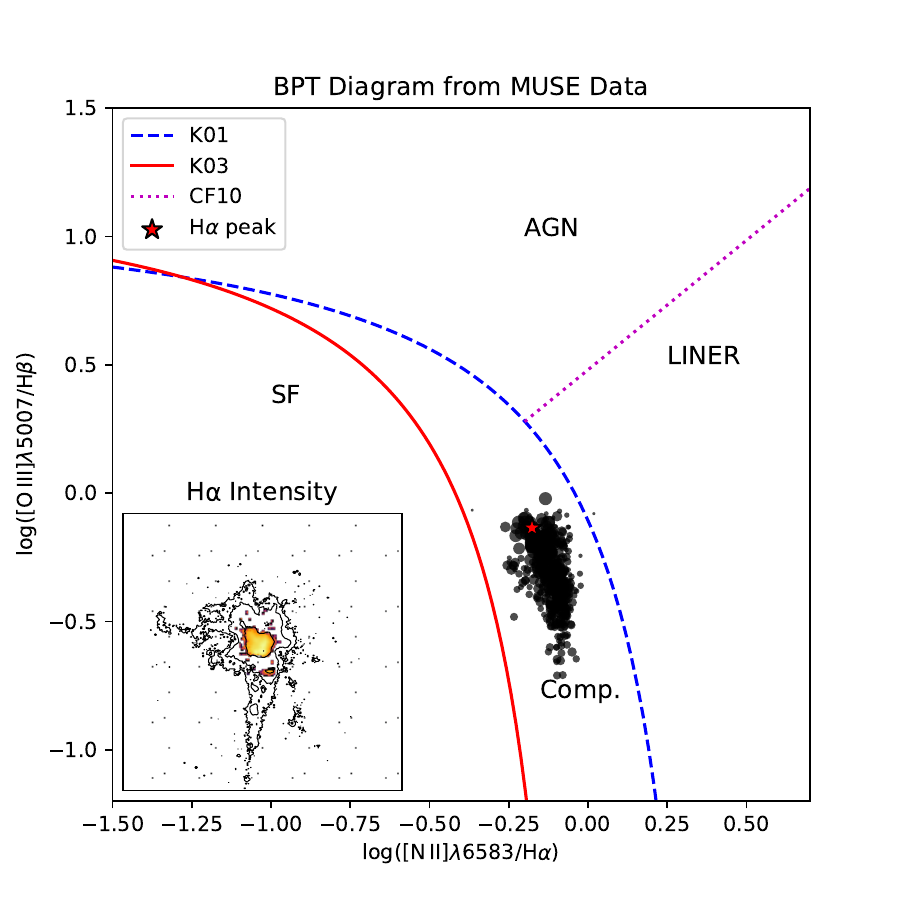}
    \includegraphics[width=0.245\linewidth, trim={0.2cm 0 1cm 1.7cm}, clip]{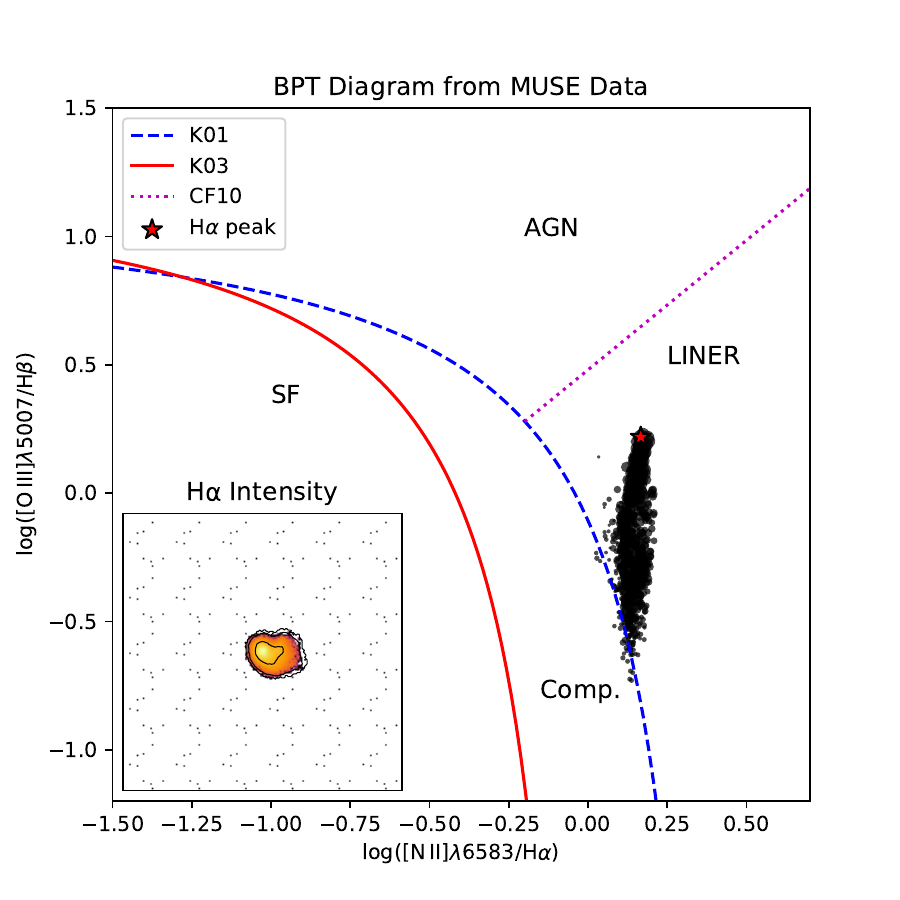}
    \includegraphics[width=0.245\linewidth, trim={0.2cm 0 1cm 1.7cm}, clip]{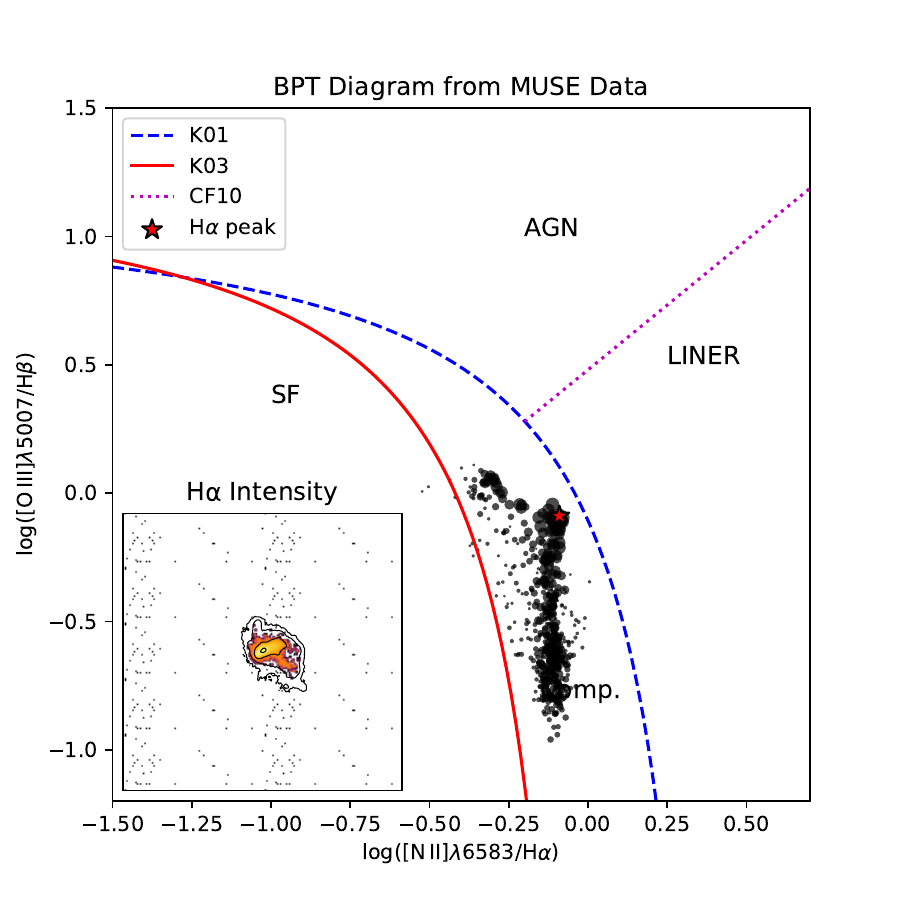}\\
    \includegraphics[width=0.245\linewidth, trim={0.2cm 0 1cm 1.7cm}, clip]{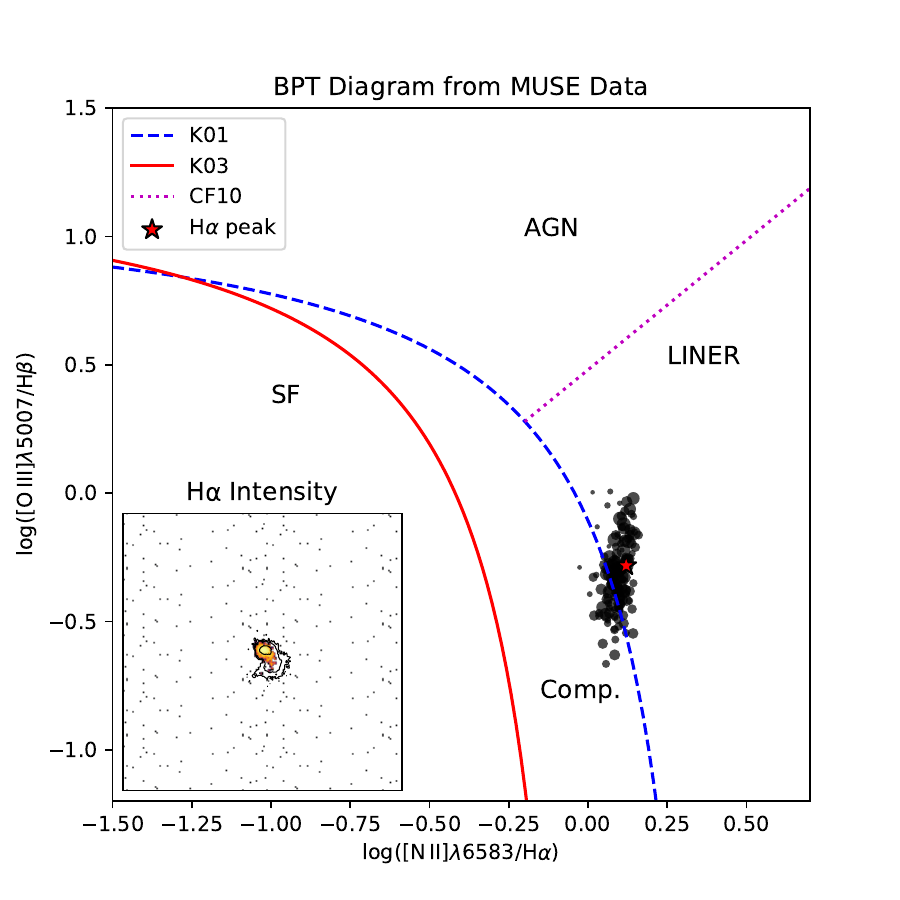}
    \includegraphics[width=0.245\linewidth, trim={0.2cm 0 1cm 1.7cm}, clip]{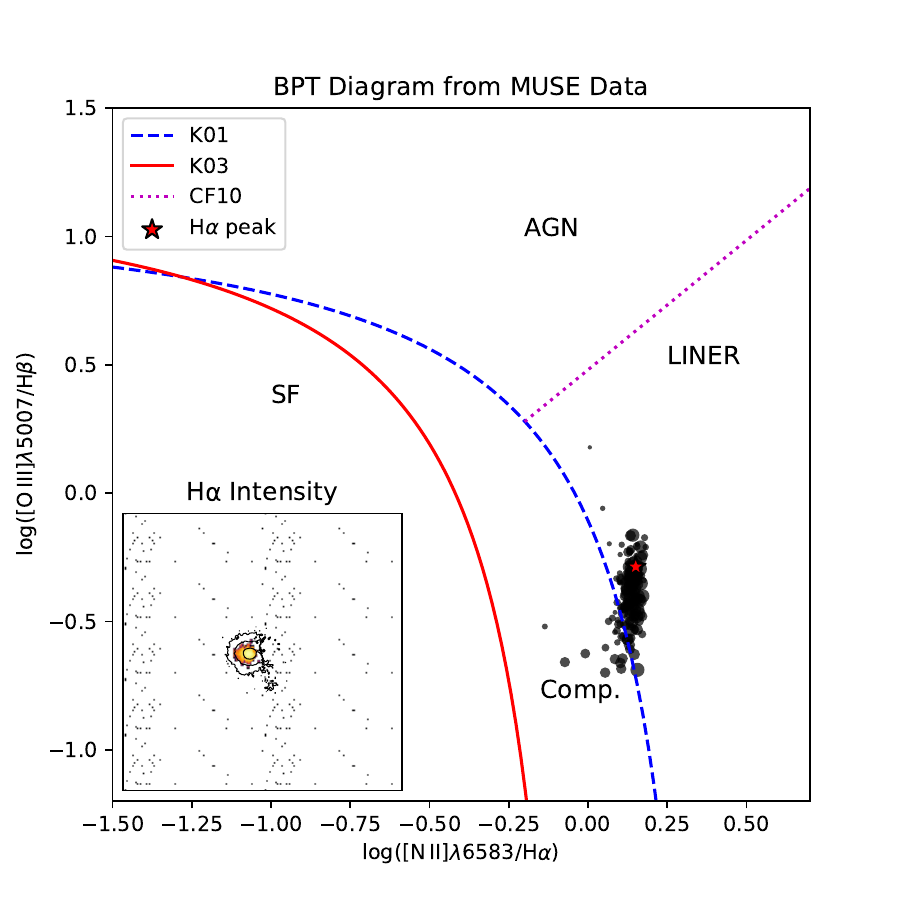}
    
    \caption{Continued. From left to right, top to bottom: RX~J0821.0+0752, Abell~1664, RX~J1350.3+0940, MCXC~J2014.8-2430, Abell~383, MCXC~J0331.1-2100.}
    \label{fig:bpt2}
\end{figure}


\bibliography{sample7}{}
\bibliographystyle{aasjournal}



\end{document}